\documentclass[3p,12pt]{elsarticle} 
\usepackage{amsmath}
\usepackage{bm}
\usepackage{amssymb}
\usepackage{color}
\usepackage{comment}
\usepackage{subcaption}
\usepackage{multirow}
\usepackage{caption}
\usepackage{url}
\usepackage{float}
\usepackage[titletoc]{appendix}

\usepackage{placeins}

\begin{document}
\title{Hemodynamics of the heart's left atrium based on a \\ Variational Multiscale-LES numerical method}

\author[polimi]{Alberto~Zingaro\corref{cor}}
\ead{alberto.zingaro@polimi.it}

\author[polimi]{Luca~Dede'}

\author[fm]{Filippo~Menghini}

\author[polimi,cmcs]{Alfio~Quarteroni}

\address[polimi]{MOX, Dipartimento di Matematica, Politecnico di Milano,
Piazza Leonardo da Vinci 32, 20133, Milan, Italy}

\address[fm]{Casale SA, Via Giulio Pocobelli 6, CH-6900 Lugano, Switzerland}

\address[cmcs]{Chair of Modelling and Scientific Computing (CMCS), Institute of Mathematics, \'Ecole Polytechnique F\'ed\'erale de Lausanne, Station 8, Av. Piccard, CH-1015 Lausanne, Switzerland}

\cortext[cor]{Corresponding author}

\journal{}

\begin{abstract}

In this paper, we investigate the hemodynamics of a left atrium (LA) by proposing a computational model suitable to provide physically meaningful fluid dynamics indications and detailed blood flow characterization. In particular, we consider the incompressible Navier-Stokes equations in Arbitrary Lagrangian Eulerian (ALE) formulation to deal with the LA domain under prescribed motion. A Variational Multiscale (VMS) method is adopted to obtain a stable formulation of the Navier-Stokes equations discretized by means of the Finite Element method and to account for turbulence modeling based on Large Eddy Simulation (LES). The aim of this paper is twofold: on one hand to improve the general understanding of blood flow in the human LA in normal conditions; on the other, to analyse the effects of the turbulence VMS-LES method on a situation of blood flow which is neither laminar, nor fully turbulent, but rather transitional as in LA. \color{black} Our results suggest that if relatively coarse meshes are adopted, the additional stabilization terms introduced by the VMS-LES method allow to better predict transitional effects and cycle-to-cycle blood flow variations than the standard SUPG stabilization 	method. \color{black}
\end{abstract}

\begin{keyword}
Left Atrium Hemodynamics \sep Finite Element method \sep VMS-LES \sep SUPG \sep Transition to Turbulence
\end{keyword}

\maketitle

\newpage 

\section{Introduction}
\label{INTRO}
\color{black}
In Western Countries, cardiovascular related diseases represent nowadays the first cause of death in the adult population \cite{MB_heart}. Non-invasive experimental techniques, such as phase-contrast magnetic resonance
imaging (PC-MRI) and computational tomography (CT) scans, allow to inspect the blood fluid-dynamics and displacement of blood vessels. These methods are widely used to better understand the complex physiology of the cardiovascular system as well as to investigate pathological conditions \cite{KWPP_mri,KAT_kinetic}. 
Cardiovascular diseases diagnosis can also be assessed through 4D flow magnetic resonance imaging (4D flow MRI) \cite{4DflowMRI}, a tool which provides 3D visualization of the blood flow along time. Differently from standard experimental techniques  \cite{ZPL_mri,CMN_leftheart}, 4D flow MRI allows to measure hemodynamics indicators as the wall shear stress (WSS) \cite{4DflowMRI_vs_CFD1}. However, such imaging based techniques - both standard and more advanced - do not allow to recover the spatial and temporal fine scales of these flows. Hence, they might not accurately catch typical flows features as small coherent structures, recirculation regions and possible regions of transition to turbulence, as pointed out in \cite{4DflowMRI_vs_CFD1}. For the aforementioned reasons, mathematical modeling and numerical simulations are largely employed to complement the available imaging techniques in an effort to better understand the physiology and pathology of the cardiovascular system \cite{QMC_review,QLRR_review}.

Literature is abundant concerning the fluid dynamics of the whole circulatory system, the study of heart valves, specific arteries and biomedical devices \cite{QMC_review,QLRR_review,MKKDL_ventr,WSKH_fsi,NMP_heartvalve, KS_dynamicsheartvalve, TDG_cerebral, Griffith_LVAD, Griffith_LVAD_2020, Peskin_2001}. 
By far, the most studied part of the heart is the left ventricle (LV), that has been considered from the electro-mechanical and fluid dynamical viewpoints,
both for idealized and patient-specific data \cite{MKKDL_ventr,SSB_el-mech,FPS_electro, mittal_diseased, mittal_mitralvalve}.
The LA is far less investigated, at least in normal conditions \cite{VGY_leftatrium,KFH_leftatrium, MAFM_atrium, MAFM_atrium_2, MAFM_atrium_3, Nordsletten_2018}.
Understanding the blood flow behavior in the LA can shed light on its functioning in physiological conditions and can be also regarded as a valuable step towards the study of the complete left heart. 

Idealized geometries for the numerical simulation of blood flows
offer the possibility of building a parametrized model that allows to obtain medical indicators for several patients without the need of performing expensive patient-specific simulations. To take into account the large geometrical inter-patient variability, an accurate idealized computational model of the LA can be parametrized based on patient-specific image acquisitions.  Another motivation behind the use of an idealized geometry with a prescribed kinematics, which we deduce from the Wiggers diagram \cite{Wiggers, Wiggers_Mitchell}, lays in the fact that patient-specific data for the atria in normal (physiological) conditions are scarce. Moreover, even if good quality kinematics images of the LA may become available, these would be typically acquired in individuals affected by pathological conditions, such as atrial fibrillation \cite{MAFM_atrium, MAFM_atrium_2, MAFM_atrium_3}.

An open issue in the blood fluid dynamics is whether a transition to turbulence occurs whenever the blood velocity increases and the interactions among vortices are strong. The Navier-Stokes equations are in principle suitable to model both transitional and turbulent flows. However, the spatial and temporal resolutions required to fully capture the details of the flow features through a Direct Numerical Simulation (DNS) for the discretized Navier-Stokes equations would require prohibitive computational resources \cite{WIL}.
For this reason, usually a turbulence model is employed, like e.g. 
the Reynolds Averaged Navier-Stokes equations (RANS models) and the Large Eddy Simulation (LES models) \cite{WIL,MEN,dynamicles}. 

From a theoretical point of view, in a fluid flow, it is possible to distinguish the eddies on the basis of their kinetic energy \cite{WIL, POPE}. The distribution of the kinetic energy as a function of the eddy length scale (or wave number ${k}$, when a Fourier transform is applied to the
energy spectrum) follows some well established findings in  homogeneous and  isotropic turbulence,
such as the $k^{-5/3}$ rule for the energy spectrum in the inertial range \cite{WIL,POPE}.
\color{black}
A DNS would allow to solve the whole range of spatio and temporal scales down to the Kolmogorov scales. 
\color{black}
In RANS models one solves for an average flow field in which only the large scale eddies containing the highest energy are considered, while the effect of the inertial range and of the fine scales is accounted by an extra term, called Reynolds stress, to be added to the momentum balance equation of the Navier-Stokes equations \cite{WIL, POPE}. When using isotropic models, the overall effect of the Reynolds stress term is to increase the viscosity of the  fluid with a turbulent viscosity that is added to the physical one. RANS models may become too dissipative and yielding to unrealistic flows when used in transitional or even laminar conditions. 
\color{black}
On the other hand, LES models aim at explicitly solving the large eddies of the flow by reaching the inertial range, while modeling the effect of the smallest eddies by exploiting self-similarity properties of the flow \cite{WIL, POPE}.
\color{black}
Stabilization methods of the Navier-Stokes equations to obtain a solution inf-sup stable and free of numerical instabilities evolved towards the formulation of a Variational Multiscale (VMS) framework, contextually yielding a LES model \cite{BCC_vmsles, Hughes_1995, HCS_2005, HMJ_2000, HOM_2001, HSF_2004}. 

In this work, we develop a computational model of the human LA based on the incompressible Navier-Stokes equations expressed in the ALE formulation; specifically, we prescribe a law of contraction and relaxation of the LA coherent with the features of the cardiac cycle. Our numerical study allows a full characterization of blood flow in the LA in normal conditions. Several meaningful fluid dynamics indicators are also provided.  We purposely use the VMS-LES method developed in \cite{BCC_vmsles} and later extended in \cite{FD_vmsles} to stabilize the numerical solution of the Navier-Stokes equations in ALE formulation and to simultaneously account for turbulence modeling, see e.g. \cite{ALE_Baz}. In particular, the formulation of \cite{FD_vmsles} considers space discretization based on Finite Element Method (FEM) \cite{Hughes_book, AQ_book}, time discretization based on (Backward Differentiation Formula) BDF  \cite{AQ_FS_RS} and quasi-static approximation of the fine scale solutions. We generate a reference solution on a very fine grid and we compare these results with those obtained on coarser mesh levels with the standard Streamline Upwind Petrov-Galerkin (SUPG) and with the VMS-LES stabilization method \cite{FD_vmsles}. \color{black} We show that the two methods yield similar results in terms of total kinetic energy and enstrophy based on the phase-averaged velocity field; moreover, as the mesh is refined, the effects of the LES model become less evident, as expected. However, especially for the coarsest mesh used in this study, remarkable differences are observed on the fluctuating kinetic energy, a suitable indicator of transitional effects and a proper measure of cycle-to-cycle variations: the VMS-LES method better captures these variations and impacts on the ability of the turbulence model to better predict the total kinetic energy peaks based on the instantaneous velocity field. \color{black} 

The outlook of the paper is as follows: in Section \ref{MATH} we recall the mathematical model, the numerical methods and the LA model that we propose based on physiological data. In Section \ref{GRIDGENERATION} we present the three mesh levels adopted, while in Section \ref{RESU} we report and discuss the numerical results obtained from the simulation run on the fine mesh (reference solution) in terms of phase-averaged flow properties. Moreover, we perform a mesh converge study along with a comparison between SUPG and VMS-LES stabilization methods. Finally, conclusions are drawn in Section \ref{CONCLU}.

\section{Mathematical model and numerical methods}
\label{MATH}
In this section we first review the Navier-Stokes equations in ALE framework, then we introduce our numerical methods and the turbulence models. Finally, we discuss the boundary conditions and the LA volume variation in time based on physiological data.

\subsection{The Navier-Stokes equations in ALE formulation and its numerical approximation}
In large vessels, as well as in the heart chambers, blood behaves as a Newtonian incompressible
fluid and the presence of small particles suspended and carried by the plasma can be neglected.
In moving domains the Navier-Stokes equations can be reformulated in an Arbitrary Lagrangian Eulerian (ALE)
framework with a mesh-moving technique \cite{DGH_ale,JT_ale}. 
In this work, we do not study the interactions between the 
fluid and the endocardium, but we consider that the solid-fluid interface has a prescribed velocity, which is equal to the fluid one with no-slip conditions on the wall. Moreover, we use a standard harmonic extension of the
displacement in the fluid domain in order to maintain a good mesh quality while moving it without the need of remeshing \cite{JT_ale}.

\subsubsection{The Navier-Stokes equations in ALE framework} \label{section_NS_ALE}
Let $\Omega_t \subset \mathbb{R}^d$ be the fluid domain at a specific time instant $t>0$, provided with a sufficiently regular boundary  $\Gamma_t$ oriented by outward pointing normal unit vector $\hat {\bm {n}}$. We denote as $\Gamma_{t}^D$ and $\Gamma_{t}^N$ the portions of the boundary where respectively Dirichlet and Neumann type boundary conditions are prescribed, with $\Gamma_t = \overline{\Gamma_t^D} \cup \overline{\Gamma_t^N}$ and $\overset{\circ}{\Gamma_t^D} \cap \overset{\circ}{\Gamma_t^N} = \emptyset$.  Let $\bm u$ be the fluid velocity and $p$ be the pressure field. The incompressible Navier-Stokes equations in ALE framework read:
\begin{align}
\nabla \cdot \bm u & =   0  &  \quad \text{ in } \Omega_t \times (0, T], \label{eq_div}\\
\rho \frac{\hat \partial \bm u}{\partial t} + \rho \left( \left( \bm{u} - \bm{u}^{\text{ALE}} \right) \cdot \nabla \right) \bm{u} -  \nabla \cdot \bm{\sigma} (\bm u, p) & =  \bm f & \quad \text{ in } \Omega_t \times (0, T],  \label{eq_ns} \\
\bm u & = \bm g & \quad \text{ on } \Gamma_t^D \times (0, T], \label{eq_dirichletbc} \\
\bm \sigma (\bm u, p) \bm{\hat n} & = \bm h & \quad \text{ on } \Gamma_t^N \times (0, T], \label{eq_neumannbc} \\
\bm u & = \bm u_0 & \quad \text{ in } \Omega_0 \times \{0\}.
\end{align}
In particular, $\frac{\hat \partial \bm u }{\partial t} =  \frac{\partial \bm u }{\partial t} + (\bm u_{\text{ALE}} \cdot \nabla ) \bm u  $ is the ALE derivative, $\rho$ the fluid density and $\bm{\sigma} (\bm u, p)$ the total stress tensor defined for Newtonian, incompressible and viscous fluids as
$
\bm \sigma (\bm u, p)=-p\bm I + 2 \mu \bm \varepsilon (\bm u),
$
being $\mu$ the dynamic viscosity and $\bm \varepsilon (\bm u)$ the strain-rate tensor defined as
$
\bm \varepsilon (\bm u) = \frac{1}{2} \left ( \nabla \bm u + \left ( \nabla \bm u \right )^T\right).
$
The function $\bm f$ is the forcing term, $\bm g$ and $\bm h$ are Dirichlet and Neumann data,  $\bm u_0$  the initial condition. 
We prescribe a velocity $\bm{g}^{\text{ALE}}$ on the whole boundary $\Gamma_t$ and we recover $\bm{u}^{\text{ALE}}$ in the whole domain at each time through an harmonic extension:
\begin{equation}
\begin{aligned}
- \nabla \cdot \left( \bm K \nabla \bm u^{\text {ALE}}\right) = & \,   \bm 0 &  \quad \text{ in } \Omega_t \times (0, T], \\
\bm u^{\text{ALE}} = & \, \bm g^{\text{ALE}} & \quad \text{ on } \Gamma_t \times (0, T],
\end{aligned} \label{ALE_laplacian}
\end{equation}
where $\bm{K}$ is a positive-definite tensor that can be properly set to better tune the harmonic extension operator, for example
depending on the local spatial scales as done in \cite{JT_ale}. Finally, the domain displacement $\bm d (\bm x, t)$ is obtained integrating over time the ALE velocity: 
$
\bm d (\bm x, t) = \int_{0}^t \bm u^{\text{ALE}} (\bm x, \tau) d \tau \, .
$
\noindent
We introduce the infinite dimensional function spaces:
\begin{align}
 \mathcal V_{\bm g} := & \,  \{ \bm v \in [H^1(\Omega_t)]^d: \bm v = \bm g \text{ on } \Gamma_t^D\}, \\
 \mathcal Q := & \,  L^2(\Omega_t),
\end{align}
to define the weak formulation of the Navier-Stokes equations in ALE framework, which reads: 

given $\bm u_0$, for any $t \in (0, T]$, find $(\bm u, p) \in \mathcal V_{\bm g} \times \mathcal Q$ such that: 
\begin{equation}
\begin{split}
\left ( \bm v, \rho \frac{\hat \partial \bm u}{\partial t}\right ) + 
\left ( \bm v, \rho (\bm u - \bm u^{\text{ALE}}) \cdot \nabla \bm u\right) + 
\left (\nabla \bm v, \mu \nabla \bm u \right) -
\left ( \nabla \cdot \bm v, p \right) + 
\left ( q, \nabla \cdot \bm u \right ) = 
\\
\left (\bm v, \bm f\right ) +
\left (\bm v, \bm h\right )_{\Gamma_t^N}, 
\quad
\text{ for all } (\bm v, q) \in \mathcal V_{\bm 0} \times \mathcal Q. 
\end{split}
\label{eq_ns-weak}
\end{equation}   
We have denoted with $(\cdot, \cdot)$ and $(\cdot, \cdot)_{\Gamma_t^N}$ the $L^2$ inner product with respect to $\Omega_t$ and $\Gamma_t^N$ respectively. \color{black} Equivalently, in a more compact form, by introducing the function space $ \bm{\mathcal V}_{\bm g} := \mathcal{V}_{\bm g} \times \mathcal{Q}$: 

given $\bm u_0$, for any $t \in (0, T]$, find $\bm U = \bm U (t) = \{\bm u, p\} \in \bm{\mathcal V}_{\bm g}$ such that: 
\begin{equation}
	A(\bm V, \bm U) = F(\bm V), \quad \text{ for all } \bm V = \{\bm v, q\} \in \bm{\mathcal V}_{\bm 0},
	\label{weak_compact}
\end{equation}
being
\begin{subequations}
	\begin{align}
		A (\bm V, \bm U) & = A_1(\bm V, \bm U) + A_2(\bm V, \bm U, \bm U), \label{formA}\\ 
		A_1(\bm V, \bm U) & =  \left ( \bm v, \rho \frac{\hat{\partial} \bm u}{\partial t}\right ), + 
		\left (\nabla \bm v, \mu \nabla \bm u \right) -
		\left ( \nabla \cdot \bm v, p \right) + 
		\left ( q, \nabla \cdot \bm u \right ), \\
		A_2(\bm U, \bm V, \bm W) & =  \left ( \bm u, \rho  \left ( \bm v - \bm u^\text{ALE}\right ) \cdot \nabla \bm w\right), \\
		F(\bm V) & = \left (\bm v, \bm f\right ) +
		\left (\bm v, \bm h\right )_{\Gamma_N}.
	\end{align}
\end{subequations}   

\color{black}

\subsubsection{Numerical methods and turbulence modeling} \label{SEC:turbulence}
For the space discretization of Eq. \eqref{eq_ns-weak}, we introduce a finite element (FE) discretization with piecewise Lagrange polynomials of degree $r \geq 1$. The function space of the FE is $X_r^h = \{ v^h \in C^0(\overline{ \Omega}_t):\,  v^h|_K \in \mathbb{P}_r,\,  \forall K \in \mathcal T_h\}$, being $\mathcal T_h$ a triangulation of $\Omega_t$ and $h$ the diameter of the grid element $K \in \mathcal T_h$. 
\color{black}
In the VMS method, one assumes a direct sum decomposition of both trial and test function spaces into coarse and fine scales subspaces as
$ \bm{\mathcal V}_{\bm g} = \bm{\mathcal V}_{\bm g}^h \oplus \bm{\mathcal V}_{\bm g}'\, , 
\bm{\mathcal V}_{\bm 0} = \bm{\mathcal V}_{\bm 0}^h \oplus \bm{\mathcal V}_{\bm 0}'$ \cite{Hughes_1995, HCS_2005, HMJ_2000, HOM_2001, HSF_2004}.  
Specifically, $\bm{\mathcal V}_{\bm g}^h = \mathcal V_{\bm g}^h \times \mathcal Q^h$, $\bm{\mathcal V}_{\bm 0}^h = \mathcal V_{\bm 0}^h \times \mathcal Q^h$ are the coarse scale function spaces, with  $\mathcal V_{\bm g}^h = \mathcal V_{\bm g} \cap [X_r^h]^d$, $\mathcal V_{\bm 0}^h = \mathcal V_{\bm 0} \cap [X_r^h]^d$ and $\mathcal Q^h = \mathcal Q \cap X_r^h$. While $\bm{\mathcal V}_{\bm g}' = \mathcal V_{\bm g}' \times \mathcal Q'$, $\bm{\mathcal V}_{\bm 0}' = \mathcal V_{\bm 0}' \times \mathcal Q'$ are the infinite-dimensional function spaces which represent the fine scale solution.

In this way, we introduce an a priori splitting of the solution (and test functions) into coarse (resolved) and fine (or subgrid, modelled) scales as :
\begin{equation}
		\bm U = \bm U ^ h + \bm U ', \quad
		\bm V  = \bm V ^ h + \bm V ' \, .
	\label{splitting}
\end{equation}
Accordingly, the superscripts $(\cdot)^h$ and $(\cdot)'$ denote the projections of $\bm U$ and $\bm V$ on coarse scale and fine scale solution spaces, respectively, with 
$\bm U^h \in \bm{\mathcal{V}}_{\bm g}^h$, 
$\bm U' \in \bm{\mathcal{V}}_{\bm g}'$ 
$\bm V^h \in \bm{\mathcal{V}}_{\bm 0}^h$ and 
$\bm V' \in \bm{\mathcal{V}}_{\bm 0}'$.
Using the decomposition \eqref{splitting} in Eq. \eqref{weak_compact}, one gets the following coupled coarse-scale and fine-scale equations:
\begin{align}
	A(\bm V^h , \bm U^h + \bm U') = F(\bm V^h),
	\label{coarse_eq}
	\\
	A(\bm V' , \bm U^h + \bm U') = F(\bm V').
	\label{fine_eq}
\end{align}
Following \cite{BCC_vmsles}, it can be shown that Eq. \eqref{coarse_eq} reduces to:

given $\bm u_0$, for any $t \in (0, T]$, find $\bm U^h = \bm U^h (t) = \{\bm u^h, p^h\} \in \bm{\mathcal V}_{\bm g}^h$ such that: 
\begin{equation}
	\begin{split}
		\left ( \bm v^h, \rho \frac{\partial \bm u^h}{\partial t}\right ) + 
		\left ( \bm v^h, \rho ( (\bm u^h - \bm u^\text{ALE}) \cdot \nabla) \bm u^h\right) + 
		\left (\nabla \bm v^h, \mu \nabla \bm u^h \right) -
		\left ( \nabla \cdot \bm v^h, p^h \right) + 
		\left ( q^h, \nabla \cdot \bm u^h \right ) 
		\\
		\underbrace{- \left ( \rho \bm u^h \cdot \nabla \bm v^h + \nabla q^h, \bm u'\right)
			- \left(\nabla \cdot \bm v^h, p'\right)}_{\text{(I)}}
		\underbrace{- \left(\rho \bm u^h \cdot (\nabla \bm w^h)^T, \bm u' \right)}_{\text{(II)}}
		\underbrace{- \left(\rho \nabla \bm v^h, \bm u' \otimes \bm u' \right)}_{\text{(III)}}
		\\
		= 
		\left (\bm v^h, \bm f\right ) +
		\left (\bm v^h, \bm h\right )_{\Gamma_N}, 
		\quad
		\text{ for all } \bm V^h = \{\bm v^h, q^h\} \in \bm{\mathcal V}_{\bm 0}^h. 
	\end{split}
	\label{coarse_withfinesolution}
\end{equation}  

In Eq. \eqref{coarse_withfinesolution}, the first and last rows contain standard terms of the Navier-Stokes equation. The second row contains additional stabilization terms, namely (I) the Streamline Upwind Petrov Galerkin (SUPG) term, (II) an additional stabilization term arising from VMS modeling and (III) the LES term which models the Reynolds stress \cite{BCC_vmsles, FD_vmsles}. We also observe that
the fine-scale solution $\bm U' = \{ \bm u', p'\}$ is still defined in an infinite dimensional function space. 

\begin{figure}[t!]
	\centering%
	\includegraphics[trim={0 0 0 0},clip,width=1.0\textwidth]{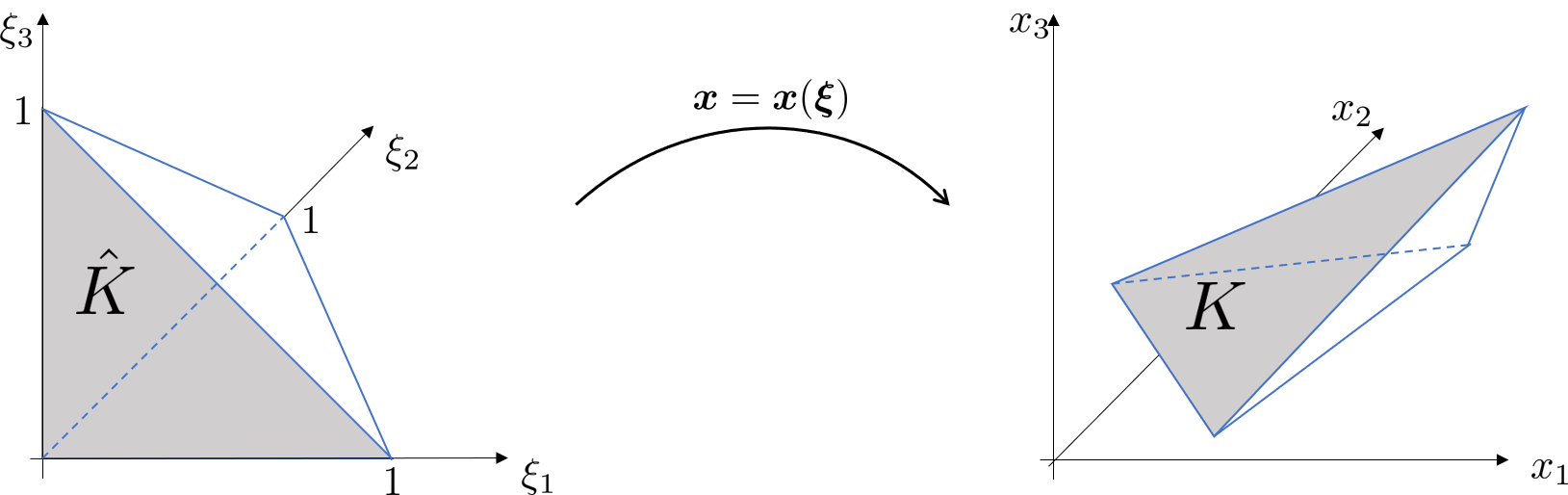}
	\caption{\label{parametric_global} \color{black} Mapping $\bm x = \bm x (\bm \xi ): \, \hat{K} \to K$ from the parametric element $\hat K$ to the physical one $K$. \color{black}}
\end{figure}

Following analogous arguments adopted for the coarse-scale equation, the solution $\bm U'$ of the fine scale equation (Eq. \eqref{fine_eq}) can be represented in terms of the coarse-scale solution $\bm U^h$ and the residual $\bm R(\bm U^h)$ of the coarse-scale equation projected onto the fine-scale space $\bm{\mathcal{V}}_{\bm 0}'$ \cite{BCC_vmsles}:
\begin{equation}
	\bm U ' = \bm{\mathcal{F}'}(\bm U^h,\bm{R}(\bm U^h) ).
\end{equation}
The latter can be inserted in Eq. \eqref{coarse_eq} to close finally the coarse-scale equation:
\begin{equation}
	\text{find } \, \bm U^h \in \bm{\mathcal{V}}_{\bm g}^h \, : \quad A(\bm V^h , \bm U^h +\bm{\mathcal{F}'}(\bm U^h,\bm{R}(\bm U^h) ) = F(\bm V^h), \quad \text{for all } \bm V^h \in \bm{\mathcal{V}}_{\bm 0}^h.
	\label{coarse_eq_withres}
\end{equation}
In order to find a numerical solution of Eq. \eqref{coarse_eq_withres}, one needs to approximate the differential operator with $\widetilde{\bm{\mathcal{F}'}} \approx \bm{\mathcal{F}'}$, which will lead to an approximation of both coarse and fine-scale solutions, namely $\widetilde{\bm U^h} \approx \bm U^h$ and $\widetilde{\bm U'}\approx \bm U'$. However, for the sake of simplicity, from now on we will refer to differential operator and solutions without the superscript $\sim$.  In particular, for the approximation of $\bm{\mathcal{F}'}$, we choose a \textit{quasi-static} approach yielding to the following approximation of fine velocity and pressure scales (or subgrid scales) \cite{BCC_vmsles,FD_vmsles}:\color{black}
\begin{align}
\bm u' & \simeq - \tau_{\text M}(\bm u^h) \bm r_{\text M} (\bm u^h, p^h) \label{eq_fine_v}\\
p' & \simeq - \tau_{\text C} (\bm u^h) r_{\text C}(\bm u^h), \label{eq_fine_p}
\end{align}
being 
$\bm r_{\text M} (\bm u^h, p^h)$ and $r_{\text C}(\bm u^h)$ the strong residuals of (\ref{eq_ns}) and (\ref{eq_div}) defined respectively as:
\begin{align}
\bm r_{\text M} (\bm u^h, p^h) & =  \rho \frac{\hat \partial \bm u^h}{\partial t} + \rho \left( \left( \bm{u}^h - \bm{u}^{\text{ALE}} \right) \cdot \nabla \right) \bm{u}^h -  \nabla \cdot \bm{\sigma} (\bm u^h, p^h)  - \bm f \\
r_{\text C}(\bm u^h) & = \nabla \cdot \bm{u}^h
\end{align}
The stabilization parameters are chosen as in \cite{BCC_vmsles,FD_vmsles}:
\begin{align}
\tau_{\text M}(\bm{u}^h) &= \left( \frac{\rho^2 }{\Delta t^2} + 
\rho^2 \, (\bm u^h - \bm u^{\text{ALE}}) \cdot \tilde{\bm{G}}(\bm u^h - \bm u^{\text{ALE}})  + 
C_r \mu^2 \tilde{\bm{G}} :\tilde{\bm{G}} \right)^{-\frac{1}{2}} \,, \\
\tau_{\text C}(\bm{u}^h) &= \left ( \tau_M(\bm{u}^h) \tilde{\bm{g}} \cdot \tilde{\bm{g}} \right )^{-1} \,,
\end{align}
being $\Delta t$ the time step that will be used for the time discretization and $C_r=15\cdot{2^{r}}$ is a constant obtained by an inverse inequality depending on the polynomial degree $r$ \cite{BCC_vmsles, FD_vmsles}. Moreover,  $\tilde{\bm{G}}$ is the metric tensor and $\tilde{\bm{g}}$ the metric vector:
\begin{equation}
\tilde{G}_{ij} = \sum_{k=1}^d \frac{\partial \xi_k }{\partial x_i} \frac{\partial \xi_k }{\partial x_j}, \quad \tilde{g}_i = \sum_{j=1}^{d} \frac{\partial \xi_j}{\partial x_i},
\label{metric}
\end{equation}
whereas\color{black}, as depicted in Figure \ref{parametric_global}\color{black}, we denote with $\bm x = \left \{ x_i\right \}_{i=1}^d$ the coordinates of the mesh element $K$ in the physical space and with $\bm \xi = \left \{ \xi_i\right \}_{i=1}^d$ the coordinates of element $\hat{K}$ in the parametric space. Let $\bm x = \bm x (\bm \xi ): \, \hat{K} \to K$ be a continuous and differentiable mapping from the parametric to the physical space, with a continuously differentiable inverse. $\frac{\partial \bm \xi}{\partial \bm x}$ in Eq. \eqref{metric} is the inverse Jacobian of the mapping \cite{BCC_vmsles}. 
The semi-discrete variational multiscale formulation with LES modeling of the Navier-Stokes equations in ALE framework reads: 

given $\bm u_0$, for any 
$t \in (0, T]$, find $ (\bm {u}^h,p^h) \in \mathcal {V}_{\bm g}^h \times Q^{h}$  such that:
\begin{equation}
\begin{aligned}
& \left ( \bm v^h, \rho \frac{\hat \partial \bm u^h}{\partial t}\right ) + 
\left ( \bm v^h, \rho \left ( (\bm u^h - \bm u^{\text{ALE}}) \cdot \nabla \right ) \bm u^h\right) + 
\left (\nabla \bm v^h, \mu \nabla \bm u^h \right) -
\left ( \nabla \cdot \bm v^h, p^h \right) + 
\left ( q^h, \nabla \cdot \bm u^h \right ) 
\\
& + {\left ( \rho (\bm u^h - \bm u^{\text{ALE}}) \cdot \nabla \bm v^h + \nabla q^h , \, \tau_{\text M}(\bm u^h ) \bm r_{\text M} (\bm u^h, p^h)\right ) +
 \left (\nabla \cdot \bm v^h, \tau_{\text C}(\bm u^h)r_{\text C}(\bm u^h)\right)}
\\
& + {\left(\rho \bm u^h \cdot (\nabla \bm v^h)^T ,  \tau_{\text M}(\bm u^h)\bm r_{\text M}(\bm u^h, p^h) \right )}
\\
& - {\left( \rho \nabla \bm v^h,  \tau_{\text M}(\bm u^h)\bm r_{\text M}(\bm u^h, p^h) \otimes  \tau_{\text M}(\bm u^h)\bm r_{\text M}(\bm u^h, p^h) \right ) }
= 
\\
& \left (\bm v^h, \bm f\right ) +
\left (\bm v^h, \bm h\right )_{\Gamma_t^N}, 
\quad
\text{for all } (\bm v^h, q^h) \in \mathcal V_{\bm 0}^h \times \mathcal Q^h. 
\end{aligned}\label{semidiscretevmsles}
\end{equation}

\color{black}
We use the Backward Euler Method to discretize the problem in time 
and we extrapolate $\bm{u}^h$ in the non-linear terms by means of the Newton-Gregory backward polynomials of order one. This yields a single linear problem at each time step. For more details on this implementation and on its strengths and limitations, the interested reader can see \cite{FD_vmsles}.

We partition the time interval into $N_t$  subintervals of equal size $\Delta t = \frac{T}{N_t}$, with $t_n = n \Delta t $ and we denote with the subscript $n$ quantities related to the time step $n$, with $n=0, \dots, N_t$.  The fully discretized linearized semi-implicit VMS-LES formulation of the Navier-Stokes equations in ALE framework with Backward Euler Method as time integration method reads: 

Given $\bm u^h_n$, for any $n = 0, \dots, N_t -1$, find $(\bm u^h_{n+1}, p^h_{n+1}) \in \mathcal{V}_g^h \times \mathcal{Q}^{h} $ such that:  
\begin{equation}
\begin{aligned}
& \left ( \bm v^h, \rho \frac{\bm u^h_{n+1}}{\Delta t }\right )_{\Omega_{n+1}} + 
\left ( \bm v^h, \rho (\bm u^h_n - \bm u^{\text{ALE}}_{n+1}) \cdot \nabla \bm u^h_{n+1}\right)_{\Omega_{n+1}}  + 
\left (\nabla \bm v^h, \mu \nabla \bm u^h_{n+1} \right)_{\Omega_{n+1}} 
\\
& - \left ( \nabla \cdot \bm v^h, p^h_{n+1} \right)_{\Omega_{n+1}}  + 
\left ( q^h, \nabla \cdot \bm u^h_{n+1} \right )_{\Omega_{n+1}}  
\\
& + \underbrace{\left ( \rho (\bm u^h_n - \bm u^{\text{ALE}}_{n+1}) \cdot \nabla \bm v^h + \nabla q^h , \, \tau_{\text M}(\bm u^h_{n+1} ) \bm r_{\text M} (\bm u^h_{n+1}, p^h_{n+1})\right )_{\Omega_{n+1}} + \left (\nabla \cdot \bm v^h, \tau_{\text C}(\bm u^h_{n+1})r_{\text C}(\bm u^h_{n+1})\right)_{\Omega_{n+1}}}_\text{SUPG}
\\
& + \underbrace{\left(\rho (\bm u^h_n - \bm u^{\text{ALE}}_{n+1}) \cdot (\nabla \bm v^h)^T, \tau_{\text M}(\bm u^h_{n+1})\bm r_{\text M}(\bm u^h_{n+1}, p^h_{n+1}) \right )_{\Omega_{n+1}}}_\text{VMS}
\\
& - \underbrace{\left( \rho \nabla \bm v^h,  \tau_{\text M}(\bm u^h_{n+1})\bm r_{\text M}(\bm u^h_n, p^h_n) \otimes  \tau_{\text M}(\bm u^h_{n+1})\bm r_{\text M}(\bm u^h_{n+1}, p^h_{n+1}) \right )_{\Omega_{n+1}}}_\text{LES}
\\
& = \left (\bm v^h, \bm f_{n+1}\right )_{\Omega_{n+1}}  +
\left (\bm v^h, \bm h_{n+1}\right )_{\Gamma_{n+1}^N} +
\left ( \bm v^h, \rho \frac{\bm u^h_{n}}{\Delta t }\right )_{\Omega_{n}}
\quad
\text{for all } (\bm v^h, q^h) \in \mathcal{V}_0^h \times \mathcal{Q}^h.
\label{discretevmsles}
\end{aligned}
\end{equation}
The strong residuals, after time discretization, read 
\begin{align}
\bm r_{\text M} (\bm u^h_{*}, p^h_{*}) = & \, \rho \frac{ \bm u^h_{*}-\bm u^h_n}{\Delta t} + \rho \left( \bm{u}^h_{n} - \bm{u}^{\text{ALE}}_{n+1} \right) \cdot \nabla \bm{u}^h_{*} - \mu \Delta \bm u^h_{*}+ \nabla p^h_{*} - \bm f_{n+1}, \label{rm_space_time} \\
r_{\text C} (\bm u^h_{n+1}) = & \,  \nabla \cdot \bm u^h_{n+1} \, .
\label{rc_space_time}
\end{align}
where the subscript $*$ denotes either the time step $n$ or $n+1$,  as the residuals appear in Eq. \eqref{discretevmsles}.  

\color{black} As for Eq. \eqref{coarse_withfinesolution}, the first, second and last rows in Eq. \eqref{discretevmsles} contain integrals of the standard Navier-Stokes equations in ALE framework (see Eq. \eqref{eq_ns-weak}), while on the remaining rows the additional stabilization and turbulence modeling terms, namely the standard SUPG term, the VMS term and, finally, the LES term. From this point of view,  the standard SUPG stabilization method can be considered as a step towards the fully stabilized formulation \cite{BCC_vmsles}. In this paper, we  will adopt either the VMS-LES method, i.e. the whole formulation in Eq. \eqref{discretevmsles}, and the SUPG method, i.e. the formulation in Eq. \eqref{discretevmsles} without the additional terms VMS and LES.

We recall that, on the one hand, both the SUPG and VMS-LES methods allow to control instabilities in the velocity field arising from convection-dominated (i.e. high Reynolds number) regimes and instabilities due to the fact that equal FE spaces $\mathbb{P}_k - \mathbb{P}_k$ would not satisfy the \textit{inf-sup} (or LBB) condition, yielding to numerical oscillations of the pressure field \cite{LBB, BCC_vmsles, FD_vmsles}. On the other hand, the VMS-LES method - as the name itself emphasizes and differently from SUPG - also yields to a LES-type modeling \cite{BCC_vmsles, Hughes_1995, HCS_2005, HMJ_2000, HOM_2001, HSF_2004, FD_vmsles} to account for the transitional-nearly turbulent flow regime that typically occurs in cardiac haemodynamics.
\color{black}

\subsection{Left atrium model} 

\begin{figure}[t!]
	\centering%
	\includegraphics[trim={2 2 2 2},clip,width=0.9\textwidth]{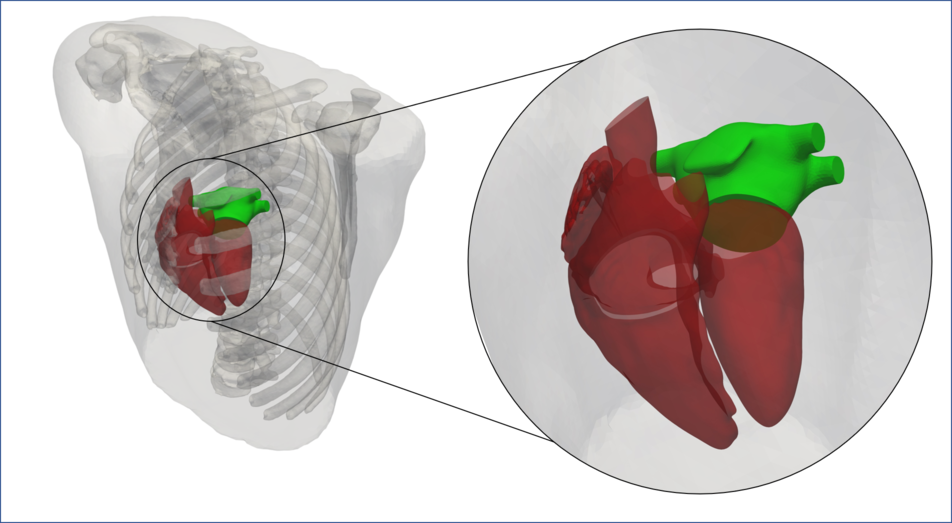}
	\caption{\label{torso}  Position of the LA inside the torso. The idealized LA geometry is in green and the remaining heart's chambers in red. The 3D torso model is taken for visualization purposes from the repository 
		 \cite{commlab,Ferrer_2015}.
	}
\end{figure}

\begin{figure}[t!]
	\centering%
	\includegraphics[trim={1cm 1cm 1cm 1cm},clip,width=0.49\textwidth]{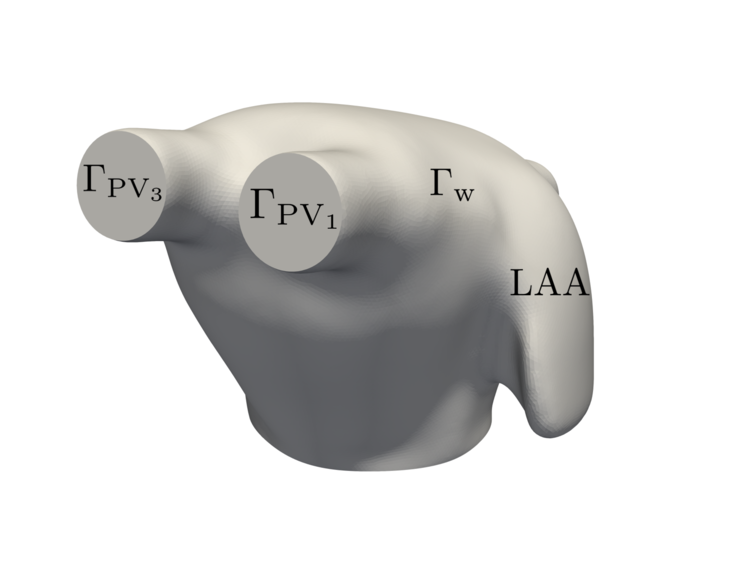}
	\includegraphics[trim={1cm 1cm 1cm 1cm},clip, width=0.49\textwidth]{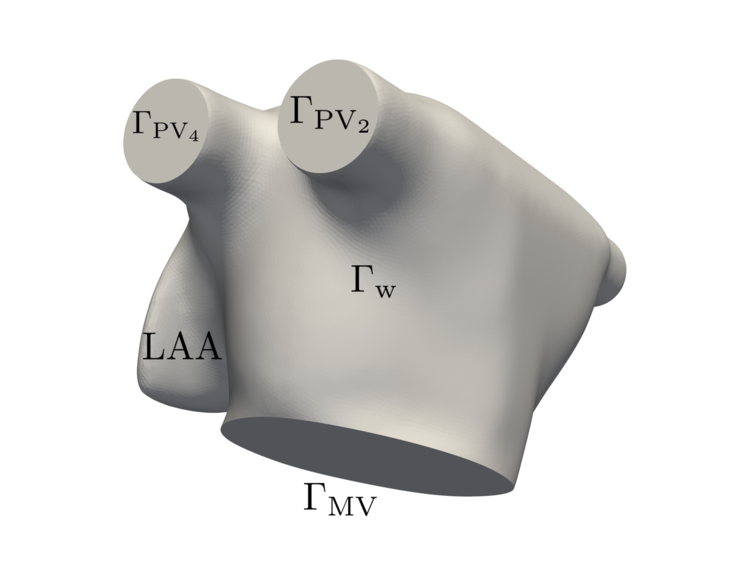}
	\caption{\label{fig_lageom} The idealized LA geometry from two different angles. The domain boundary is $\Gamma_t = \Gamma_{\text w} \cup \Gamma_{\text{MV}} \cup \left (\bigcup_{i=1}^4 \Gamma_{\text{PV}_i}\right )$.
	}
\end{figure}

The LA is a chamber located in the left part of the heart anchored on the top of the LV, connected to the pulmonary circulation system through the pulmonary veins (PVs) and to the LV through the mitral valve (MV). 
The position, size and even the number of PVs is specific to the individual, but there are usually four veins situated in the upper part of the LA in a perpendicular direction with respect to the MV axis. The left atrial appendage (left auricle) is a small secondary cavity located on one side of the LA and connected to the main cavity through an orifice. In Figure \ref{fig_lageom} we report the geometry of the idealized LA that is used for the numerical simulations, while in Figure \ref{torso} we highlight the position of the chamber inside a human torso. 

 The LA boundary $\Gamma_t$ is split into six portions: four PVs sections $\Gamma_{\text{PV}_i}, \, i= 1, \dots, 4$,  the MV section $\Gamma_{\text{MV}}$ and the LA endocardium $\Gamma_\text{w}$. The PVs are considered equal sized and the left atrial appendage is labelled as LAA. The section area of the MV is 6.74 cm$^2$, while the area of each PV is 0.78 cm$^2$, if the former were to be considered circular, their diameters would be 2.93 cm and 1 cm respectively.

In physiological conditions, during diastole, blood is ejected from the LA into the LV through the open MV with a first strong ejection and a second weaker one, strengthened by the LA contraction known also as atrial kick. This process is characterized by a volume reduction of about $25\%$ of the initial volume. The first blood ejection from the LA is called Early wave (E-wave) while the atrial kick is also known as After wave (A-wave). During systole the MV closes and the LA is filled with blood coming from the PVs, enlarging to reach the original volume.
\begin{figure}[t!]
	\centering
	\includegraphics[trim={0cm 0cm 0.5cm 0.5cm},clip, width=0.49\textwidth]{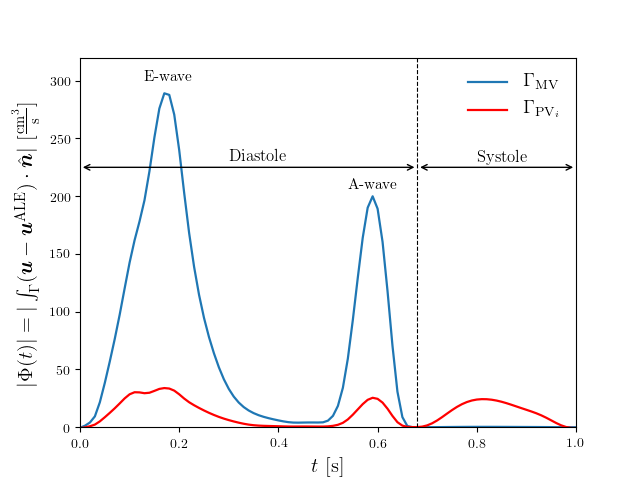}
	\includegraphics[trim={0cm 0cm 0.5cm 0.5cm},clip, width=0.49\textwidth]{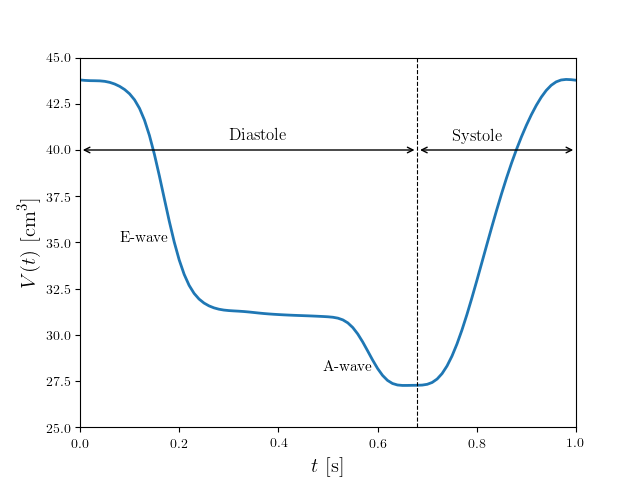}
	\caption{\label{fig_flow} Blood flow through the MV
		section ($\Gamma_{\text{MV}}$) and in each PV ($\Gamma_{\text{PV}_i}, \, i = 1, \dots, 4$) vs. time (left).
		Idealized LA volume vs. time (right).}
\end{figure}

In literature, the MV flow has been
studied and measured in both physiological and pathological conditions
\cite{KWPP_mri,CMN_leftheart,QMC_review,TDQ_3d, mittal_mitralvalve, VGY_leftatrium}.
In Figure \ref{fig_flow} (left) we report the inlet (PVs section) and outlet (MV section) flow rates against time. The first peak during diastole is the E-wave, while the second one is the A-wave. During
systole the flow through the MV is zero because the valve is closed. The heart cycle considered in
this work corresponds to a rest condition at 60 bpm, i.e. the period is equal to $T_\text{HB}=1$ s. The diastole lasts
for $T_{\text{dias}}=0.68$ s and the systole for the remaining $T_{\text{syst}}=0.32$ s; a whole heartbeat lasts $T_\text{HB} = T_{\text{dias}} + T_{\text{syst}}$. We simulate respectively diastole and systole, so that the initial time corresponds to the end systolic phase.
The volume variation of
the LA is based on the ejection phases, so the volume decrease is modeled in two phases
corresponding to the E and A-waves. The LA filling phase is shorter and is accomplished with a continuous
rise of the volume. The LA volume as a function of time $V(t)$ is reported in Figure \ref{fig_flow} (right) \cite{KFH_leftatrium}. 

\begin{figure}[!t]
	\begin{subfigure}{.5\textwidth}
		\centering
		\includegraphics[width=0.9\textwidth]{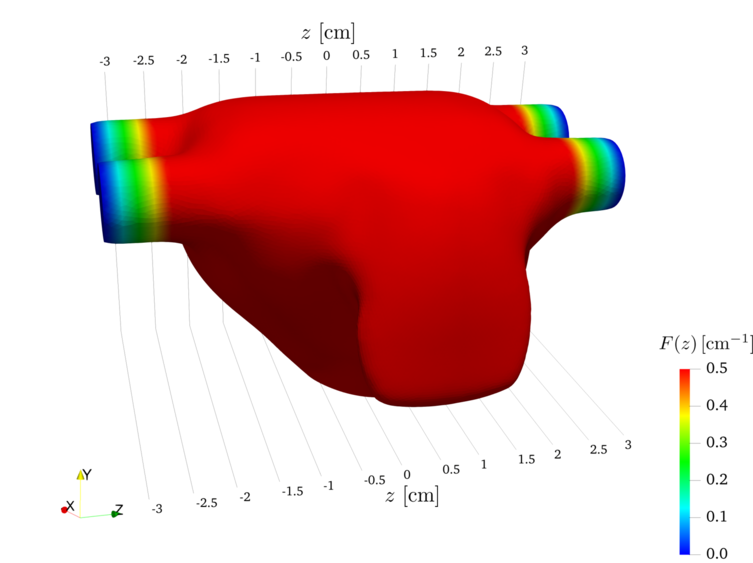}
	\end{subfigure}
	\begin{subfigure}{.5\textwidth}
		\includegraphics[width=0.9\textwidth]{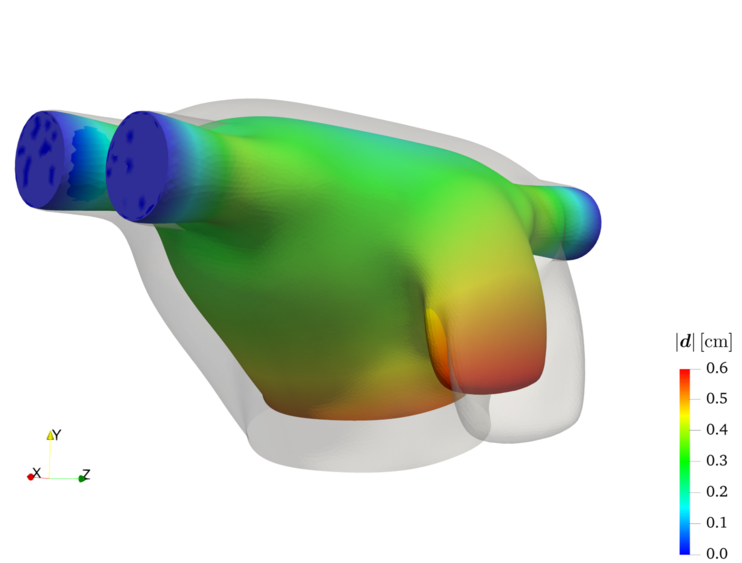}
	\end{subfigure}
	\caption{ \color{black}Function $F(z)$ on the LA surface (left). LA geometry at its maximum contraction at end diastole (right): the colors in the deformed geometry highlight the magnitude of the displacement field $|\bm d|$. \color{black}}
	\label{coefficientC_displacement}
\end{figure}

As explained in Section \ref{section_NS_ALE}, we prescribe a velocity $\bm g^{\text{ALE}}$ on the boundary $\Gamma_t$ and we extend it harmonically to the whole domain to get the ALE velocity $\bm u^{\text{ALE}}$ (see Eq. \eqref{ALE_laplacian}). In particular, we compute the ALE velocity on the LA boundary by assuming separation of variables as:
\begin{equation}
\bm g^{\text{ALE}} (\bm x, t) = \bm f^{\text{ALE}}(\bm x)\, g^{\text{ALE}}(t) \quad \text{on } \Gamma_t,  \label{gALEdef}
\end{equation} 
where $\bm f^{\text{ALE}}(\bm x)$ contains the directions of $\bm g^{\text{ALE}}$ and $g^{\text{ALE}}(t)$ is a time-dependent function. We design $\bm f^{\text{ALE}} $ to decrease the wall velocity near the PVs ($\bm g^{\text{ALE}}= \bm 0$ on ${\Gamma_{\text{PV}_i}}, \, i = 1, \dots, 4$). Let $x_G, \, y_G, \, z_G$ be the coordinates of the LA center of mass (units in cm), we define the function $\bm f^{\text{ALE}}$ as:
\begin{equation}
\bm f^{\text{ALE}} (\bm x) = F(z) ((x-x_G) \hat{\bm x} + (y-y_G) \hat{\bm y} + 0.6(z - z_G) \hat{\bm z} ), 
\label{fALE}
\end{equation}
with
\begin{equation}
F(z)= \left\{ \begin{array}{l l l} 
0.5  & \text{ if } |z-z_G| \in \left[0,2.5\right]\text{ cm}, \\
\\
0.5 \left(\dfrac{2.5 - |z-z_G|}{0.72} + 1 \right) & \text{ if }|z-z_G| \in \left[2.5, 3.22\right]\text{ cm}, \\
\\
0  & \text{ if } |z-z_G| \in \left[3.22, 10\right]\text{ cm}. 
\end{array} \right.
\label{Fz}
\end{equation}
The function $F$ is represented in Figure \ref{coefficientC_displacement} (left). In order to get the time variation of the prescribed ALE velocity $g^{\text{ALE}}(t)$, we consider the volume variation and we exploit the Reynolds transport theorem (RTT) and Eq. \eqref{gALEdef}:
\begin{equation}
\frac{dV(t)}{dt} = \frac{d}{dt} \int_{\Omega_t} d \Omega \overset{\text{RTT}}{=} \int_{\Gamma_t} \bm g^{\text{ALE}} \cdot \hat{\bm n} d \Gamma \overset{\text{Eq. \eqref{gALEdef}}}{=} g^{\text{ALE}}(t) \int_{\Gamma_t} \bm f^{\text{ALE}} \cdot \hat{\bm n} d \Gamma, 
\end{equation}
which gives the following definition of ${g}^{\text{ALE}}$: 
\begin{equation}
\everymath{\displaystyle}
{g}^{\text{ALE}}(t) = \dfrac{1}{\int_{\Gamma_t} \bm f^{\text{ALE}} \cdot \hat{\bm n} d \Gamma} \dfrac{dV(t)}{dt}.
\end{equation}

\begin{figure}[!t]
	\centering
	\includegraphics[width=0.5\textwidth]{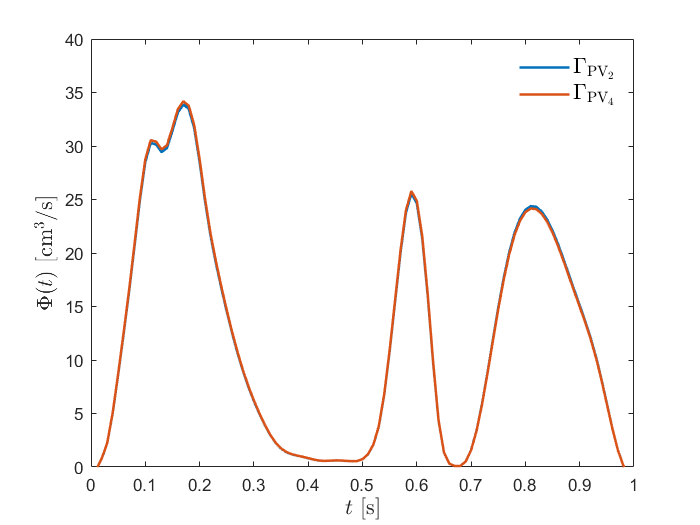}
	\caption{\color{black} Inlet flow rates on $\Gamma_{\text{PV}_2}$ and $\Gamma_{\text{PV}_4}$. \color{black}}
	\label{flux_PV2_PV4}
\end{figure} 

To better appreciate the LA deformation, in Figure \ref{coefficientC_displacement} (right) we overlap the geometry of the LA in its relaxed  and contracted configurations, at the beginning and at the end of diastole, where the maximum LA contraction is met, respectively. 

In terms of boundary conditions, during diastole (MV is open), we set a homogeneous Neumann boundary condition on the MV section and we 
prescribe Poiseuille profiles on the PVs. We do this using, for each vein, a parabolic velocity profile and imposing the inlet flow rate $\Phi_{\text{PV}_i}(t), \, i = 1, \dots, 4$ that fulfils the mass balance:
\begin{equation}
\sum_{i=1}^4\Phi_{\text{PV}_i}(t) + \Phi_{\text{MV}}(t) + \frac{d V(t)}{dt} = 0,
\label{eq_incompr}
\end{equation}
whereas the flow rates are defined as:
\begin{equation}
\Phi_{\text{PV}_i}(t) = \int_{\Gamma_{\text{PV}_i}} \left ( \bm u - \bm u ^ {\text{ALE}}\right )\cdot \hat{\bm n} d\Gamma, \quad i = 1, \dots, 4, 
\end{equation}
\begin{equation}
\Phi_{\text{MV}} (t)= \int_{\Gamma_{\text{MV}}} \left ( \bm u - \bm u ^ {\text{ALE}}\right )\cdot \hat{\bm n} d\Gamma.
\end{equation}
During systole, the MV is closed ($\Phi_{\text{MV}}(t)=0$), so we switch the boundary condition on $\Gamma_{\text{MV}}$ to a Dirichlet one to model the closed behaviour of the valve: $\bm u = \bm g^{\text{ALE}}$. The sudden switch of boundary conditions from natural to essential and viceversa -- aimed at replicating the rapid closing and opening stages of the MV -- may potentially introduce some artifacts on the numerical solution, even if these are  negligible in our experience. 
\color{black}
However, we observed that, during systole, numerical oscillations would arise by keeping Dirichlet boundary conditions on all the inlet sections (as done during diastole) and on the MV section. For this reason, unlike diastole, we use a homogeneous Neumann boundary condition on one of the PVs (specifically, we choose $\Gamma_{\text{PV}_4}$) while keeping a Dirichlet boundary condition with assigned flow rate given by (\ref{eq_incompr}) on the other three. As a matter of fact, from the simulation results, we found that numerical oscillations are strongly reduced using this boundary setting. In addition, we have found that during systole the flow is always entering in the domain (as for the remaining inlet sections), with a flow rate on $\Gamma_{\text{PV}_4}$ almost equal to the remaining sections where Dirichlet boudary is prescribed. We show this in Figure \ref{flux_PV2_PV4} where we compare the flow rate prescribed on inlet portion $\Gamma_{\text{PV}_2}$, where a Dirichlet boundary condition is set for the whole heartbeat, and the flow rate computed on $\Gamma_{\text{PV}_4}$, where the homogeneous Neumann boundary condition is prescribed in systole and a Dirichlet one in diastole. Mass balance condition in Eq. \eqref{eq_incompr} is hence always satisfied. By setting the boundary conditions as explained, we finally obtain the fluxes through the MV section and through each PV as reported in Figure \ref{fig_flow}. 
\color{black}

Moreover, a backflow stabilization is introduced in all the homogeneous Neumann-type boundary conditions in order to weakly penalize the reverse flow \cite{BGHMZ_backflow}: 
\begin{equation}
\bm{\sigma}(\bm u, p) \bm{\hat n}= \rho (\{ \left (\bm u -\bm u^\text{ALE} \right )\cdot \hat{\bm n }\}_{-})\left (\bm u -\bm u^\text{ALE} \right )\quad \text{ on } \Gamma^N_t, 
\end{equation}
being $\{\left (\bm u -\bm u^\text{ALE} \right ) \cdot \hat{\bm n }\}_{-}$ the negative part of $\left (\bm u -\bm u^\text{ALE} \right )\cdot \hat{\bm n}$:
\begin{equation}
\{\left (\bm u -\bm u^\text{ALE} \right ) \cdot \hat{\bm n }\}_{-} = 
\begin{cases}
\left (\bm u -\bm u^\text{ALE} \right ) \cdot \hat{\bm n} & \text{ if } \left (\bm u -\bm u^\text{ALE} \right ) \cdot \hat{\bm n} < 0, \\
0 & \text{ if } \left (\bm u -\bm u^\text{ALE} \right )\cdot \hat{\bm n} \geq 0.
\end{cases}
\end{equation}
Finally, we summarize in Eq. \eqref{wholemodel} the whole set of boundary and initial conditions for the modelling of blood flow in the LA. 

\begin{equation}
\begin{aligned}
&\bm u = -\frac{|\Phi_{\text{PV}_i}(t)|}{4|\Gamma_{\text{PV}_i}|}\left (1-\frac{r(\bm{x})^2}{R_i^2} \right )  {\bm{ \hat n}}_i & \quad \quad & \text{ on } \Gamma_{\text{PV}_i} \times (0, T_{\text{dias}} ) , \, i=1 \dots, 4 \text{, } 
\\
&\bm{\sigma}(\bm u, p) \bm{\hat n}= \rho (\{ \left (\bm u -\bm u^\text{ALE} \right )\cdot \hat{\bm n }\}_{-})\left (\bm u -\bm u^\text{ALE} \right ) && \text{ on } \Gamma_{\text{MV}}  \times (0, T_{\text{dias}}], 
\\
&\bm u = \bm g^{\text{ALE}} && \text{ on } \Gamma_{\text w}  \times (0, T_{\text{dias}}], 
\\
&\bm u = -\frac{|\Phi_{\text{PV}_i}(t)|}{4|\Gamma_{\text{PV}_i}|}\left (1-\frac{r(\bm{x})^2}{R_i^2} \right )  {\bm{ \hat n}}_i && \text{ on } \Gamma_{\text{PV}_i}   \times (T_{\text{dias}}, T_\text{HB}] , \, i=1 \dots, 3, 
\\
&\bm{\sigma}(\bm u, p) \bm{\hat n}= \rho (\{ \left (\bm u -\bm u^\text{ALE} \right )\cdot \hat{\bm n }\}_{-})\left (\bm u -\bm u^\text{ALE} \right ) && \text{ on } \Gamma_{\text{PV}_4}  \times (T_{\text{dias}}, T_\text{HB}],\\
&\bm u = \bm g^{\text{ALE}} && \text{ on } \Gamma_{\text w} \cup \Gamma_{\text{MV}} \times (T_{\text{dias}}, T_\text{HB}],\\
&\bm u = \bm 0 && \text{ in } \Omega_0  \times \{0\}, \\
\end{aligned}
\label{wholemodel}
\end{equation}
in the Dirichlet inflow boundary condition, $r(\bm x)=|\bm x|$, $R_i$ is the radius of the $i$--th PV section and $\hat{\bm n}_i$ its outward directed unit vector normal to.

\section{Mesh generation} \label{GRIDGENERATION}
\begin{table}[!t]
	\centering
	\begin{tabular}{|c|c|c|c|c|}
		\cline{2-5}
		\multicolumn{1}{c|}{} & Mesh & $\mathcal{T}_{h_1}$ & $\mathcal{T}_{h_2}$ & $\mathcal{T}_{h_3}$ \\
		\cline{2-5}
		\multicolumn{1}{c|}{}& \# elements & 575'220 & 1'711'622 & 8'344'030 \\
		\hline
		&$\bm u^h$ & 291'561 & 830'517 & 4'030'227 \\
		\multirow{1}{*}{\# DOFs ($\mathbb P1-\mathbb P1$)} & $p^h$ & 97'187 & 276'839 & 1'343'409 \\
		\cline{2-5}
		& total    & 388'748 & 1'107'356 & 5'373'636 \\
		\hline
		\multirow{2}{*}{Inner elements} & $h_\text{min}$[cm] & 0.05 & 0.05 & 0.05 \\
		& $h_\text{max}$ [cm] & 0.2 & 0.1 & 0.05 \\
		\hline
		& $\delta_{\text{BL}}$ [cm] & 0.05 & 0.05 & 0.05 \\
		\multirow{1}{*}{Boundary layer} & $n_{\text{layers}}$ & 3 & 4 & 5 \\
		&  $\chi_{\text{BL}}$ & 0.8 & 0.8 & 0.8 \\
		\hline
	\end{tabular}
	\caption{Details on the three meshes  $\mathcal{T}_{h_i}, \, i = 1, \dots, 3$: number of elements;  number of degrees of freedom (DOFs) using Lagrangian linear elements (for velocity, pressure and total number); minimum and maximum cell size for the inner elements of the mesh; boundary layer: boundary layer thickness $\delta_{\text{BL}}$, number of layers $n_{\text{layers}}$ and ratio among successive layers' thicknesses $\chi_{\text{BL}}$.}.
	\label{table_grids}
\end{table}

\begin{figure}[!t]
	\centering
	\begin{subfigure}{.325\textwidth}
		\centering
		\includegraphics[trim={1 1 1 1},clip, width=0.9\textwidth]{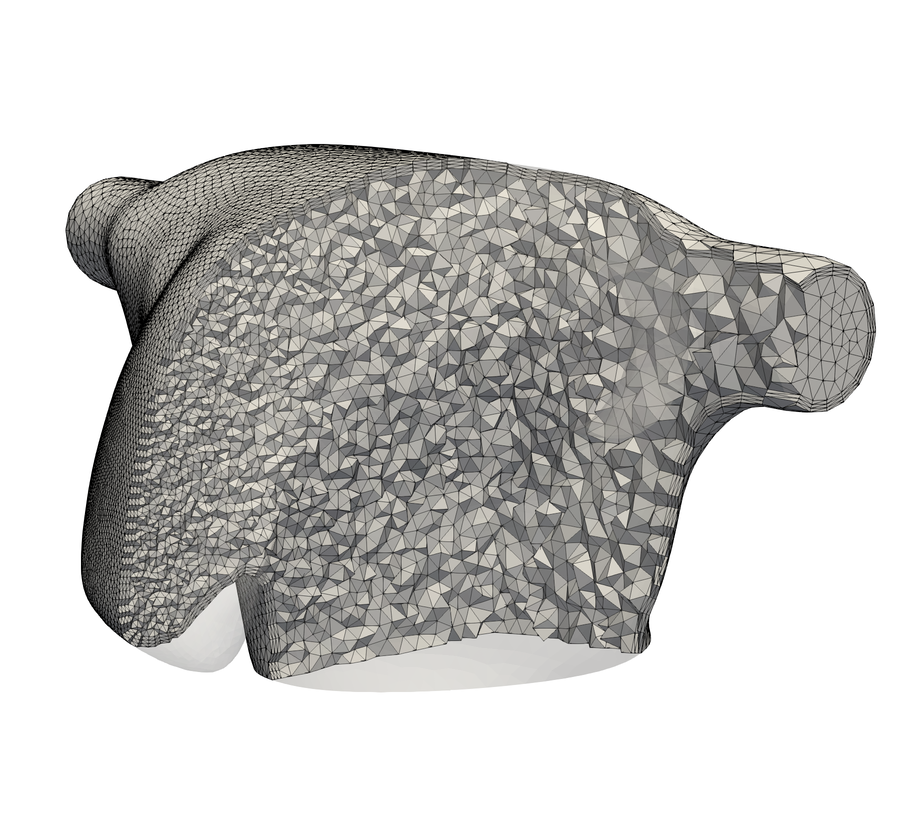}
	\end{subfigure}
	\begin{subfigure}{.325\textwidth}
		\includegraphics[trim={1 1 1 1 0},clip,width=0.9\textwidth]{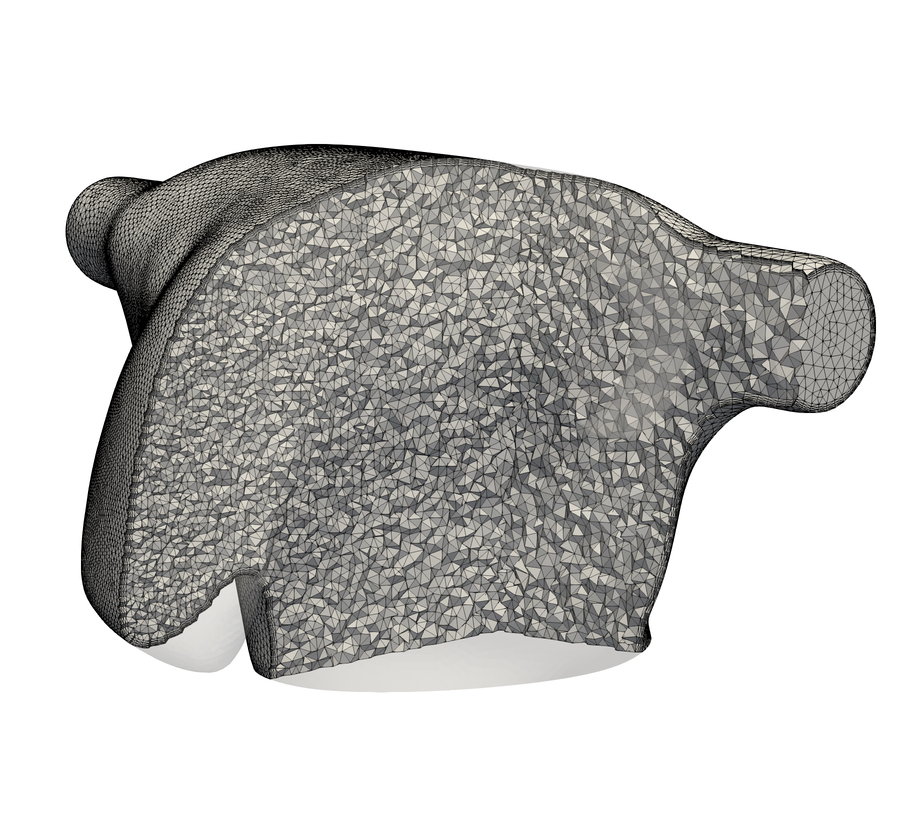}
	\end{subfigure}
	\begin{subfigure}{.325\textwidth}
	\includegraphics[trim={1 1 1 1 },clip,width=0.9\textwidth]{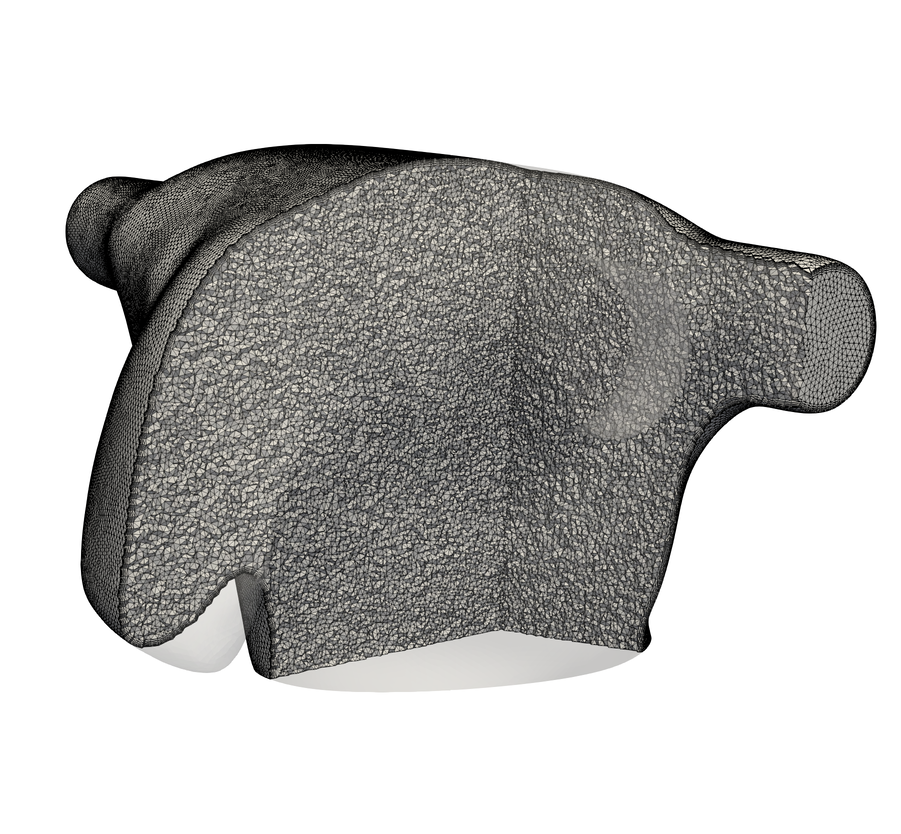}
\end{subfigure}

\begin{subfigure}{.325\textwidth}
	\centering
	\includegraphics[width=0.9\textwidth]{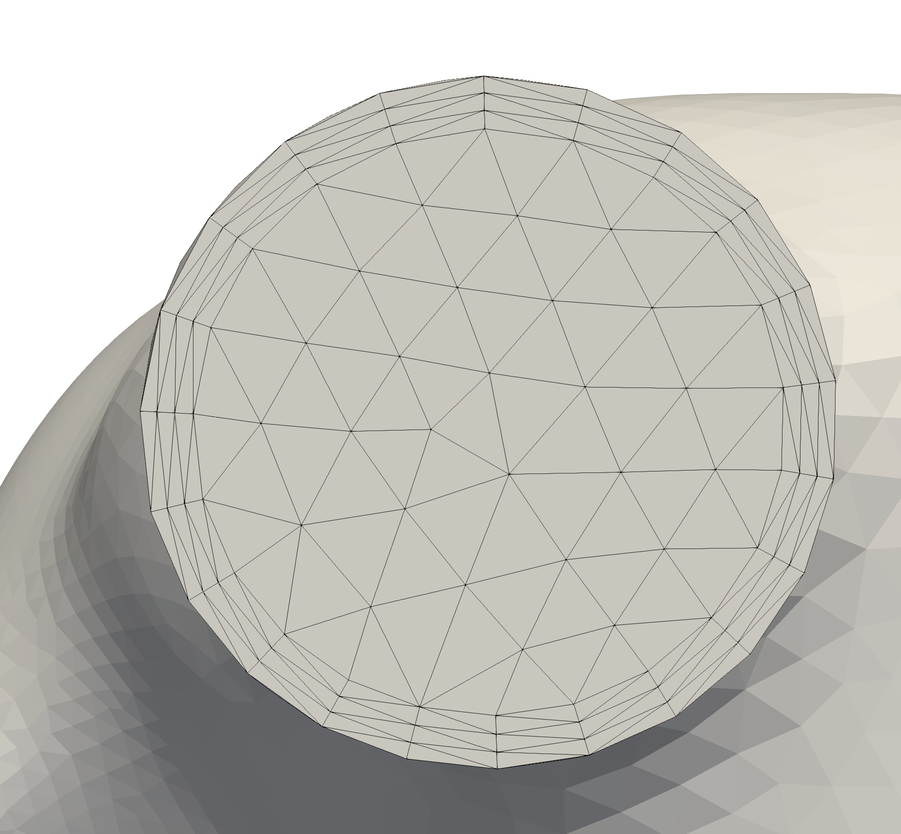}
	\caption{mesh $\mathcal{T}_{h_1}$}
	\label{la_mesh_coarse}
\end{subfigure}
\begin{subfigure}{.325\textwidth}
	\includegraphics[width=0.9\textwidth]{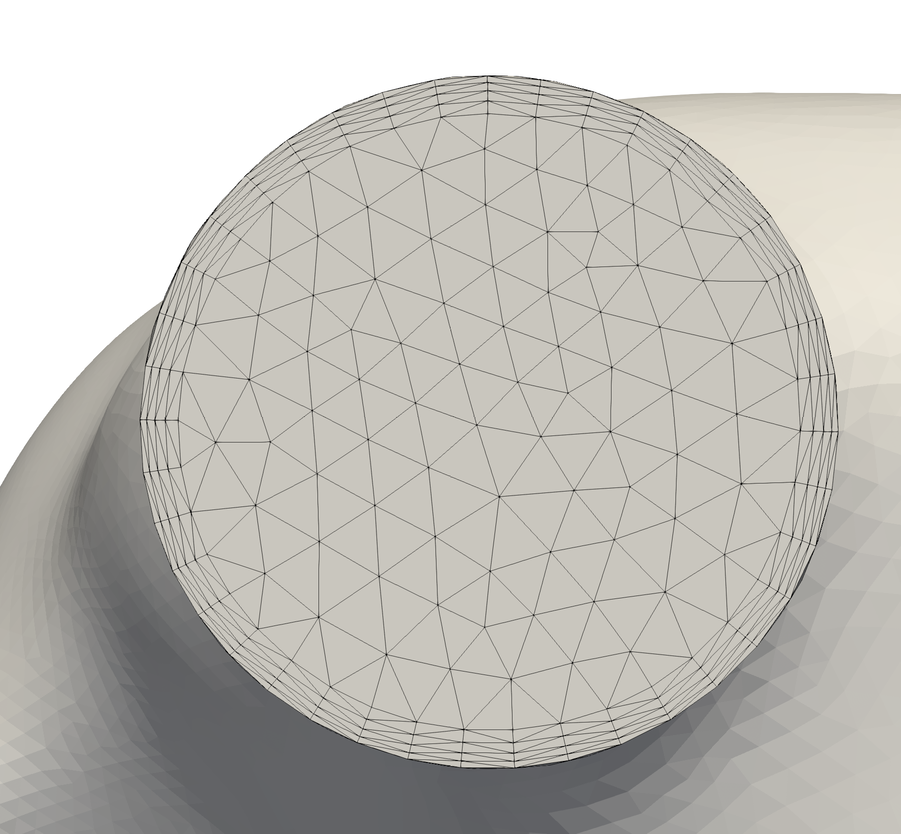}
	\caption{mesh $\mathcal{T}_{h_2}$}
	\label{la_mesh_medium}
\end{subfigure}
\begin{subfigure}{.325\textwidth}
	\includegraphics[width=0.9\textwidth]{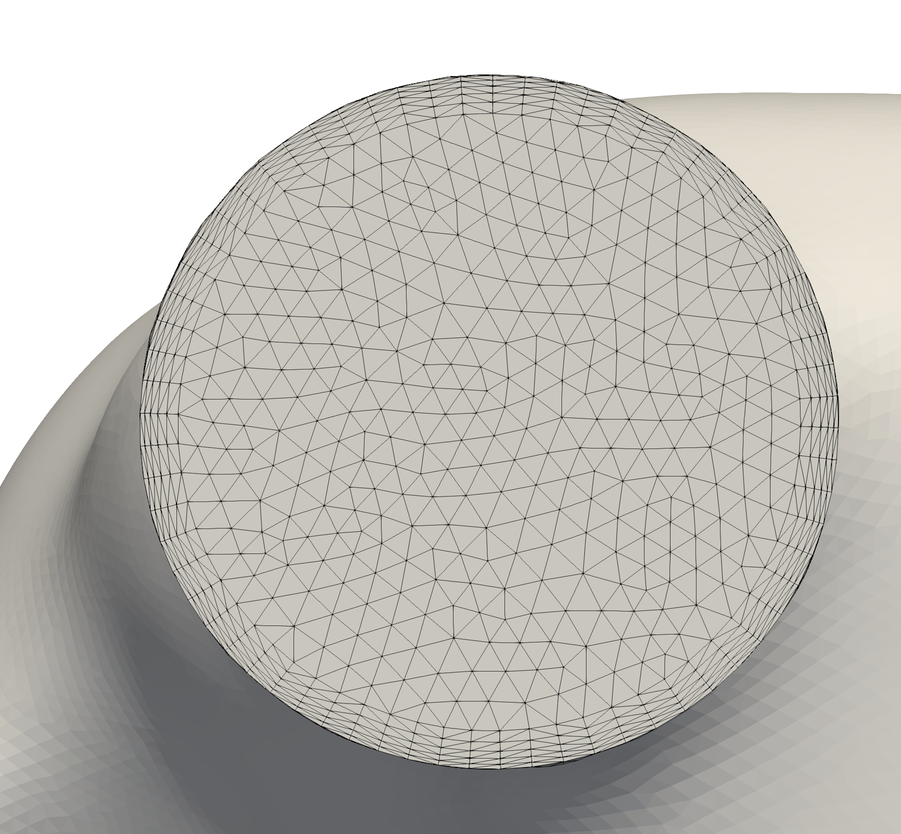}
	\caption{mesh $\mathcal{T}_{h_3}$}
	\label{la_mesh_fine}
\end{subfigure}
	\caption{The three meshes $\mathcal{T}_{h_i}, \, i = 1, \dots, 3$ adopted for the CFD simulations of the idealized LA geometry with a focus on a inlet section. }
	\label{la_mesh}
\end{figure}

\color{black} The LA endocardium was originally built by means of NURBS with the purpose of modeling the electric potential wavefront \cite{PDL_lageom}. In particular, the LA endocardium is built as a single NURBS patch starting from B-splines basis functions of degree 2; the LA fluid mesh is then obtained by filling the obtained surface. For further details on the idealized LA geometrical representation we refer the interested reader to \cite{PDL_lageom} and references there in. \color{black}
We generate three meshes, namely a coarse, medium and a fine one, denoted respectively as $\mathcal{T}_{h_1}$, $\mathcal{T}_{h_2}$ and $\mathcal{T}_{h_3}$. As shown in Figure~\ref{la_mesh} and reported in Table~\ref{table_grids}, for  $\mathcal{T}_{h_1}$ and $\mathcal{T}_{h_2}$, a non-uniform distribution of mesh element size is considered in order to have a well resolved LAA. In particular, we adopt for all mesh levels the same minimum cell-size $h_\text{min}=0.05$ cm in the lower corner of the LAA and we increase it linearly through an appropriate distance function (for $\mathcal{T}_{h_1}$ and $\mathcal{T}_{h_2}$ only). $\mathcal{T}_{h_3}$ instead keeps uniform grid cells sizes $h_\text{min} = h_\text{max} = 0.05$ cm. Furthermore, in order to accurately catch viscous effects near the wall, we introduce a boundary layer made of $n_{\text{layers}}$ layers with linearly variable element thicknesses. In particular, we adopt for all the meshes the same boundary layer thickness $\delta_{\text{BL}}=0.05$ cm, while we increase the number of layers - going from a mesh level to another - keeping the same ratio among successive layers' thicknesses $\chi_\text{BL}$. 
Table~\ref{table_grids} lists quantitative information about the three meshes. Mesh generation is performed by exploiting the VMTK library \cite{vmtk, vmtk_fedele}. Meshes are uploaded to a GitLab repository and publicly accessible \cite{repository_cfdmesh}.

\section{Numerical results and discussion}
\label{RESU}
We report the numerical results obtained performing numerical simulations\footnote{Numerical simulations were run on the cluster iHEART (Lenovo SR950 8 x 24-Core Intel Xeon Platinum 8160, 2100 MHz and
	1.7TB RAM) available at MOX, Dipartimento di Matematica, Politecnico di Milano. Furthermore, simulations on the mesh $\mathcal{T}_{h_3}$ were run on the cluster GALILEO supercomputer (IBM NeXtScale cluster, 1022 nodes (Intel Broadwell), 2 x 18-Cores Intel Xeon E5-2697 v4 at 2.30 GHz, 36 cores/node, 26.572 cores in total with 128 GB/node) by CINECA. 
} with  the FE library
LifeV \cite{LifeV, LifeV_paper} for the solution of the fluid dynamics in the idealized LA as modeled in Sections \ref{MATH} and~\ref{GRIDGENERATION}.

Blood is set as Newtonian, incompressible and viscous fluid with density $\rho=1.06$~g/cm$^3$ and dynamic viscosity $\mu=0.035$~g/(cm~s).
For each  $\mathcal{T}_{h_i}$, $i=1, \dots, 3$ , we simulate six heartbeats, starting from the initial condition $\bm u_0 = \bm 0$. Due to the periodicity in time of the boundary conditions of the problem, we analyse the output of the numerical simulations with a phase-averaging filter in order to get average quantities on one representative cycle. Furthermore, in order to remove the influence the unphysical initial condition $\bm u_0 = \bm 0$, we discard the first two heartbeats. Hence, referring to $N_{\text{HB}} = 4 $ heartbeats, with period $T_\text{HB} = 1$ s, we introduce the phase-averaging filter for the velocity as:
\begin{equation}
\langle{\bm u }(\bm x, t) \rangle = \frac{1}{N_\text{HB}} \sum_{n = 1} ^ {N_\text{HB}} {\bm u} (\bm x, t + (n-1)T_\text{HB}). 
\label{phase_averaged}
\end{equation}

First, we present the results achieved with the mesh $\mathcal{T}_{h_3}$ using the SUPG stabilization method, which will represent our reference solution. Then, we perform a mesh convergence study using both SUPG and VMS-LES methods and we compare the two methods in terms of fluid dynamics indicators with the results achieved with the reference solution.

\begin{figure}[!t]
	\centering
	\begin{subfigure}{.325\textwidth}
		\centering
		\includegraphics[trim={50 1 50 80 },clip,width=\textwidth]{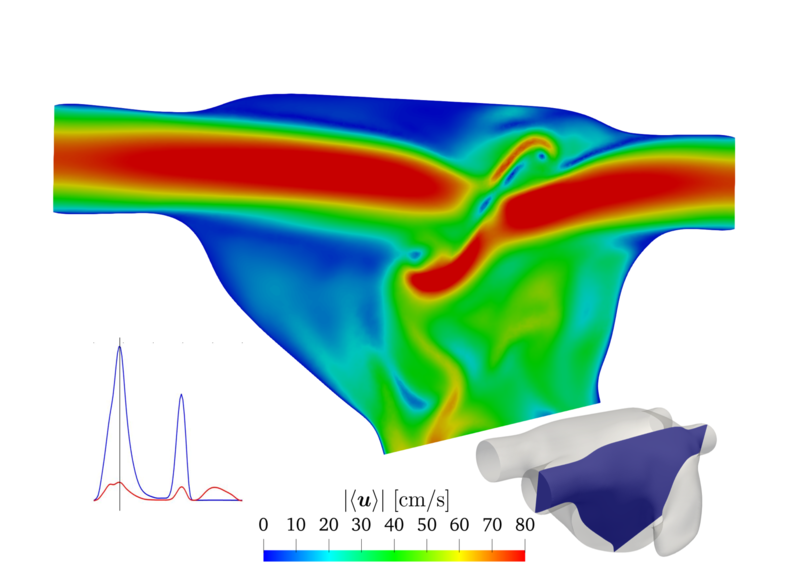}
		\caption{$t = 0.20$ s}
		\label{slice_u_20_fine}
	\end{subfigure}
	\begin{subfigure}{.325\textwidth}
		\centering
		\includegraphics[trim={50 1 50 80 },clip,width=\textwidth]{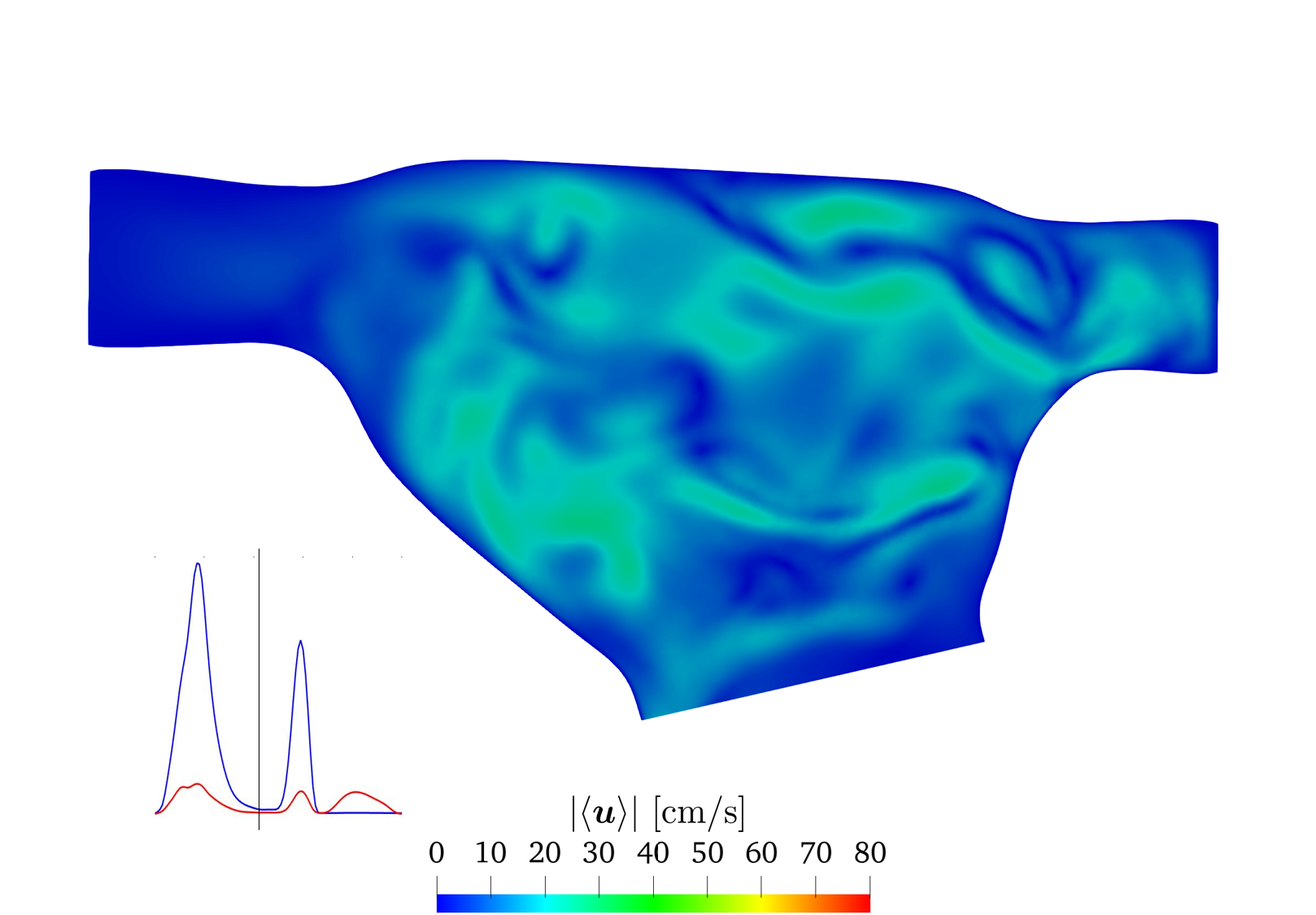}
		\caption{$t = 0.40$ s}
		\label{slice_u_40_fine}
	\end{subfigure}
	\begin{subfigure}{.325\textwidth}
		\centering
		\includegraphics[trim={50 1 50 80 },clip,width=\textwidth]{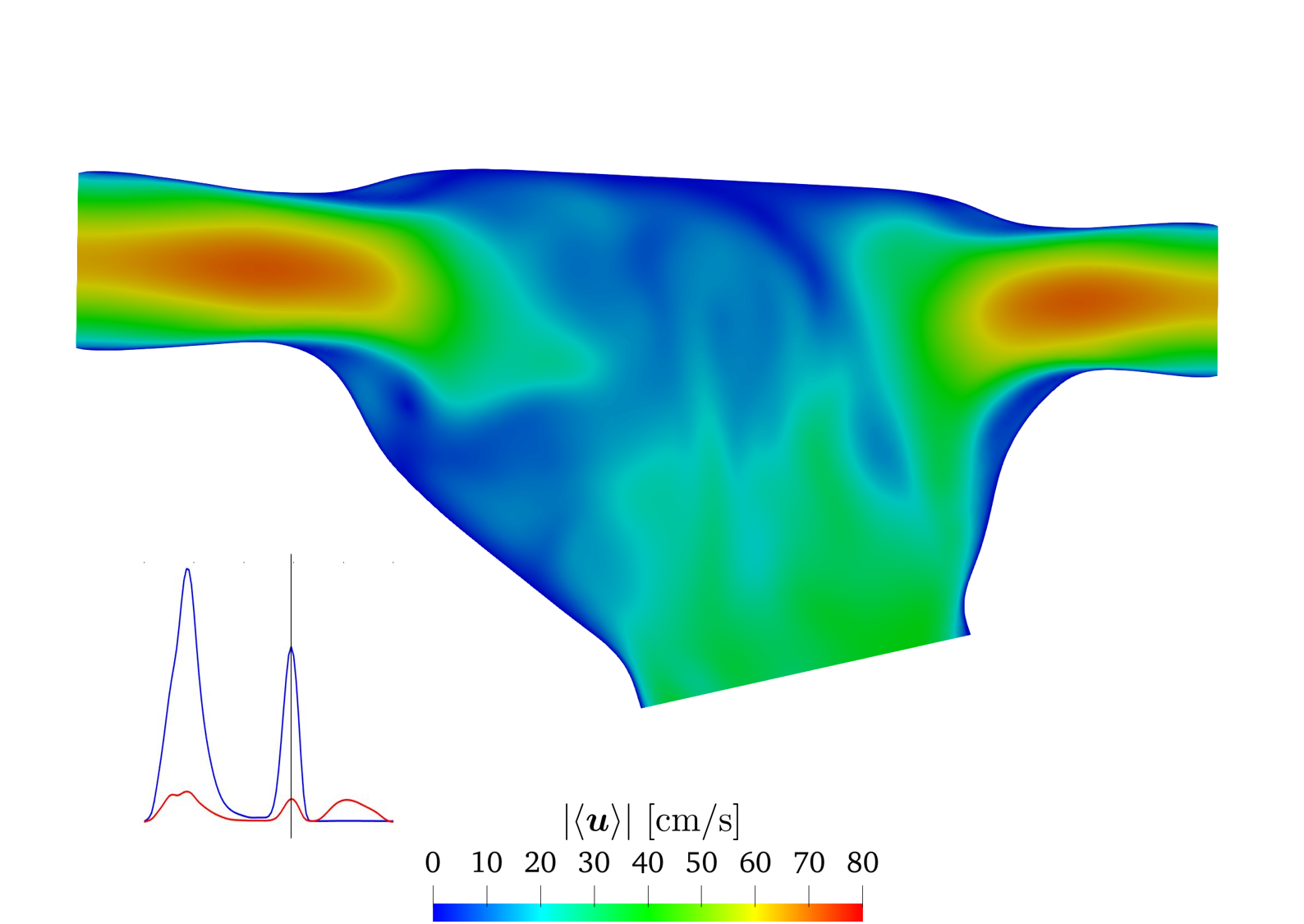}
		\caption{$t = 0.60$ s}
		\label{slice_u_60_fine}
	\end{subfigure}
	
	\begin{subfigure}{.325\textwidth}
		\centering
		\includegraphics[trim={50 1 50 80 },clip,width=\textwidth]{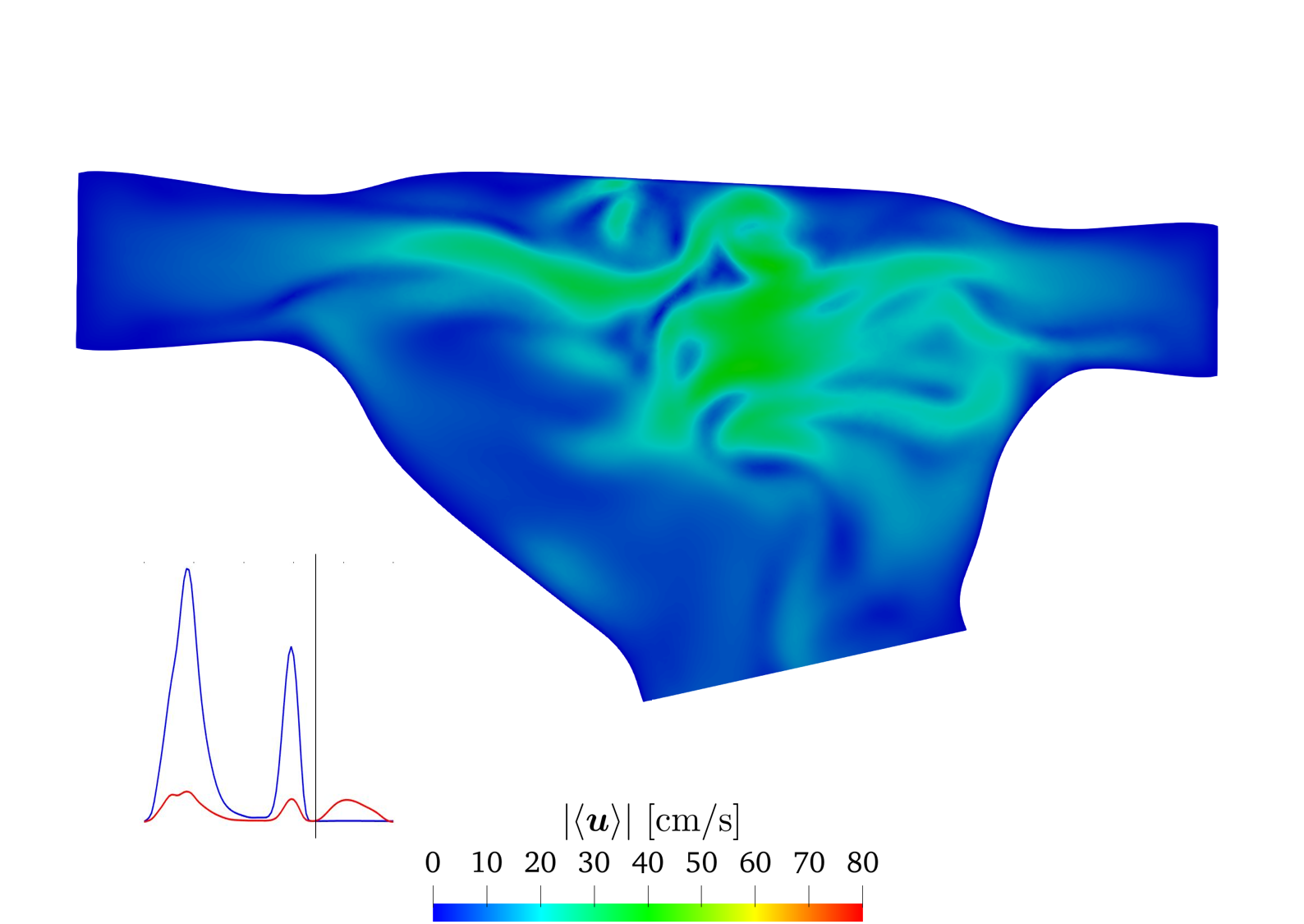}
		\caption{$t = 0.68$ s}
		\label{slice_u_20_fine}
	\end{subfigure}
	\begin{subfigure}{.325\textwidth}
		\includegraphics[trim={50 1 50 80 },clip,width=\textwidth]{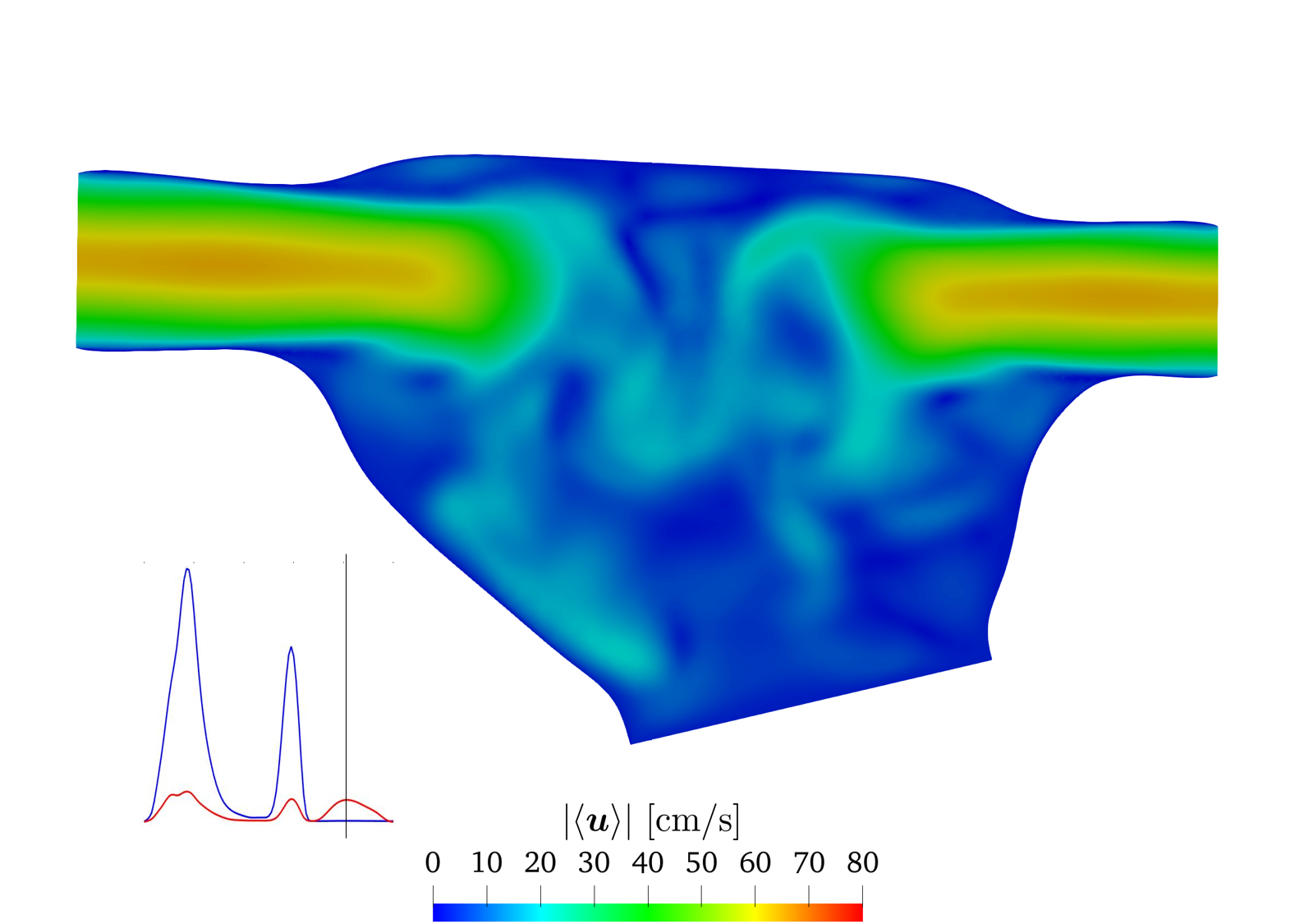}
		\caption{$t = 0.80$ s}
		\label{slice_u_80_fine}
	\end{subfigure}
	\begin{subfigure}{.325\textwidth}
		\includegraphics[trim={50 1 50 80 },clip,width=\textwidth]{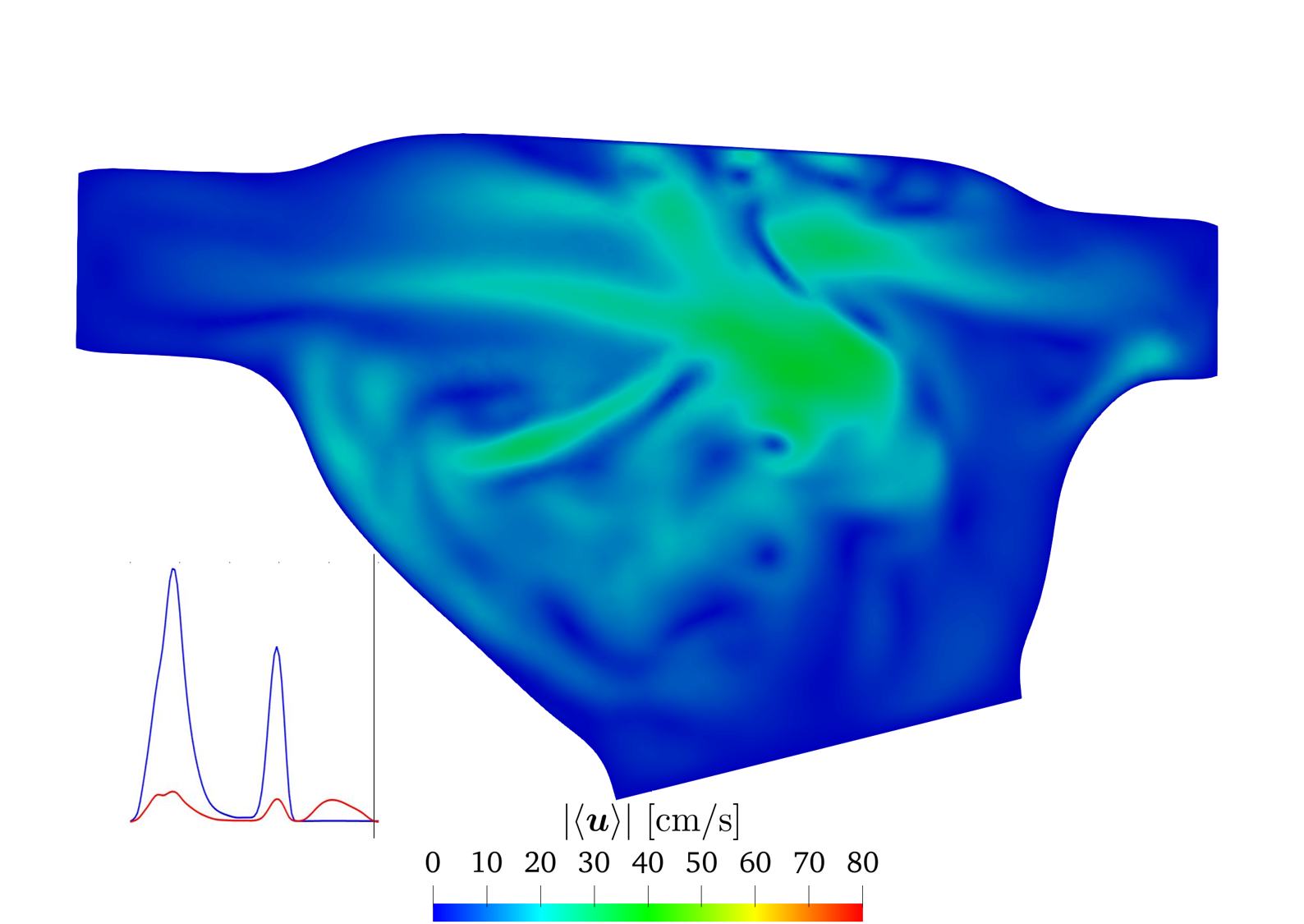}
		\caption{$t = 1.00$ s}
		\label{slice_u_100_fine}
	\end{subfigure}
	\caption{Reference solution: phase-averaged velocity magnitude $|\langle \bm u \rangle |$ on a slice cutting two PVs (top-left) at different time instants.}
	\label{fig_slice}
\end{figure}

\subsection{The reference solution}
\label{RESU_reference}
We report the results obtained with the mesh $\mathcal{T}_{h_3}$ by adopting a  SUPG stabilization method, with a time step $\Delta t = 6.25 \cdot 10^{-5}$~s. The numerical solution correspondingly obtained is denoted as our reference solution. We remark that the results that will present are referred to the phase-averaged velocity $\langle \bm u \rangle $ which is representative of a heartbeat defined in the time domain $[0, T_\text{HB}]$.

In Figure \ref{fig_slice}, we report the phase-averaged velocity magnitude of the blood on a slice cutting two PVs at six  time instants corresponding to the diastolic peak of the E-wave ($t = 0.20$ s), the plateau between E and A-waves ($t = 0.40$ s), the A-wave ($t = 0.60$ s), the beginning of systole ($t = 0.68$ s), the filling phase during systole ($t = 0.80$ s) and the end of systole ($t = 1.00$ s).
The peak velocity attained in our simulations is around 90 cm/s during the E-wave. The jets coming from the PVs impact one on each other, as it can be seen at time 0.20 s. 

\begin{figure}[!t]
	\centering
	\begin{subfigure}{.325\textwidth}
		\centering
		\includegraphics[trim={1 1 1 1 },clip,width=\textwidth]{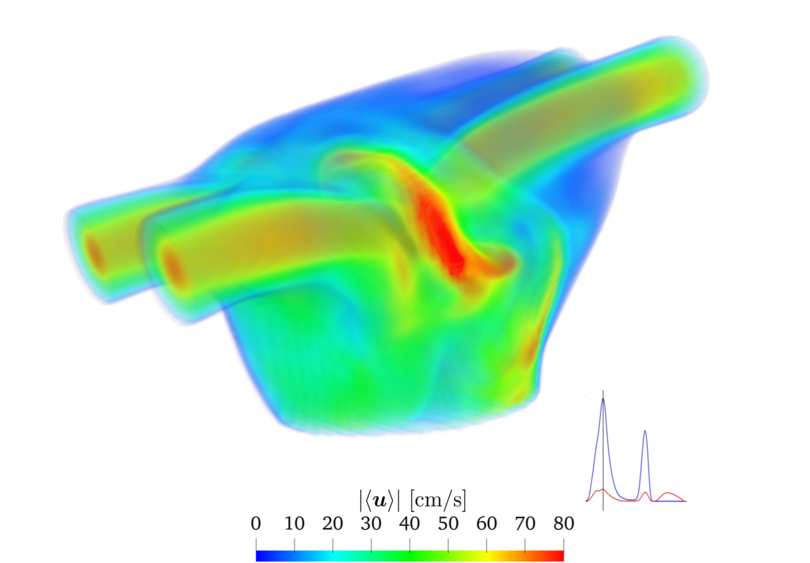}
		\caption{$t = 0.20$ s}
		\label{volume_rendering_20}
	\end{subfigure}
	\begin{subfigure}{.325\textwidth}
		\centering
		\includegraphics[trim={1 1 1 1 },clip,width=\textwidth]{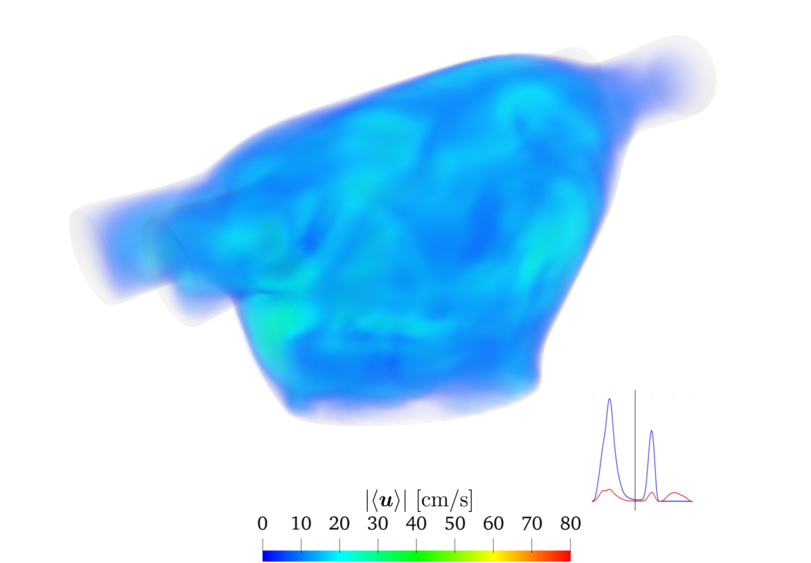}
		\caption{$t = 0.40$ s}
		\label{volume_rendering_40}
	\end{subfigure}
	\begin{subfigure}{.325\textwidth}
		\centering
		\includegraphics[trim={1 1 1 1 },clip,width=\textwidth]{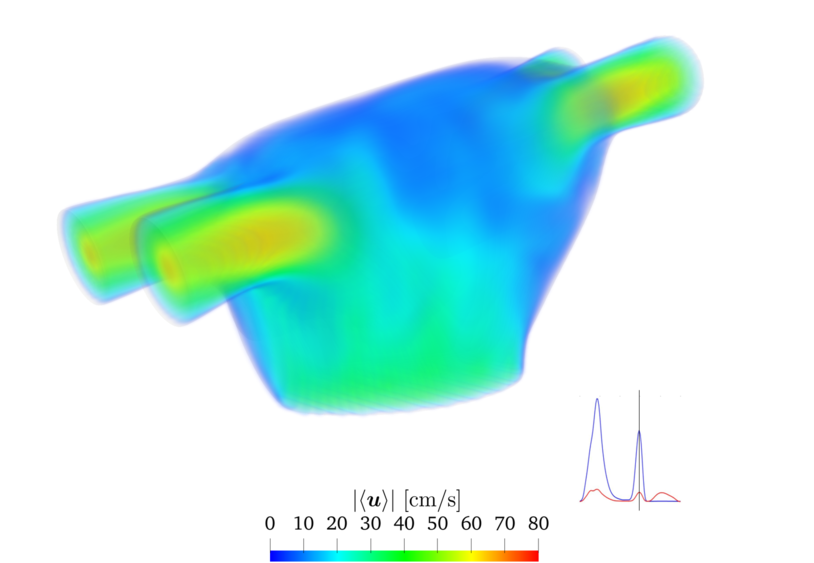}
		\caption{$t = 0.60$ s}
		\label{volume_rendering_60}
	\end{subfigure}
	
	\begin{subfigure}{.325\textwidth}
		\centering
		\includegraphics[trim={1 1 1 1 },clip,width=\textwidth]{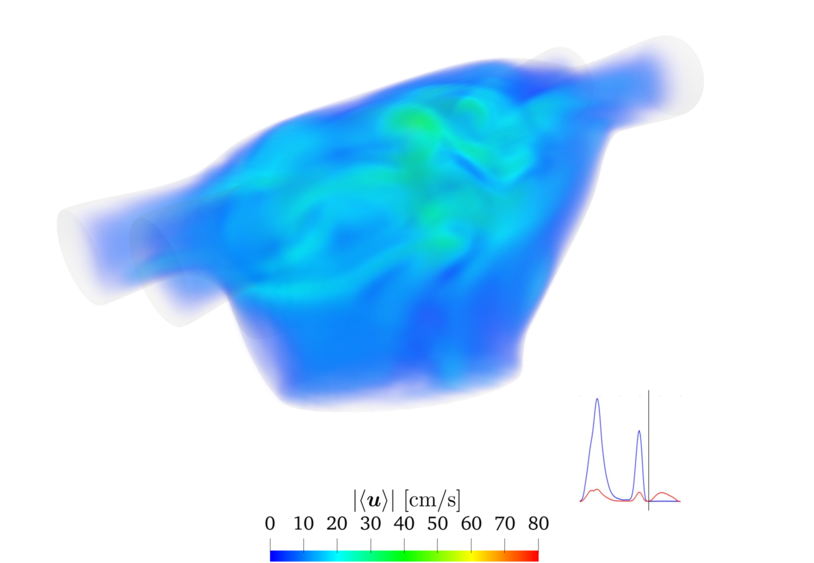}
		\caption{$t = 0.68$ s}
		\label{volume_rendering_68}
	\end{subfigure}
	\begin{subfigure}{.325\textwidth}
		\includegraphics[trim={1 1 1 1 },clip,width=\textwidth]{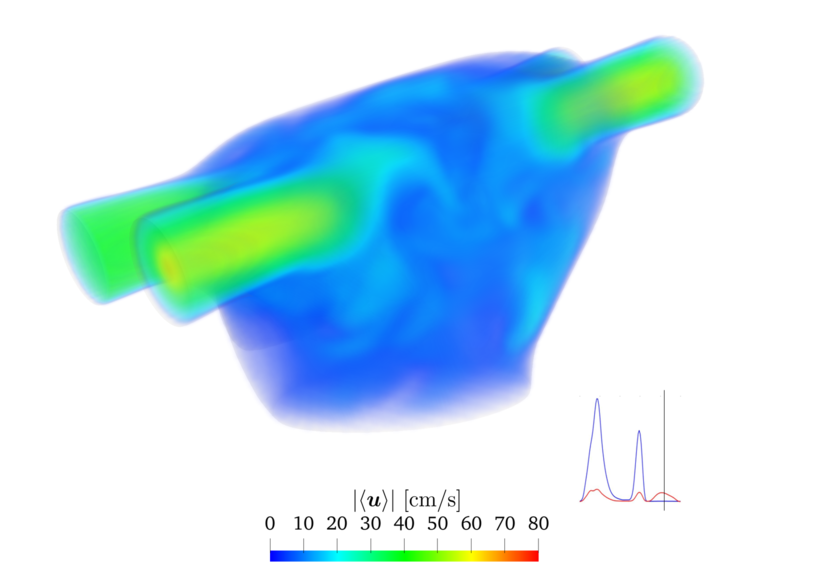}
		\caption{$t = 0.80$ s}
		\label{volume_rendering_80}
	\end{subfigure}
	\begin{subfigure}{.325\textwidth}
		\includegraphics[trim={1 1 1 1 },clip,width=\textwidth]{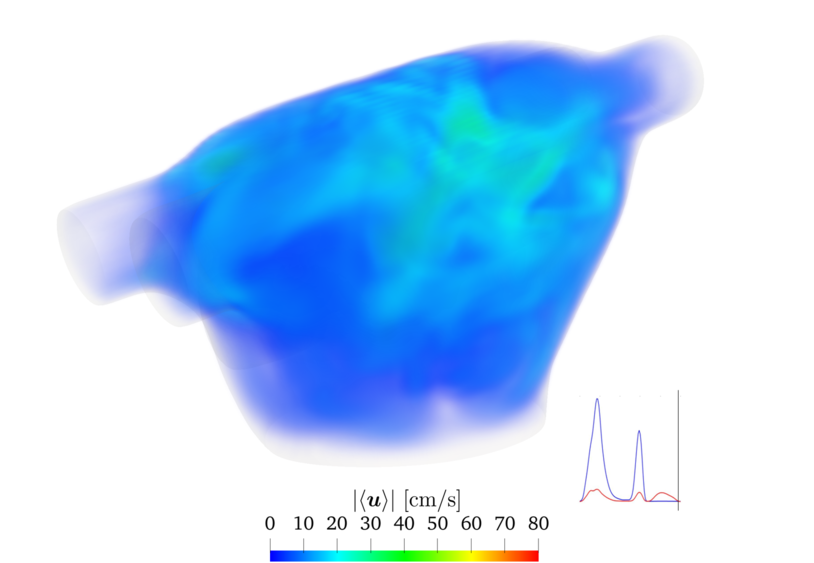}
		\caption{$t = 1.00$ s}
		\label{volume_rendering_100}
	\end{subfigure}
	\caption{Reference solution: volume rendering of phase-averaged velocity magnitude  $|\langle \bm u \rangle |$ at different time instants.}
	\label{fig_volume_rendering}
\end{figure}

In Figure \ref{fig_volume_rendering}, we report volume rendering of the phase-averaged velocity magnitude at different time instants. The flow shows quite complex features, in particular we observe that the jets impact during the heartbeat in three peculiar instants: the E-wave (Figure \ref{volume_rendering_20}), the A-wave (Figure \ref{volume_rendering_40}) and during the filling phase of systole (Figure \ref{volume_rendering_80}). 

\begin{figure}[!t]
	\centering
	\begin{subfigure}{.325\textwidth}
		\centering
		\includegraphics[trim={1 1 1 1 },clip,width=\textwidth]{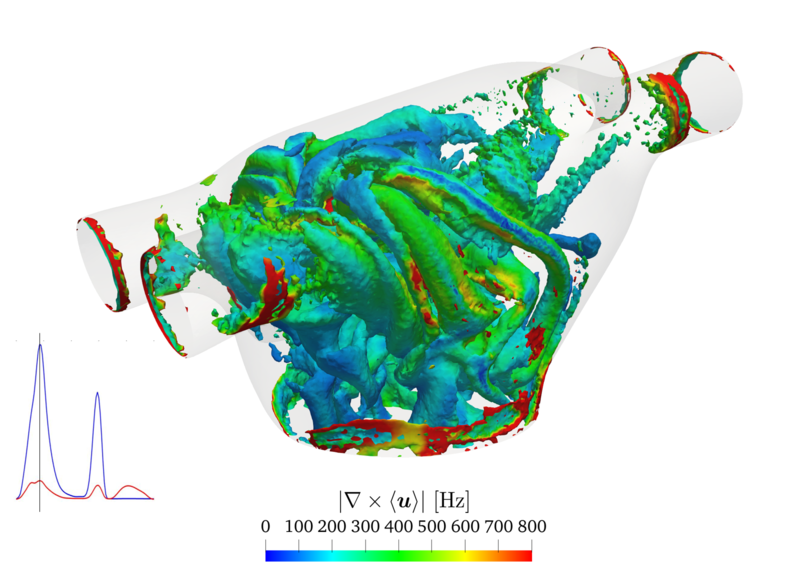}
		\caption{$t = 0.20$ s}
		\label{Qcriterion_vorticity_20}
	\end{subfigure}
	\begin{subfigure}{.325\textwidth}
		\centering
		\includegraphics[trim={1 1 1 1 },clip,width=\textwidth]{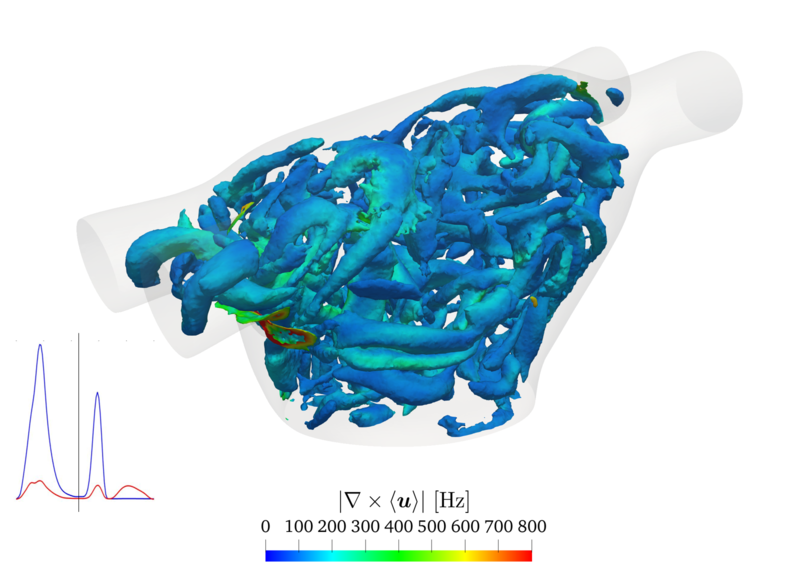}
		\caption{$t = 0.40$ s}
		\label{Qcriterion_vorticity_40}
	\end{subfigure}
	\begin{subfigure}{.325\textwidth}
		\centering
		\includegraphics[trim={1 1 1 1 },clip,width=\textwidth]{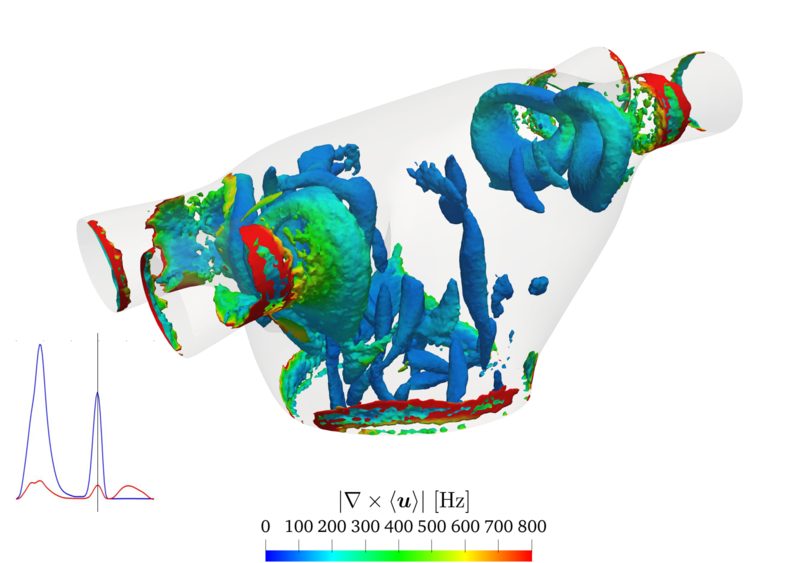}
		\caption{$t = 0.60$ s}
		\label{Qcriterion_vorticity_60}
	\end{subfigure}
	
	\begin{subfigure}{.325\textwidth}
		\centering
		\includegraphics[trim={1 1 1 1 },clip,width=\textwidth]{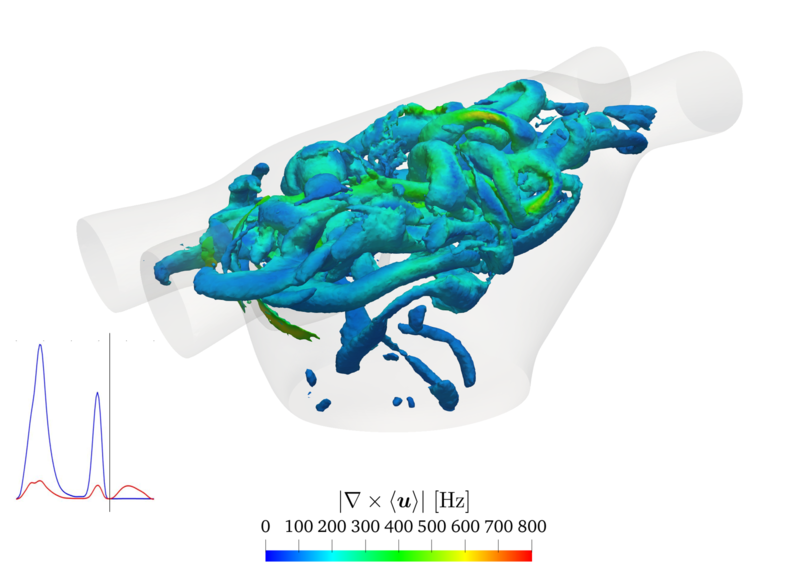}
		\caption{$t = 0.68$ s}
		\label{Qcriterion_vorticity_68}
	\end{subfigure}
	\begin{subfigure}{.325\textwidth}
		\includegraphics[trim={1 1 1 1 },clip,width=\textwidth]{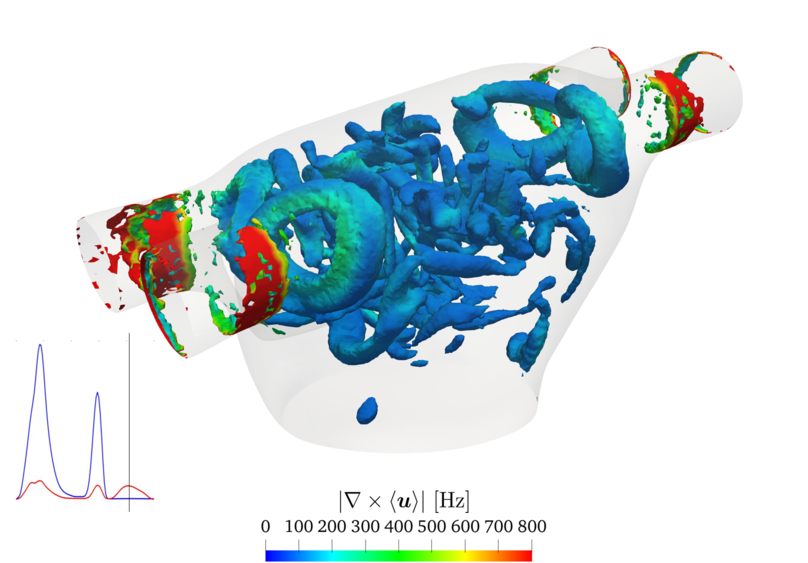}
		\caption{$t = 0.80$ s}
		\label{Qcriterion_vorticity_80}
	\end{subfigure}
	\begin{subfigure}{.325\textwidth}
		\includegraphics[trim={1 1 1 1 },clip,width=\textwidth]{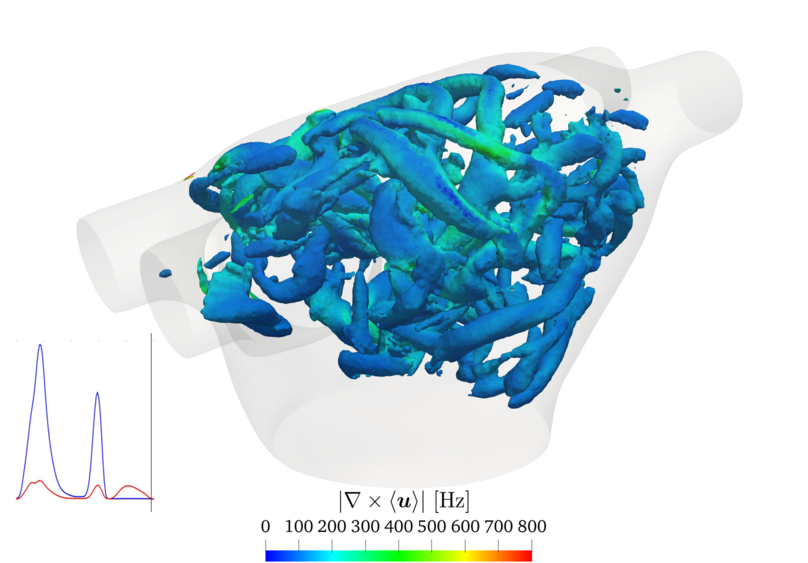}
		\caption{$t = 1.00$ s}
		\label{Qcriterion_vorticity_100}
	\end{subfigure}
	\caption{Reference solution: isosurfaces of Q-criterion $ Q = 2000$ Hz$^2$ coloured by phase-averaged vorticity magnitude  $|\nabla \times \langle \bm u \rangle |$ at different time instants.}
	\label{fig_Qcriterion}
\end{figure}
We split the velocity gradient $\nabla \langle \bm u \rangle$ into its symmetric $\bm \varepsilon (\langle \bm u \rangle)$ and skew-symmetric  $\bm \omega (\langle \bm u \rangle)$ parts: 
\begin{equation}
\nabla \langle \bm u \rangle = \frac{1}{2} \left ( \nabla \langle \bm u \rangle + \left ( \nabla \langle \bm u \rangle \right )^T\right )  + \frac{1}{2} \left ( \nabla\langle \bm u \rangle - \left ( \nabla \langle \bm u \rangle\right )^T\right ) = \bm \varepsilon ({{\langle \bm u \rangle}}) + \bm \omega (\langle \bm u \rangle),
\end{equation}
being respectively the strain-rate tensor and the rotation tensor. In order to identify coherent vortex structures, we introduce the scalar function \cite{Hunt_QCriterion}:
\begin{equation}
Q (\langle \bm u \rangle) = \frac{1}{2}\left( | \bm \omega(\langle \bm u \rangle)|_{\text{F}}^2 - |\bm \varepsilon(\langle \bm u \rangle)|_{\text{F}}^2 \right),
\end{equation}
 where $|\cdot|_{\text{F}}$ is the Frobenius norm of a tensor. If $  Q (\langle \bm u \rangle) > 0$, the rotation of a fluid element becomes dominant over its stretching: the Q-criterion consists in analysing the isosurfaces of the positive part of $  Q (\langle \bm u \rangle)$ \cite{Hunt_QCriterion}. In Figure \ref{fig_Qcriterion} we plot the isosurfaces corresponding to $ Q=2000$ Hz$^2$ coloured with the phase-averaged vorticity magnitude $|\nabla \times \langle \bm u \rangle |$. \color{black}
The main feature of this flow is the formation of vortex rings out of
the PVs when the blood enters in the LA. These rings mutually interact when the corresponding jets impact and then form structures that become smaller and smaller until disappearing by dissipating their energy. 
In Figure \ref{Qcriterion_vorticity_20}, we highlight the impact among the strong jets during the E-wave. Then, at time $t=0.40$ s (Figure \ref{Qcriterion_vorticity_40}), the structures become smaller and they have nearly disappeared as new jet enters at time $t=0.60$ s (Figure \ref{Qcriterion_vorticity_60}) forming four well visible vortex rings around the PVs sections (A-wave). In the refilling phase of systole ($t=0.80$ s), the vortex rings are again visible with some residual structures still present at the center of the chamber.

\begin{figure}[!t]
	\centering
	\begin{subfigure}{.325\textwidth}
		\centering
		\includegraphics[trim={1 1 1 1 },clip,width=\textwidth]{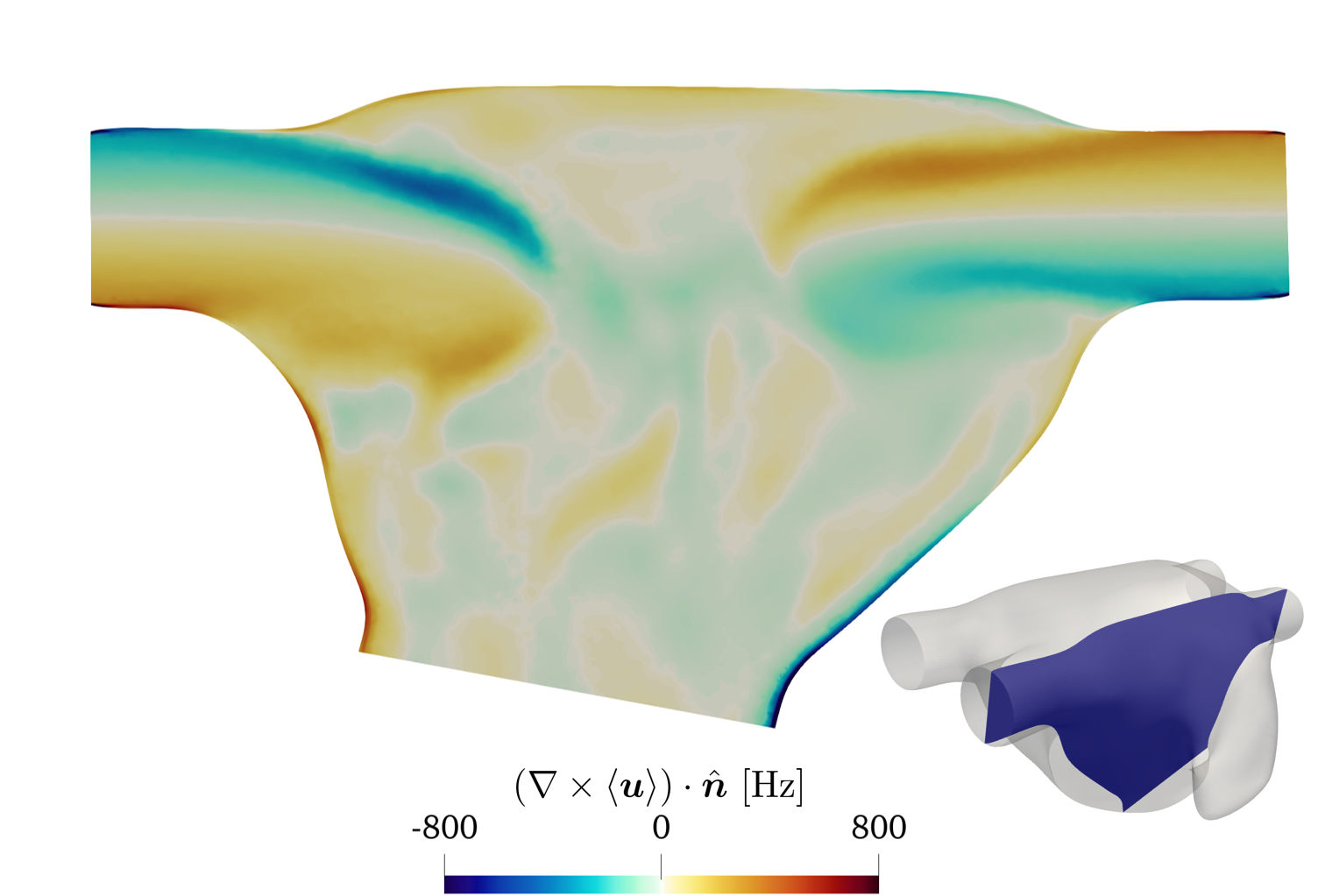}
		\caption{$t = 0.16$ s}
		\label{slice_vorticity_16}
	\end{subfigure}
	\begin{subfigure}{.325\textwidth}
		\centering
		\includegraphics[trim={1 1 1 1 },clip,width=\textwidth]{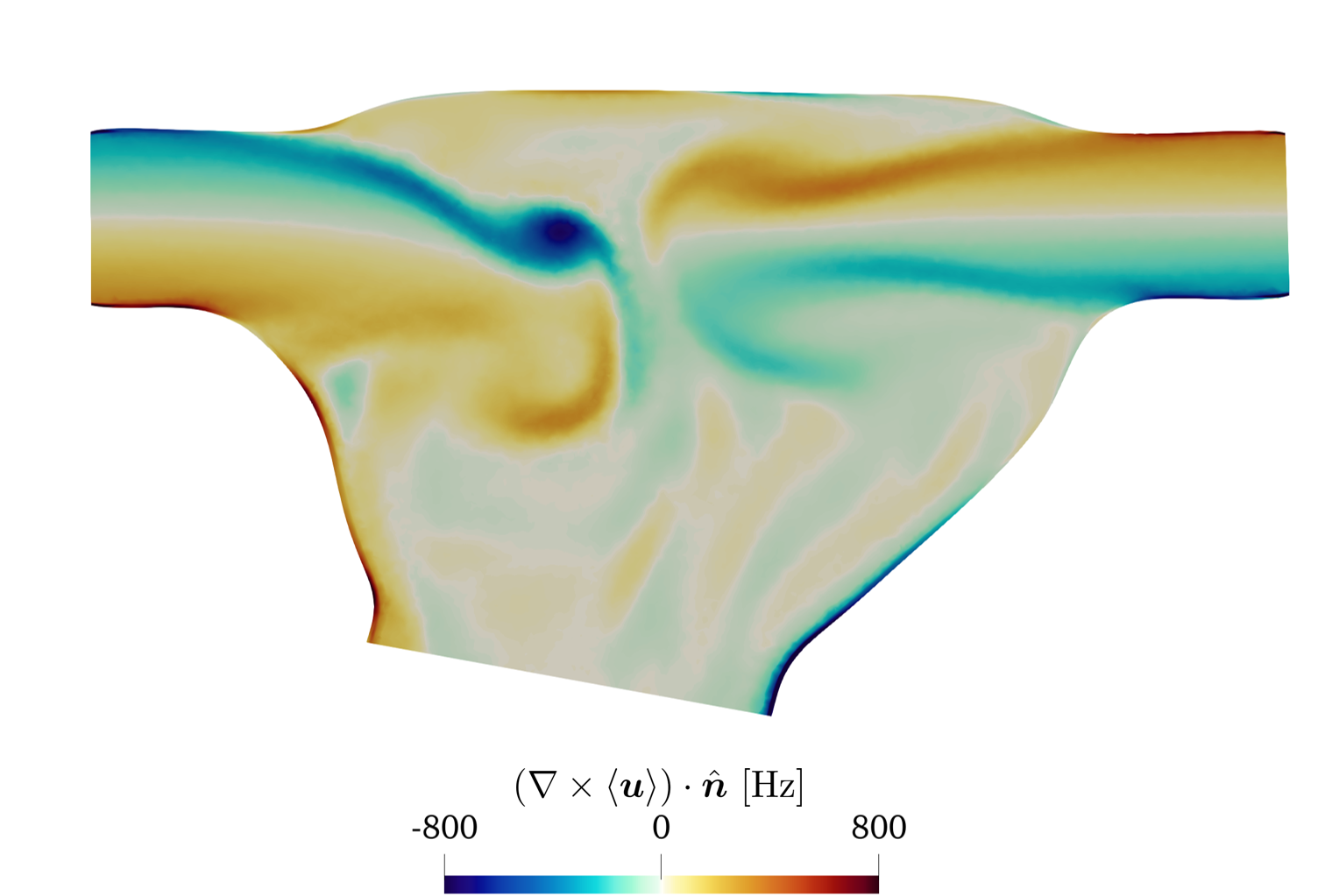}
		\caption{$t = 0.17$ s}
		\label{slice_vorticity_17}
	\end{subfigure}
	\begin{subfigure}{.325\textwidth}
		\centering
		\includegraphics[trim={1 1 1 1 },clip,width=\textwidth]{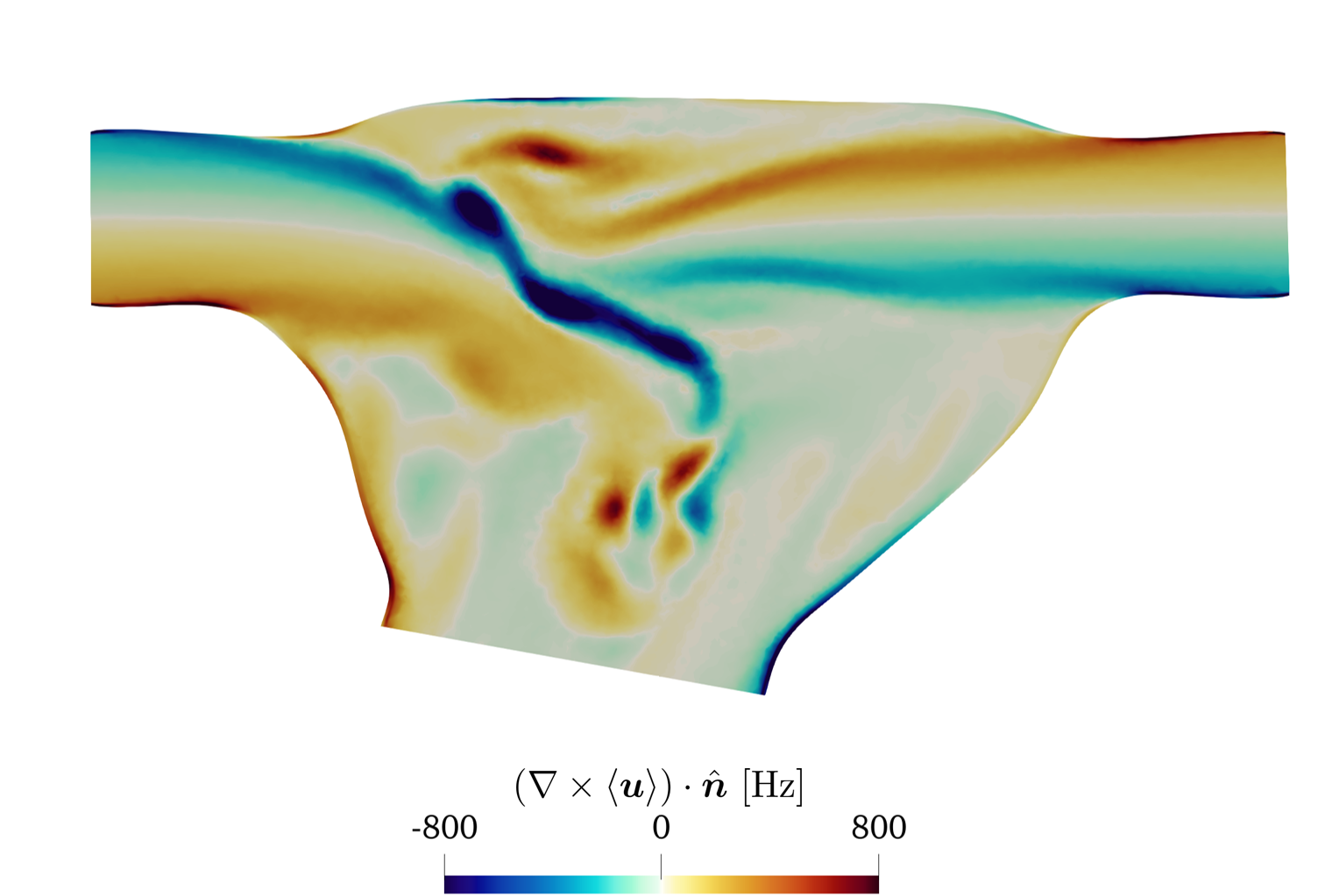}
		\caption{$t = 0.18$ s}
		\label{slice_vorticity_18}
	\end{subfigure}
	
	\begin{subfigure}{.325\textwidth}
		\centering
		\includegraphics[trim={1 1 1 1 },clip,width=\textwidth]{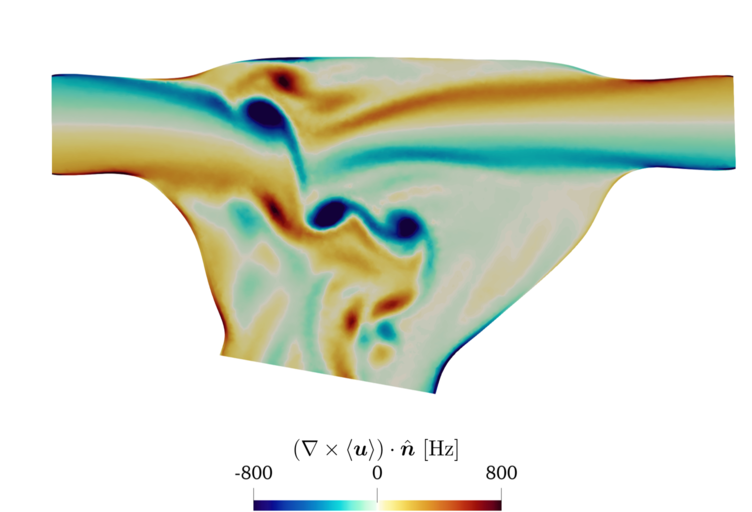}
		\caption{$t = 0.20$ s}
		\label{slice_vorticity_20}
	\end{subfigure}
	\begin{subfigure}{.325\textwidth}
		\includegraphics[trim={1 1 1 1 },clip,width=\textwidth]{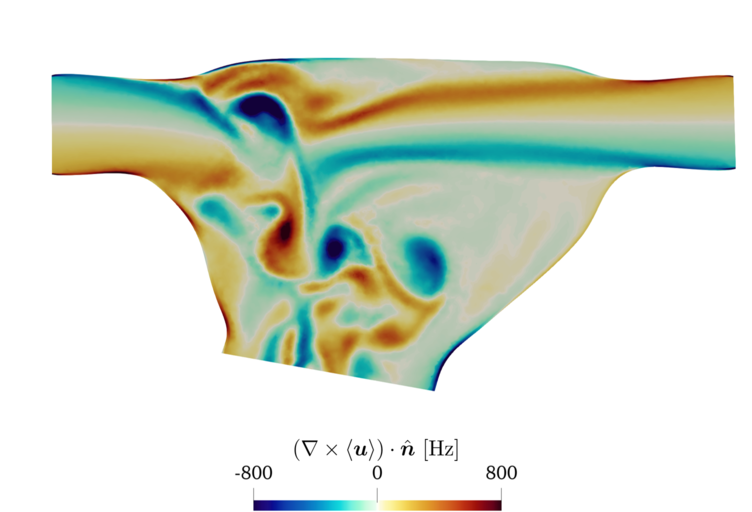}
		\caption{$t = 0.21$ s}
		\label{slice_vorticity_21}
	\end{subfigure}
	\begin{subfigure}{.325\textwidth}
		\includegraphics[trim={1 1 1 1 },clip,width=\textwidth]{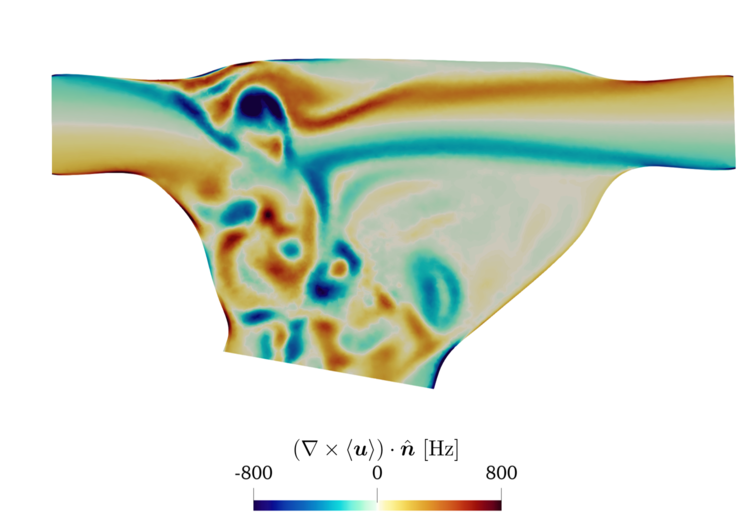}
		\caption{$t = 0.22$ s}
		\label{slice_vorticity_22}
	\end{subfigure}
	\caption{Reference solution: projection of the phase-averaged vorticity on the normal direction of a slice cutting two PVs (top-left) $(\nabla \times \langle \bm u \rangle ) \cdot \hat{\bm n}$. Results at different time instants in the proximity of the E-wave ($t=0.20$ s).}
	\label{fig_slice_vorticity}
\end{figure}

By focusing on the impact during the E-wave, in Figure \ref{fig_slice_vorticity}, we show the projection of the phase-averaged vorticity $\nabla \times \langle \bm u \rangle$ on the normal direction of a slice cutting two PVs.  We observe the formation of shear layers from the PVs (Figure \ref{slice_vorticity_16}), a early-stage interaction in Figure \ref{slice_vorticity_17} along with some recirculation regions. Then, from $t=0.18$ s, we observe perturbed shear layers with a coalescence of vortices and a dispersion of the organized flow pattern previously seen (Figures \ref{slice_vorticity_18}, \ref{slice_vorticity_20}). In particular, the vortices breakdown propagates in the rest of the chamber, towards the MV section (Figures \ref{slice_vorticity_21}, \ref{slice_vorticity_22}).

\begin{figure}[t!]
	\centering
	\begin{subfigure}{.325\textwidth}
		\centering
		\includegraphics[trim={40 1 40 10 },clip,width=\textwidth]{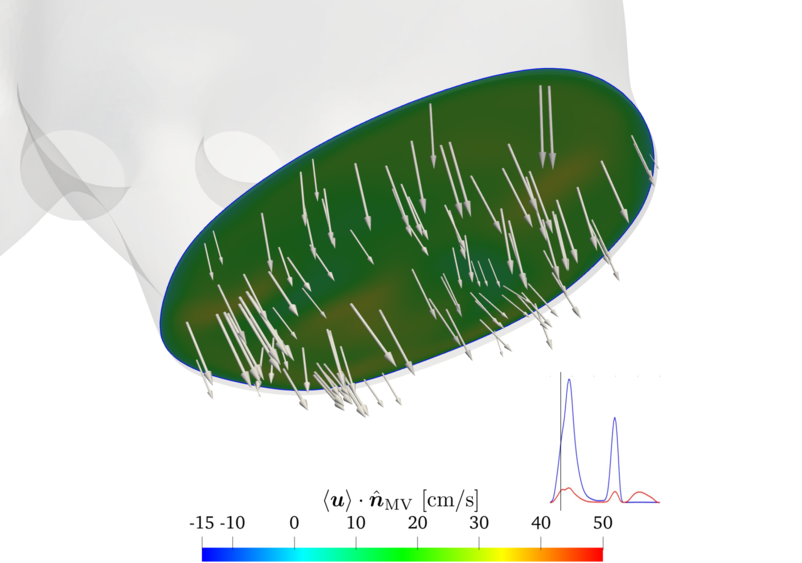}
		\caption{$t = 0.10$ s}
		\label{flux_MV_10}
	\end{subfigure}
	\begin{subfigure}{.325\textwidth}
		\centering
		\includegraphics[trim={40 1 40 10 },clip,width=\textwidth]{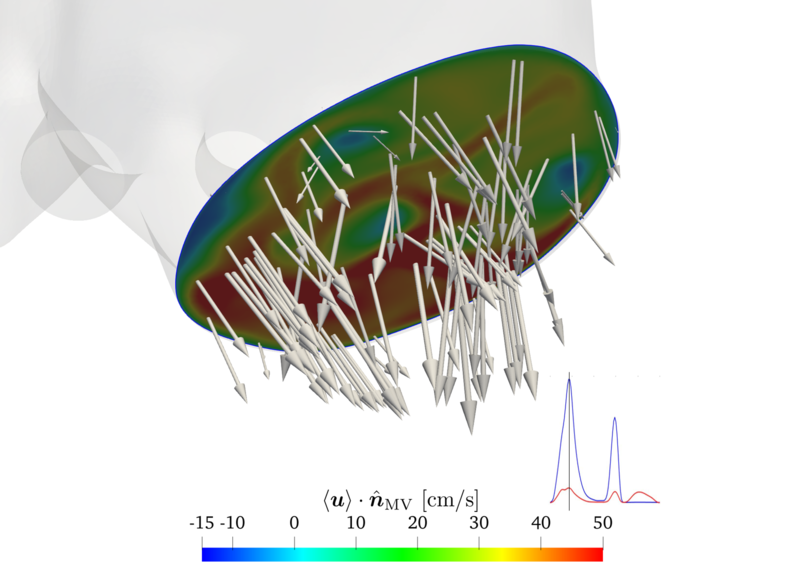}
		\caption{$t = 0.20$ s}
		\label{flux_MV_20}
	\end{subfigure}
	\begin{subfigure}{.325\textwidth}
		\centering
		\includegraphics[trim={40 1 40 10 },clip,width=\textwidth]{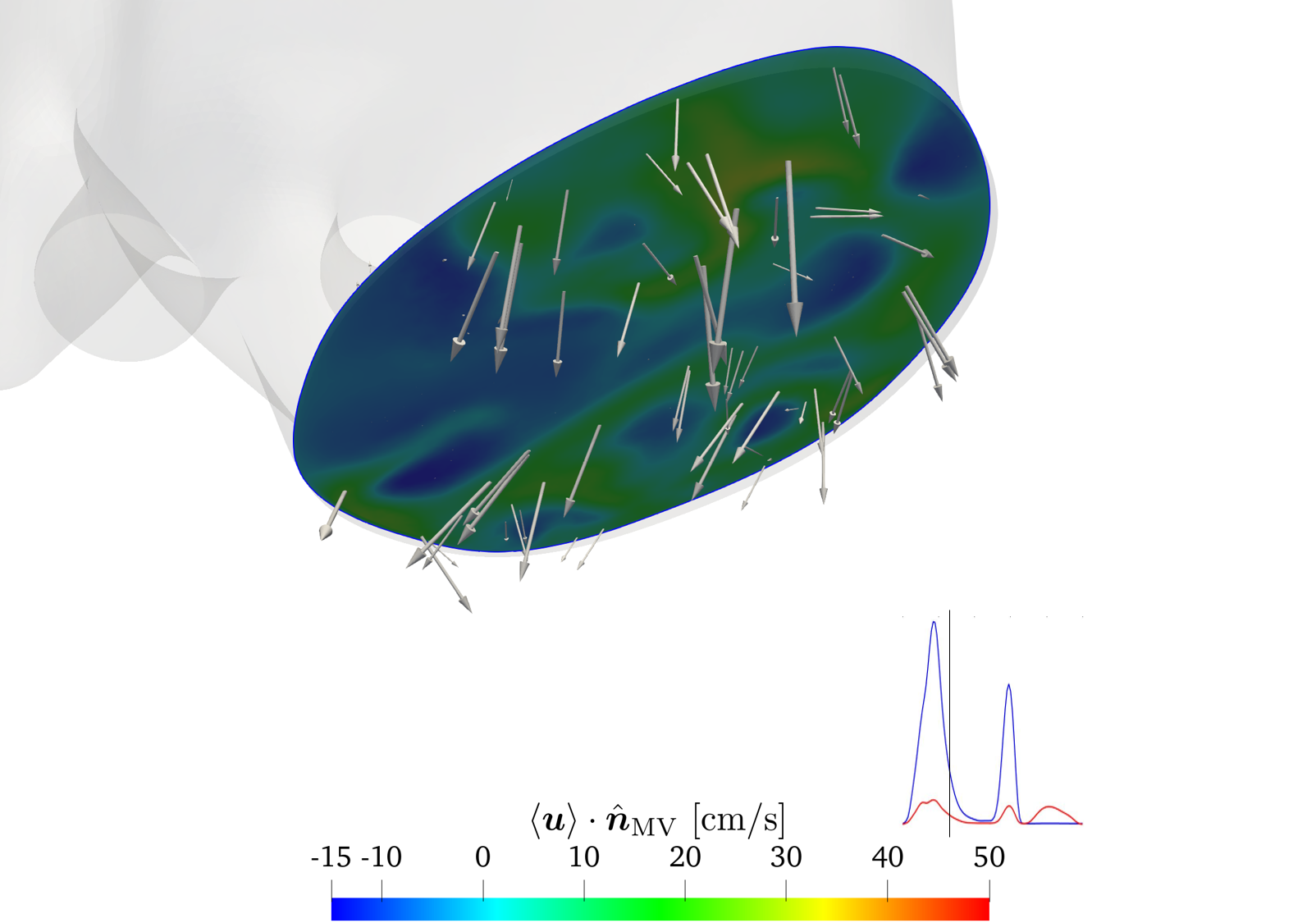}
		\caption{$t = 0.30$ s}
		\label{flux_MV_30}
	\end{subfigure}
	
	\begin{subfigure}{.325\textwidth}
		\centering
		\includegraphics[trim={40 1 40 10 },clip,width=\textwidth]{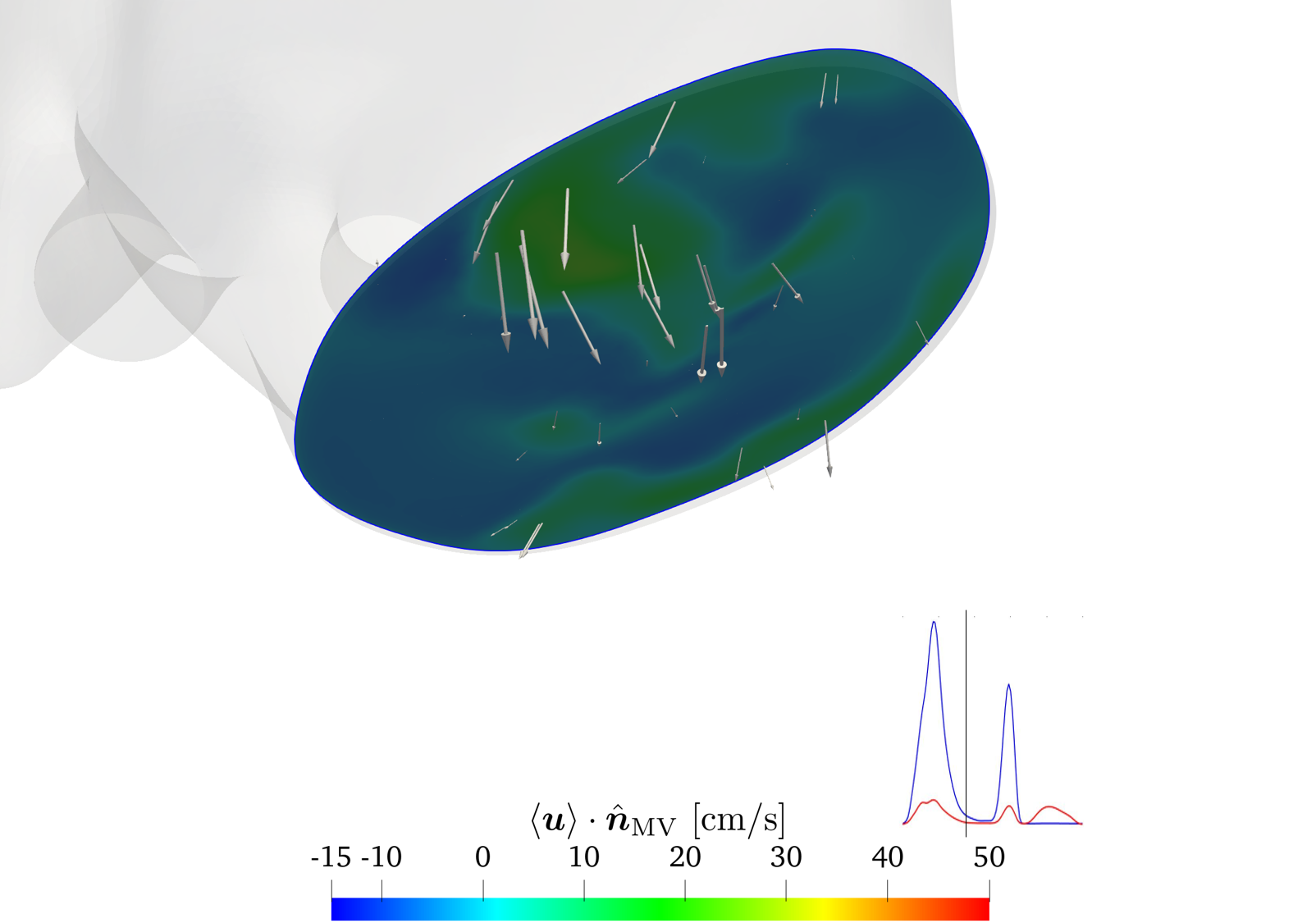}
		\caption{$t = 0.40$ s}
		\label{flux_MV_40}
	\end{subfigure}
	\begin{subfigure}{.325\textwidth}
		\includegraphics[trim={40 1 40 10 },clip,width=\textwidth]{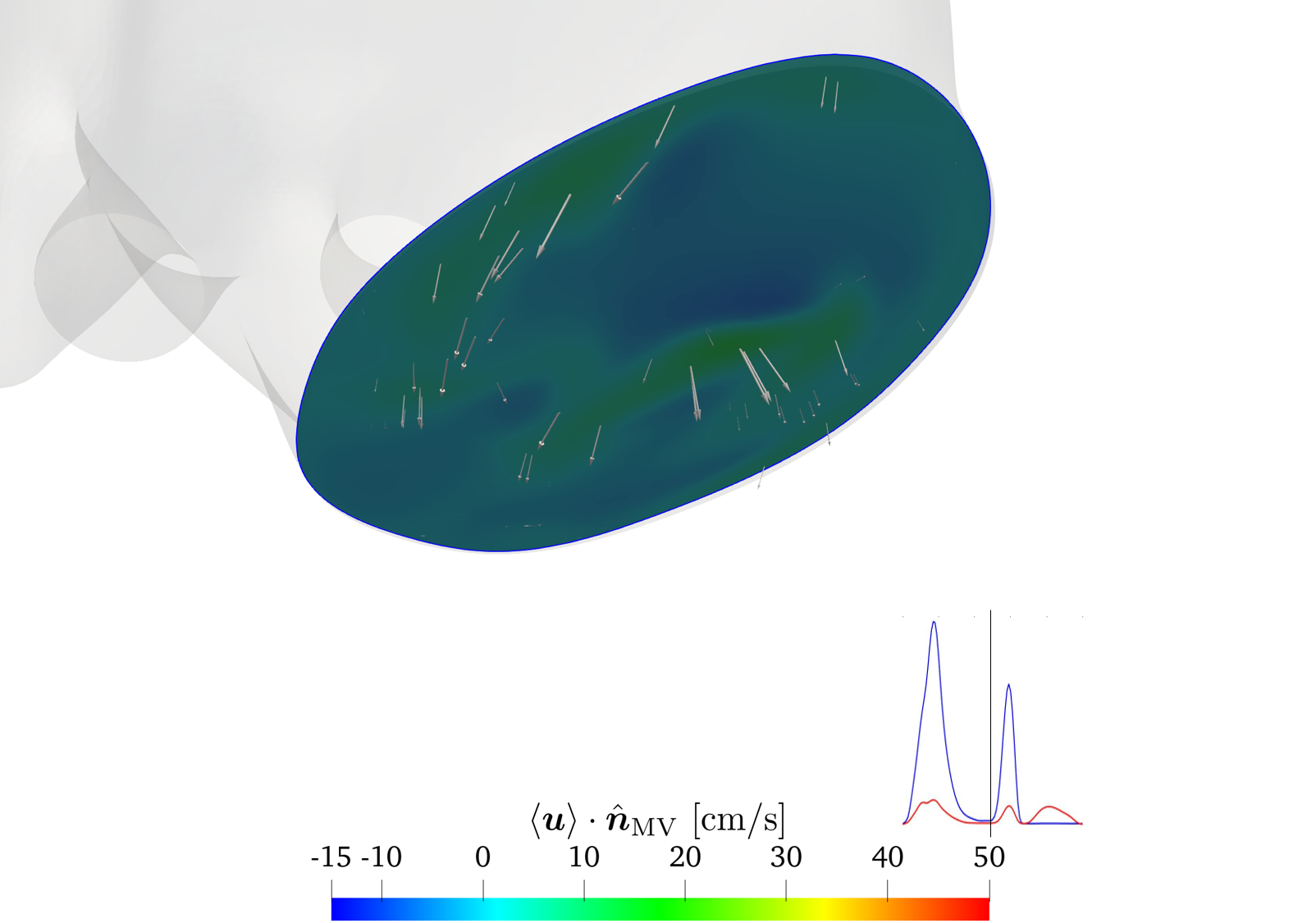}
		\caption{$t = 0.50$ s}
		\label{flux_MV_50}
	\end{subfigure}
	\begin{subfigure}{.325\textwidth}
		\includegraphics[trim={40 1 40 10 },clip,width=\textwidth]{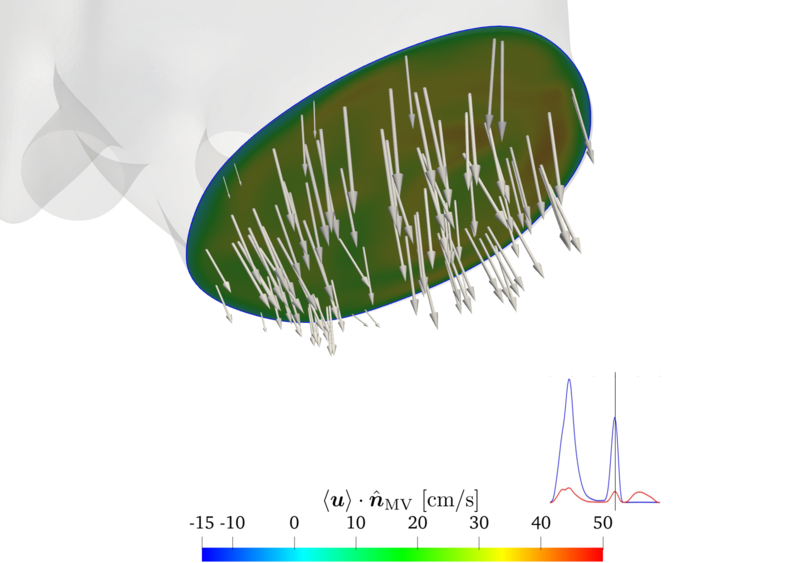}
		\caption{$t = 0.60$ s}
		\label{flux_MV_60}
	\end{subfigure}
	\caption{Reference solution: glyphs of velocity vector at the MV section during diastole on a slice coloured with $\langle \bm{u} \rangle \cdot \hat{\bm n}_{\text{MV}}$, i.e. the scalar product among the phase-averaged velocity and the outward pointing unit vector normal to the MV section.}
	\label{fig_flux_MV}
\end{figure}

\begin{figure}[t!]
		\centering
		\includegraphics[trim={1 1 1 1 },clip,width=\textwidth]{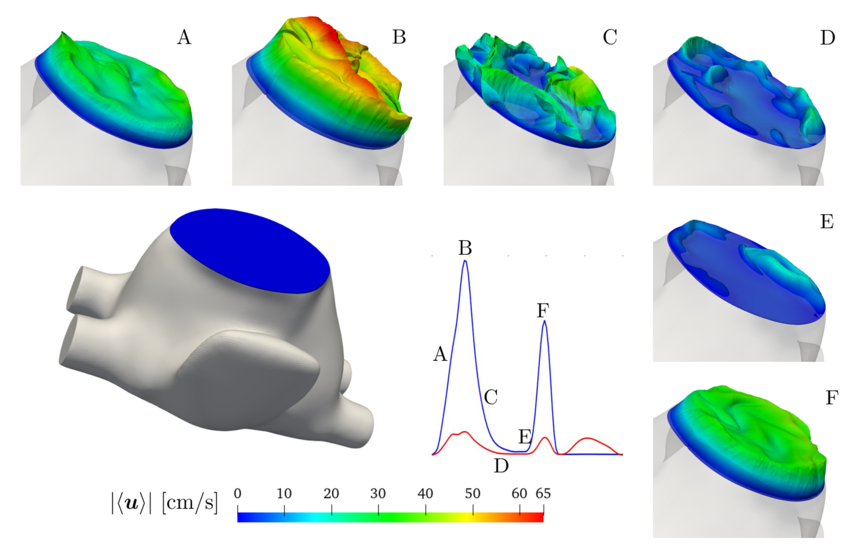}
		\caption{\color{black}Reference solution: velocity profile at the MV section at different time during diastole. \color{black}}
		\label{velocity_MV_warp}
\end{figure}

The velocity profile at the MV is an interesting output of this computation since it can be used as input for the simulation of the LV hemodynamics \cite{TDQ_2d,TDQ_3d}. 
In Figure \ref{fig_flux_MV}, we report glyphs of velocity vector at the MV section during diastole  (i.e. when the MV is open) on a slice coloured with $\langle \bm{u} \rangle \cdot \hat{\bm n}_{\text{MV}}$, i.e. the scalar product among the phase-averaged velocity and the outward pointing unit vector normal to the MV section. 
We notice that the velocity profile that we obtain is highly variable in time and, more importantly, the velocity shows a flat profile only at some specific times, such as at $t=0.10$ s (Figure \ref{flux_MV_10}). Even when the flow is intense, such as at $t=0.20$ s or $t=0.60$ s, the velocity profile is never flat but, on the contrary, the presence of vortices located above the MV section produces low velocity regions as shown in Figures \ref{flux_MV_20} and \ref{flux_MV_60}. During the time between the two waves, the flow rate is positive, as can be seen in Figures \ref{flux_MV_30}, \ref{flux_MV_40} and \ref{flux_MV_50},  but some recirculating velocities are visible in some spots reaching negative values of $\langle \bm{u} \rangle \cdot \hat{\bm n}_{\text{MV}} = - 15$ cm/s. \color{black} In Figure \ref{velocity_MV_warp}, we also report the MV velocity profile at different instants during diastole, which we remark being an output of our numerical simulations. The velocity profiles obtained significantly differ from a flat profile, a Pouiseuille profile, or, more generally, from those analytical profiles generally prescribed as inlet boundary conditions during diastole for haemodynamic simulations of the LV (i.e. on the MV section) \cite{Domenichini_Pedrizzetti_2011, Domenichini_Pedrizzetti_2005, MKKDL_ventr}.  We made these profiles publicly available at the repository \cite{repository_cfdmesh}: they can be used, after a suitable fitting in space and time, to prescribe an inflow boundary condition at MV section during diastole for the LV simulation. This boundary treatment better accounts for the effect of the flow coming from the LA, which may considerably affects the haemodynamics of the LV. \color{black}

\newpage
\begin{figure}[!t]
	\centering
	\begin{subfigure}{.325\textwidth}
		\centering
		\includegraphics[trim={1 1 1 1 },clip,width=\textwidth]{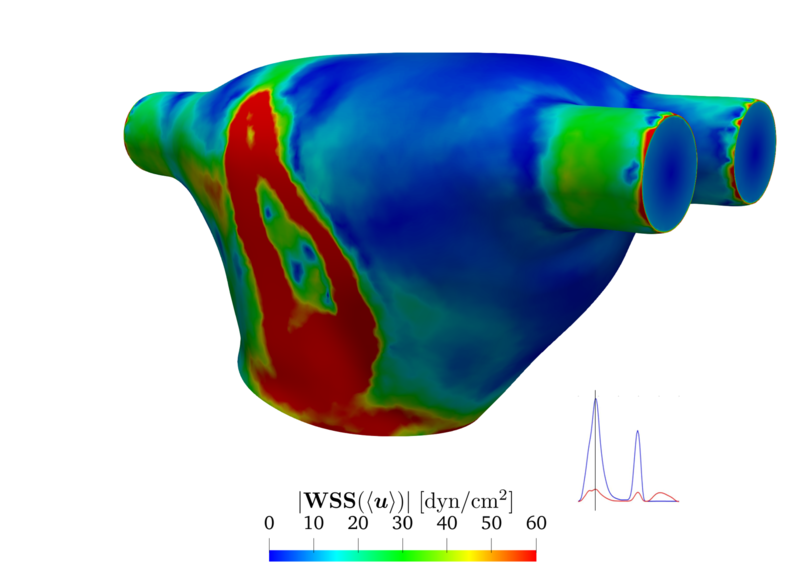}
		\caption{$t = 0.20$ s}
		\label{WSS_ref_20}
	\end{subfigure}
	\begin{subfigure}{.325\textwidth}
		\centering
		\includegraphics[trim={1 1 1 1 },clip,width=\textwidth]{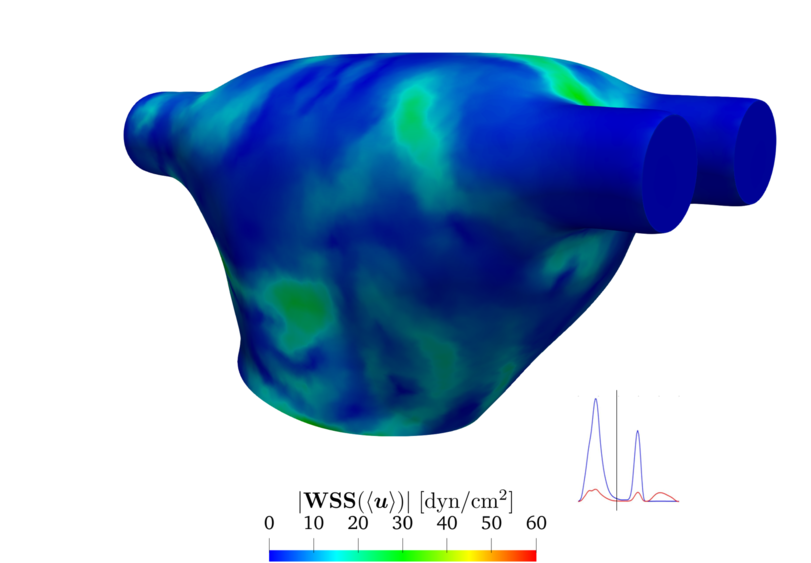}
		\caption{$t = 0.40$ s}
		\label{WSS_ref_40}
	\end{subfigure}
	\begin{subfigure}{.325\textwidth}
		\centering
		\includegraphics[trim={1 1 1 1 },clip,width=\textwidth]{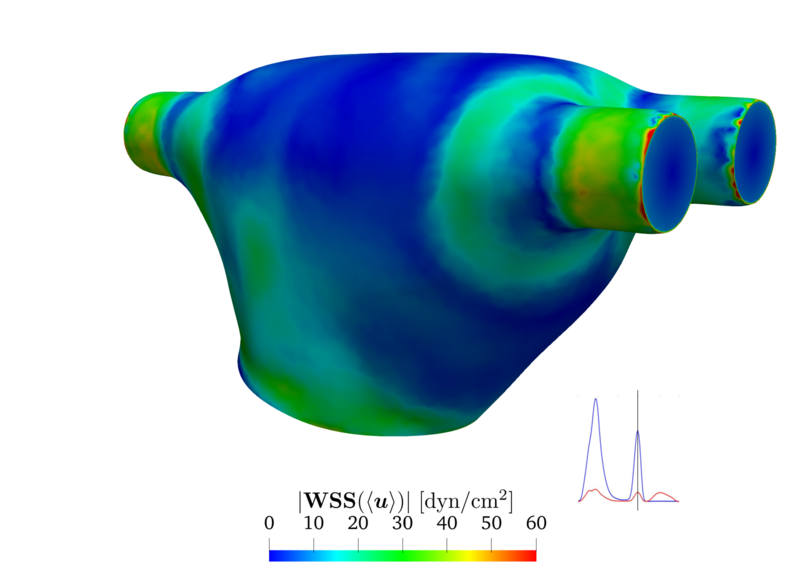}
		\caption{$t = 0.60$ s}
		\label{WSS_ref_60}
	\end{subfigure}
	
	\begin{subfigure}{.325\textwidth}
		\centering
		\includegraphics[trim={1 1 1 1 },clip,width=\textwidth]{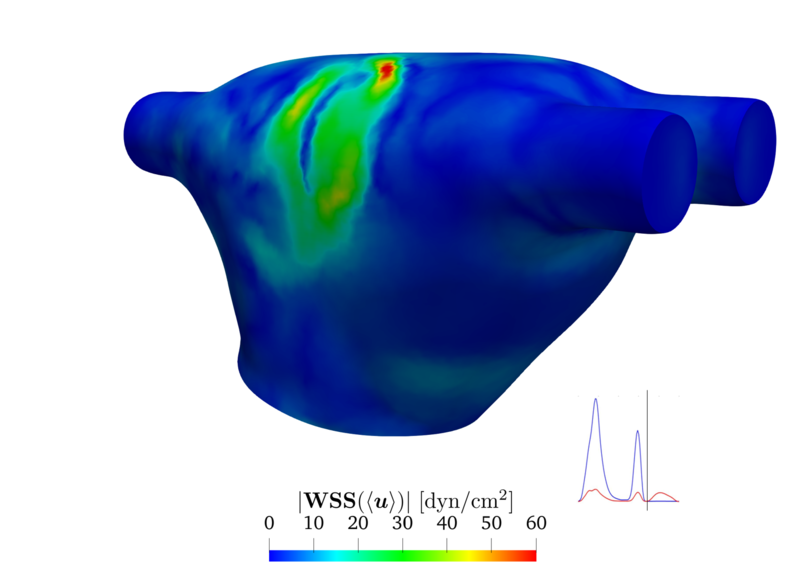}
		\caption{$t = 0.68$ s}
		\label{WSS_ref_68}
	\end{subfigure}
	\begin{subfigure}{.325\textwidth}
		\includegraphics[trim={1 1 1 1 },clip,width=\textwidth]{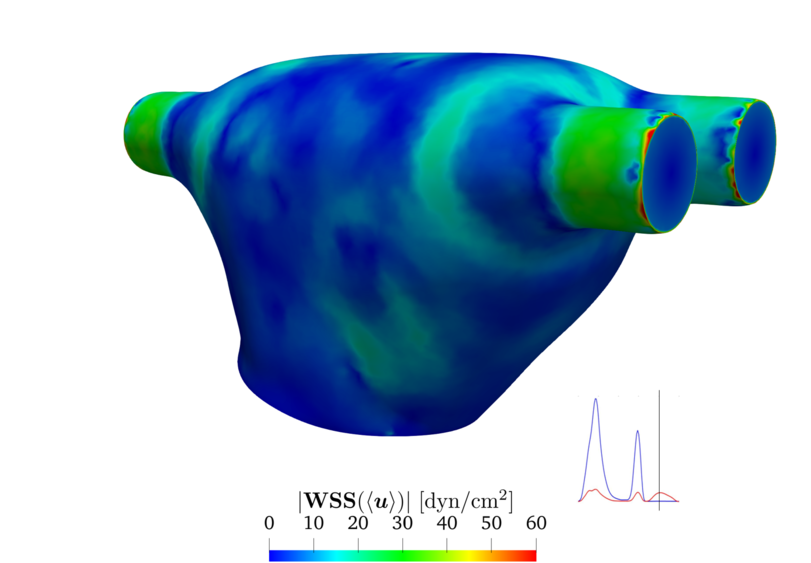}
		\caption{$t = 0.80$ s}
		\label{WSS_ref_80}
	\end{subfigure}
	\begin{subfigure}{.325\textwidth}
		\includegraphics[trim={1 1 1 1 },clip,width=\textwidth]{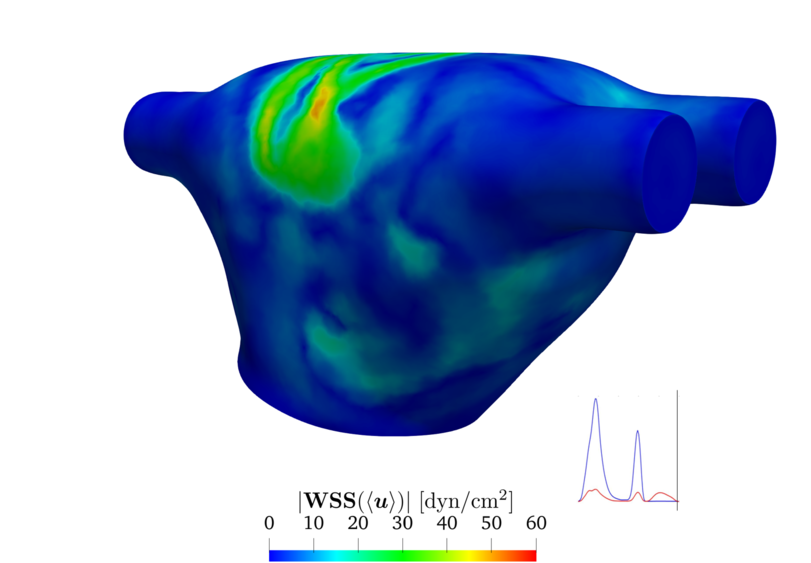}
		\caption{$t = 1.00$ s}
		\label{WSS_ref_100}
	\end{subfigure}
	\caption{Reference solution: wall shear stress (WSS) magnitude at different time instants.}
	\label{WSS_ref}
\end{figure}
In view of calculating  hemodynamic indicators, we define the viscous stress tensor related to the phase-averaged velocity field as
\begin{equation}
\bm \tau (\langle \bm u \rangle) = 2 \mu \bm \varepsilon  (\langle \bm u \rangle).
\end{equation}
We compute the vector wall shear stress (\textbf{WSS}) on the boundary of the reference configuration $\Omega_0$ (i.e. the LA at the beginning of diastole) as
\begin{equation}
\textbf{WSS} (\langle \bm u \rangle) = \bm \tau (\langle \bm u \rangle) \hat{\bm n} - \left ( \bm \tau (\langle \bm u \rangle) \hat{\bm n} \cdot \hat{\bm n}\right )\hat {\bm n } \quad \text{on } \partial \Omega_0,
\end{equation}
and the scalar fields time averaged wall shear stress (TAWSS),
oscillatory shear index (OSI) and relative residence time (RRT) (see \cite{KFH_leftatrium,KU_osi,HG_rrt}).
\color{black}These indicators can help shedding light on long-term response of endothelial cells since they are affected by both the magnitude of the WSS and its evolution in time. For this reason, they can be used to identify formation of new tissues, plaques and the promoting of neointimal hyperplasia \cite{KU_osi}.  \color{black} 
With the WSS, we compute the TAWSS as the integral over the time period of the magnitude of the \textbf{WSS},
\begin{equation}
\text{TAWSS}(\langle \bm u \rangle) = \frac{1}{T_\text{HB}} \int_{0}^{T_\text{HB}} | \mathbf{WSS} (\langle \bm u \rangle) |_2 dt  \quad \text{on } \partial \Omega_0,
\end{equation}
where $| \cdot |_2$ denotes the Euclidean norm of a vector. The OSI is defined as \cite{KU_osi}:
\begin{equation}
\text{OSI}(\langle \bm u \rangle) =  \frac{1}{2} \left( 1 - \dfrac{\left | \int_{0}^{T_\text{HB}}  \mathbf{WSS} (\langle \bm u \rangle) dt \right|_2}{ \int_{0}^{T_\text{HB}} \left| \mathbf{WSS} (\langle \bm u \rangle) \right|_2 dt } \right) \quad \text{on } \partial \Omega_0,
\end{equation}
and it is higher in regions where the WSS changes much during a heart cycle. Finally, we compute the RRT as in \cite{HG_rrt}
\begin{equation}
\text{RRT}(\langle \bm u \rangle) =  \left( \left(1 - 2\,\text{OSI} (\langle \bm u \rangle) \right) \dfrac{1}{T_\text{HB}}
\int_{0}^{T_\text{HB}} \left| \mathbf{WSS} (\langle \bm u \rangle) \right|_2 dt \right)^{-1} \quad \text{on } \partial \Omega_0.
\end{equation}
The RRT is proportional to the residence time of blood particles in the proximity of the wall, and it can be regarded as a convenient
fluid dynamics indicator to identify regions where WSS is both low and oscillatory \cite{Domanin_2017}. 

In Figure \ref{WSS_ref}, we report the WSS magnitude as computed on the surface of the LA at different time instants by using the phase averaged velocity. The largest values  are attained during the E-wave in the middle of the surface of the LA, towards the MV. This region corresponds to areas where vortices interact and are pushed towards the LA wall. During the rest of the cycle, the WSS values remain quite small; large values are attained only in the PVs and in the lower part of the LA. 

\begin{figure}[!t]
	\centering
	\begin{subfigure}{0.325\textwidth}
		\centering
		\includegraphics[trim={5 2 4 2 },clip,width=1.1\textwidth]{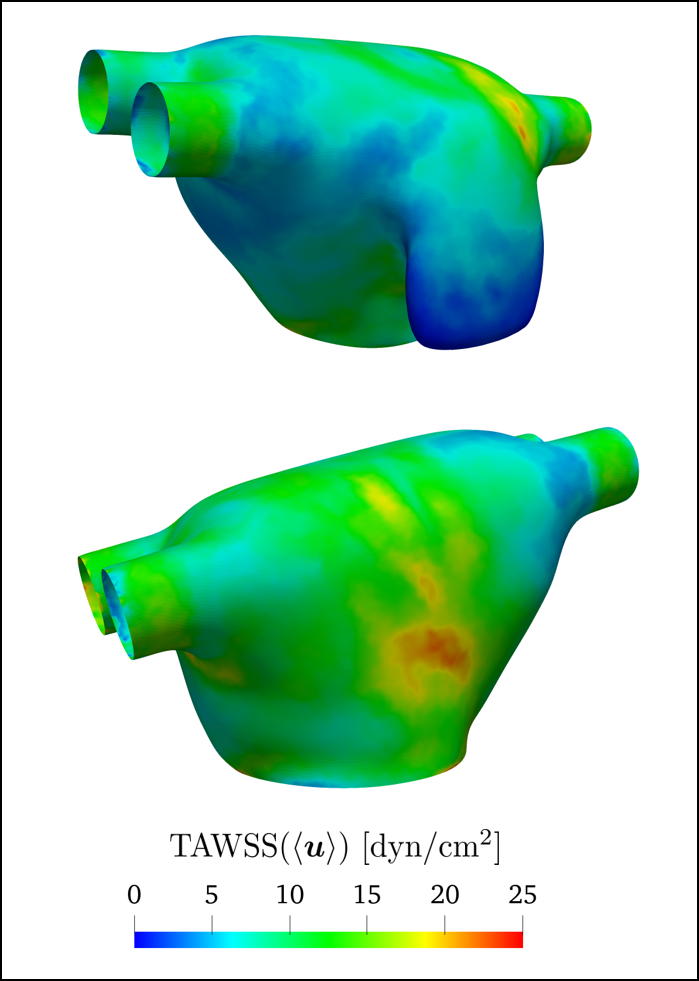}
		\caption{TAWSS}
		\label{TAWSS_ref}
	\end{subfigure}
	\begin{subfigure}{0.325\textwidth}
	\centering
	\includegraphics[trim={5 2 2 2 },clip,width=1.1\textwidth]{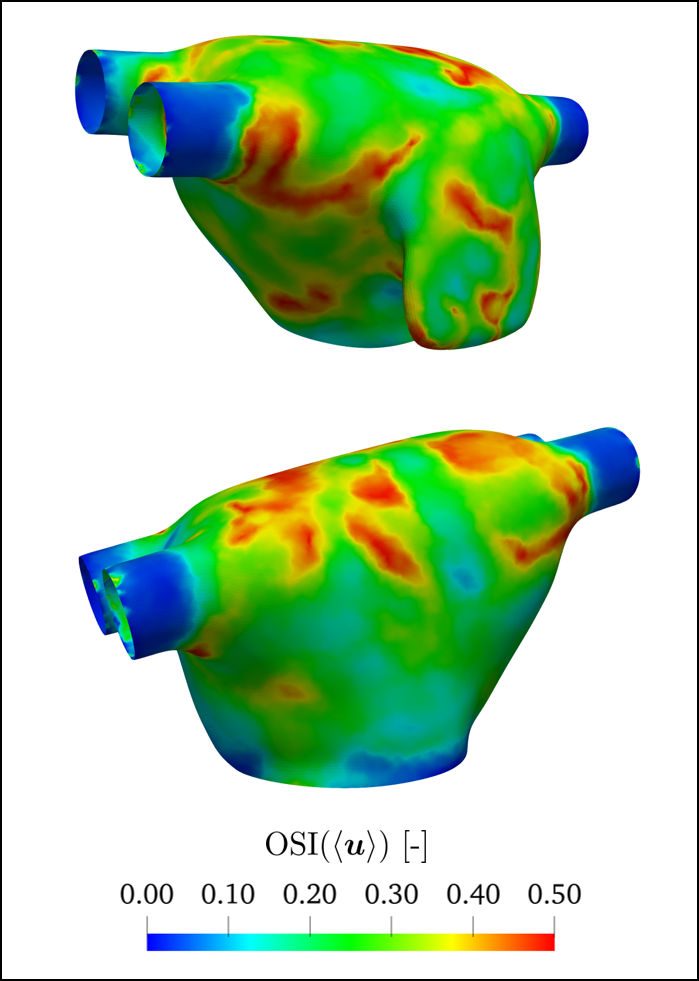}
	\caption{OSI}
	\label{OSI_ref}
\end{subfigure}
	\begin{subfigure}{0.325\textwidth}
	\centering
	\includegraphics[trim={5 2 4 2 },clip,width=1.1\textwidth]{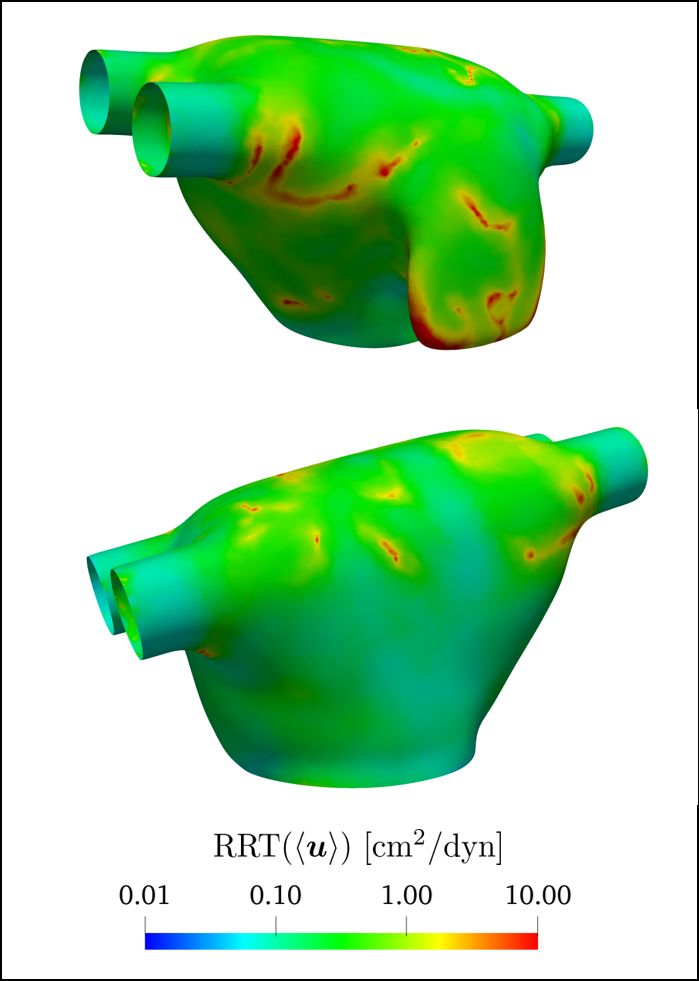}
	\caption{RRT}
	\label{RRT_ref}
\end{subfigure}
	\caption{Reference solution, haemodynamic indicators from two different perspectives: (a) TAWSS, (b) OSI, (c) RRT.}
	\label{TAWSS_OSI_RRT_ref}
\end{figure}

Figure~\ref{TAWSS_ref} shows the TAWSS on the reference configuration from two different perspectives: low values of the TAWSS are achieved in the LAA, while some peaks can be appreciated in the opposite side of the chamber, in accordance with the large values of $|\textbf{WSS}|$ previously observed due to the interaction among the vortices and the endocardium. 

In Figure \ref{OSI_ref}  we report the OSI computed in the same settings of Figure \ref{TAWSS_ref}. The OSI is large on the top of the LA where a large recirculation is present and on the bottom of the LAA, revealing hence a significant variation of the wall shear stress.

As a qualitative indication of the time that a fluid particle spends in the vicinity of the wall, we report in Figure~\ref{RRT_ref} the RRT: as expected, the largest values are attained in the bottom of the LAA. We suggest it could be related to the shape and position of the LAA, where the blood reaches very low velocities and recirculation effects are observed. Interestingly, analogous considerations in terms of all the analysed hemodynamic indicators are found in healthy patient-specific studies as highlighted in \cite{KFH_leftatrium}, both in terms of magnitude and their distribution on the LA surface.

\begin{table}[!t]
	\centering
	\begin{tabular}{|c|c|c|c|c|c|c|}
		\hline
		{time} [s] & $t = 1.00$  &  $t = 2.00$ &  $t = 3.00$  &  $t = 4.00$ &  $t = 5.00 $  &  $t = 6.00 $  \\
		\hline
		{particles} & 37'971 & 7'336 & 1'583 & 344 & 91 & 42 \\
		\hline
		{\% on total injected} & 75.23 & 14.53 & 3.14 & 0.68 & 0.18 & 0.08 \\
		\hline
	\end{tabular}
	\caption{Reference solution: particles remaining in the LA at the end of each cardiac cycle and percentage of particles on total injected. }
	\label{table_particles}
\end{table}

\begin{figure}[!t]
	\centering
	\begin{subfigure}{.48\textwidth}
		\centering
		\includegraphics[trim={1 1 1 1},clip,width=\textwidth]{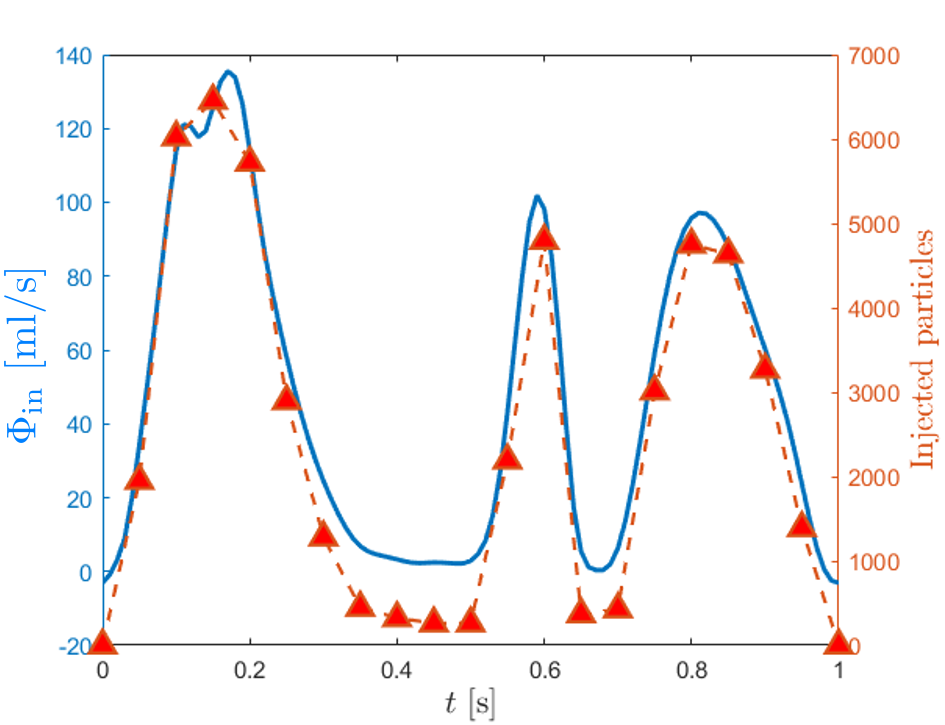}
		\caption{}
		\label{injected_particles}
	\end{subfigure}
	\begin{subfigure}{.48\textwidth}
		\centering
		\includegraphics[trim={1 1 1 1},clip,width=\textwidth]{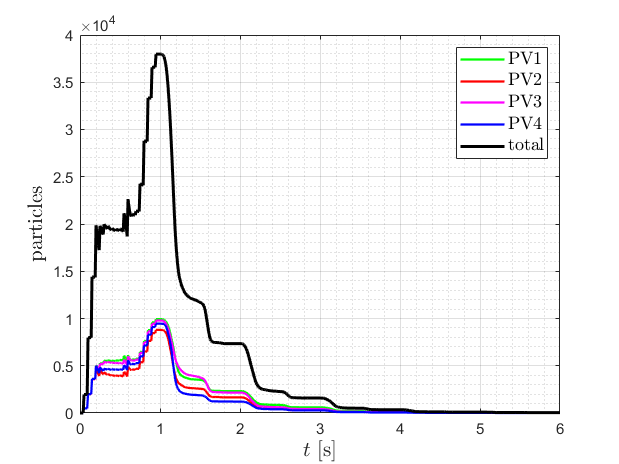}
		\caption{}
		\label{plot_particles}
	\end{subfigure}
	\caption{{Reference solution. Particles injected every 0.05 s (in red) are proportional to the inlet flow rate (in blue) (left). Number of particles inside the LA during 6 cardiac cycles, introducing particles in the first cycle only. With different colours: the number of particles in the chamber coming from different PVs (right).}}
	\label{injected_particles_and_plot_particles}
\end{figure}

\begin{figure}[!t]
	\centering
	\begin{subfigure}{.325\textwidth}
		\centering
		\includegraphics[trim={1 1 1 1 },clip,width=\textwidth]{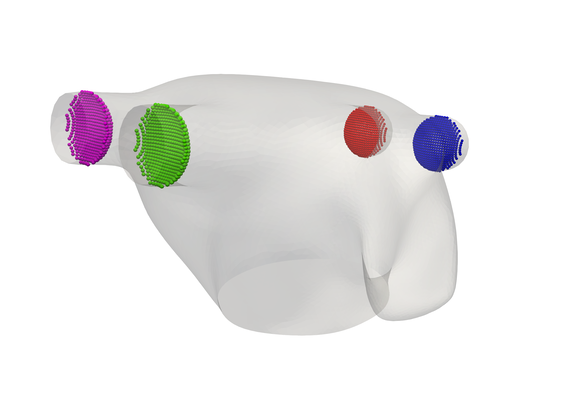}
		\caption{$t = 0.05 $ s}
		\label{frame_0p05}
	\end{subfigure}
	\begin{subfigure}{.325\textwidth}
		\centering
		\includegraphics[trim={1 1 1 1 },clip,width=\textwidth]{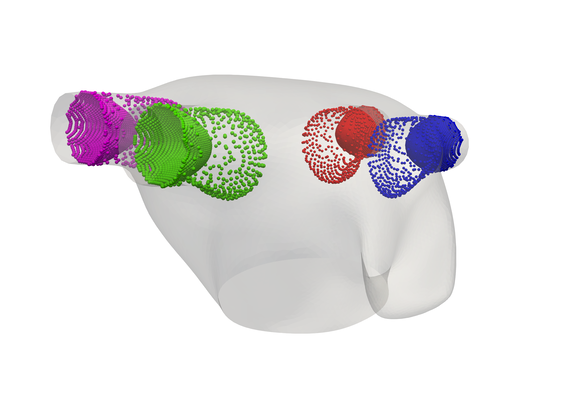}
		\caption{$t = 0.10 $ s}
	\end{subfigure}
	\begin{subfigure}{.325\textwidth}
		\centering
		\includegraphics[trim={1 1 1 1 },clip,width=\textwidth]{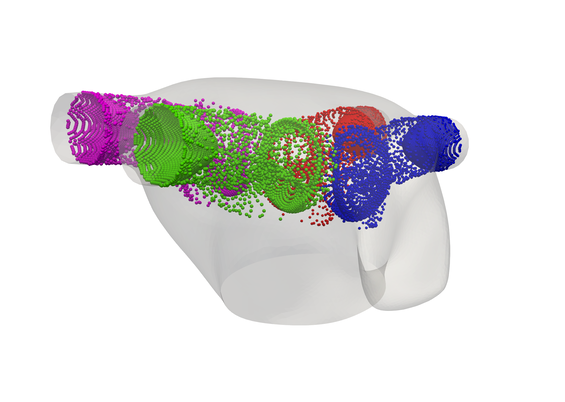}
		\caption{$t = 0.15 $ s}
	\end{subfigure}
	
	\begin{subfigure}{.325\textwidth}
		\centering
		\includegraphics[trim={1 1 1 1 },clip,width=\textwidth]{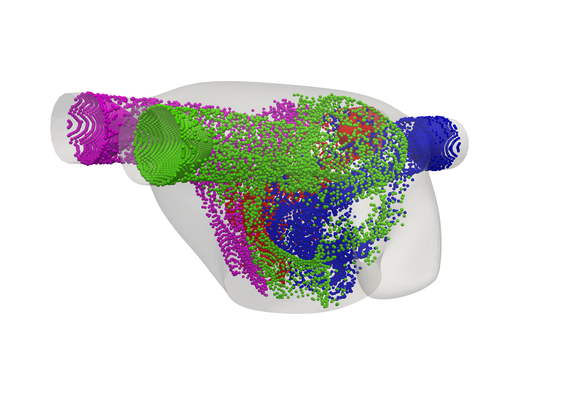}
		\caption{$t = 0.20 $ s}
	\end{subfigure}
	\begin{subfigure}{.325\textwidth}
		\centering
		\includegraphics[trim={1 1 1 1 },clip,width=\textwidth]{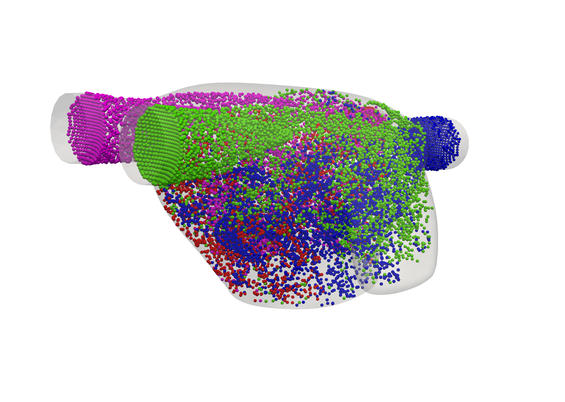}
		\caption{$t = 0.30 $ s}
	\end{subfigure}
	\begin{subfigure}{.325\textwidth}
		\centering
		\includegraphics[trim={1 1 1 1 },clip,width=\textwidth]{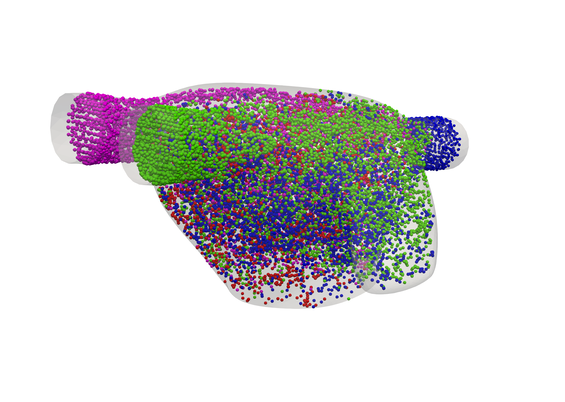}
		\caption{$t = 0.40 $ s}
	\end{subfigure}
	
	\begin{subfigure}{.325\textwidth}
		\centering
		\includegraphics[trim={1 1 1 1 },clip,width=\textwidth]{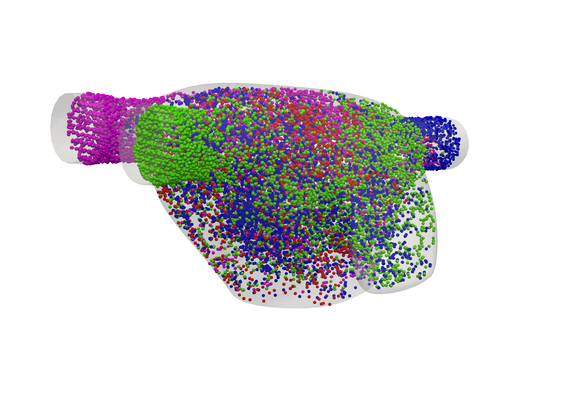}
		\caption{$t = 0.50 $ s}
	\end{subfigure}
	\begin{subfigure}{.325\textwidth}
		\centering
		\includegraphics[trim={1 1 1 1 },clip,width=\textwidth]{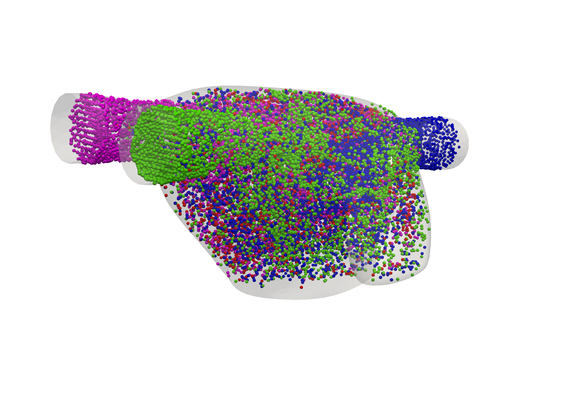}
		\caption{$t = 0.70 $ s}
	\end{subfigure}
	\begin{subfigure}{.325\textwidth}
		\centering
		\includegraphics[trim={1 1 1 1 },clip,width=\textwidth]{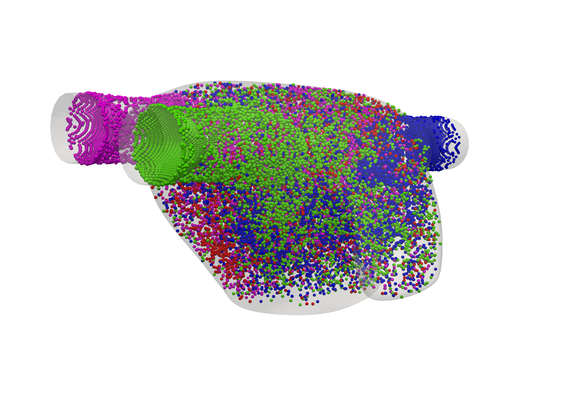}
		\caption{$t = 0.80 $ s}
	\end{subfigure}
	
	\begin{subfigure}{.325\textwidth}
		\centering
		\includegraphics[trim={1 1 1 1 },clip,width=\textwidth]{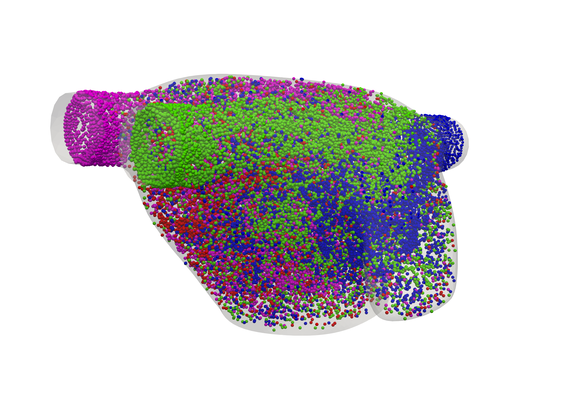}
		\caption{$t = 1.00 $ s}
		\label{frame_1p00}
	\end{subfigure}
	\begin{subfigure}{.325\textwidth}
		\centering
		\includegraphics[trim={1 1 1 1 },clip,width=\textwidth]{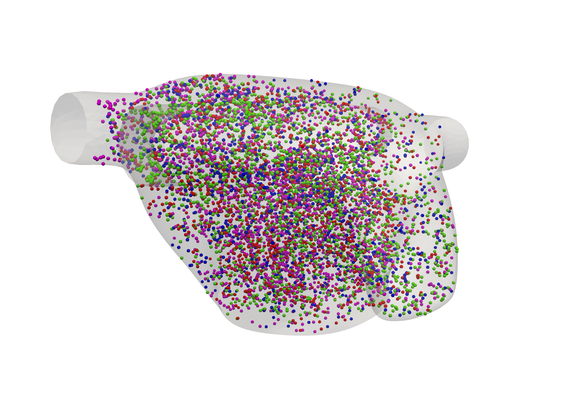}
		\caption{$t = 2.00 $ s}
		\label{frame_2p00}
	\end{subfigure}
	\begin{subfigure}{.325\textwidth}
		\centering
		\includegraphics[trim={1 1 1 1 },clip,width=\textwidth]{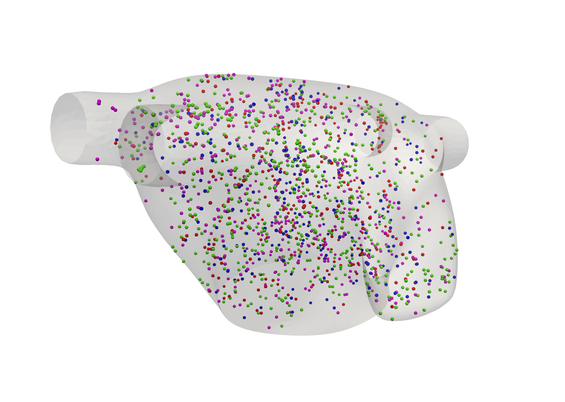}
		\caption{$t = 3.00 $ s}
	\end{subfigure}
	
	\begin{subfigure}{.325\textwidth}
		\centering
		\includegraphics[trim={1 1 1 1 },clip,width=\textwidth]{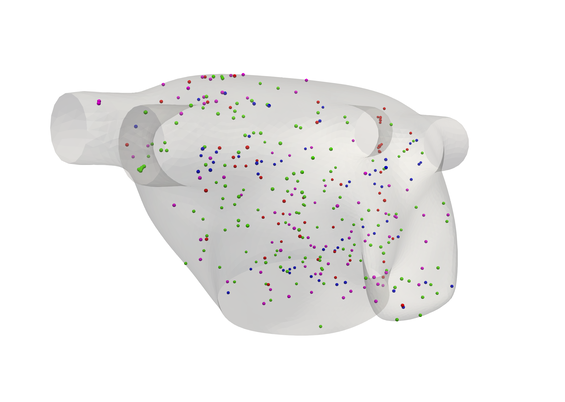}
		\caption{$t = 4.00 $ s}
	\end{subfigure}
	\begin{subfigure}{.325\textwidth}
		\centering
		\includegraphics[trim={1 1 1 1 },clip,width=\textwidth]{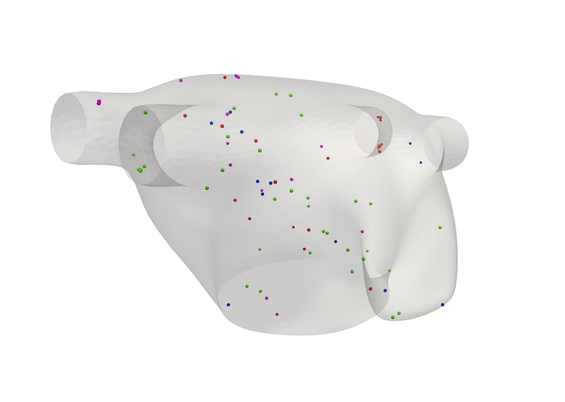}
		\caption{$t = 5.00 $ s}
	\end{subfigure}
	\begin{subfigure}{.325\textwidth}
		\centering
		\includegraphics[trim={1 1 1 1 },clip,width=\textwidth]{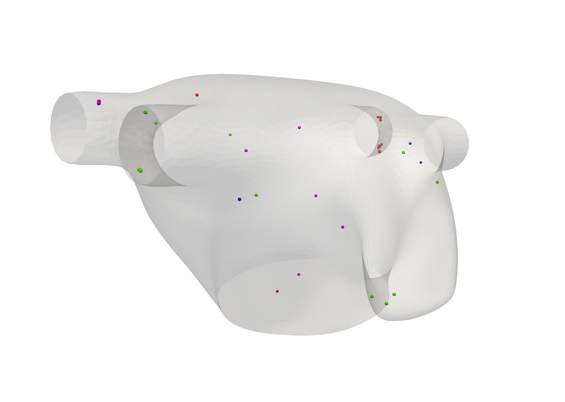}
		\caption{$t = 6.00 $ s}
		\label{frame_6p00}
	\end{subfigure}
	\caption{Reference solution: blood particles in the LA during six heartbeats, injecting particles for the first heartbeat only in a number proportional to the inlet flow rate. From (a) to (j) injection during the first cycle, from (k) to (o) particles remained inside the chamber at the end of each heart-cycle.}
	\label{particles_frames}
\end{figure}

Large values of RRT in the LAA suggest the stasis of blood particles, i.e. the coagulation of blood in low velocity regions, which may result in the formation of blood clots \cite{Gupta_2014, Nordsletten_2018}.  For this reason, we want to count the number of blood particles remaining in the LA at
the end of each heart-cycle. \color{black} Details on the methodology adopted to study the particles are provided in  \ref{appendix_particles}. \color{black} We inject particles in the four PVs every 0.05 s, proportionally to the inlet flow rate. In Figure \ref{injected_particles} we plot the number of particles injected. In Figure \ref{particles_frames} we report snapshots of the blood particles during six heartbeats, injecting in the first heart cycle only and leaving particles in the chamber for the following five cycles. We studied the contribution of particles coming from different veins representing with different colours particles from different inlets. We can observe the formation of four vortex rings coming from the PVs, with four jets  impacting in the middle of the chamber and producing hence a mixing of particles. Particles remain inside the LAA, as also confirmed by large values of RRT previously found. In order to quantify wash-out effects, we stop particles introduction at $t=1.00$ s, counting the number of particles at the end of each cycle. This result is then visualized in Figure \ref{plot_particles} and quantified in Table \ref{table_particles}. The overall number of particles introduced in the chamber during the first heartbeat is 50'471 and, at the end of each cardiac cycle, we report the percentage of particles still inside, showing that, after 5 cycles, in the LA there are the 0.08\% of the total injected particles.

\FloatBarrier

\subsection{Mesh convergence and comparison of SUPG and VMS-LES} \label{comparison_SUPG_VMSLES}
We present a comparison between VMS-LES and SUPG stabilization methods using the meshes $\mathcal{T}_{h_1}$ and $\mathcal{T}_{h_2}$. The results are compared with the reference solution of Section~\ref{RESU_reference}, which we remark has being obtained performing numerical simulation on the mesh  $\mathcal{T}_{h_3}$  with the SUPG method. In Table \ref{table_setup_simulations}, we summarize details of the five numerical simulations performed \color{black} along with the Courant numbers computed as in \cite{AQ_book} using the average mesh element size and the maximum (in space and time) velocity magnitude obtained in the numerical simulations. When the same mesh and the same time step are adopted, different Courant numbers are achieved due to different velocities obtained by the two methods. \color{black} Further features on the meshes adopted are given in Table \ref{table_grids}.  

\begin{table}[!t]
	\centering
	\begin{tabular}{|c|c|c|c|}
		\hline
		 Mesh level & $\Delta t$ [s] & Method & \color{black}Courant numbers\color{black} \\
		\hline
		$\mathcal{T}_{h_1}$ & $1.00 \cdot 10^{-3}$ & SUPG & 0.8281 \\
		$\mathcal{T}_{h_1}$ & $1.00 \cdot 10^{-3}$ & VMS-LES &  0.7666\\
		$\mathcal{T}_{h_2}$ & $2.50 \cdot 10^{-4}$ & SUPG & 0.3425\\
		$\mathcal{T}_{h_2}$ & $2.50 \cdot 10^{-4}$ & VMS-LES & 0.3871 \\
		$\mathcal{T}_{h_3}$ (reference) & $6.25 \cdot 10^{-5}$ & SUPG & 0.1560\\
		\hline
	\end{tabular}
\caption{Details on the numerical simulations used to compare SUPG and VMS-LES stabilization methods in transitional regime. In all the simulations, we adopt $\mathbb{P}1-\mathbb{P}1$ FE spaces, Backward Euler Method as time discretization scheme, and a semi-implicit treatment of the non linear terms.}
\label{table_setup_simulations}
\end{table}

\begin{figure}[!t]
	\centering
	\includegraphics[clip,width=0.9\textwidth]{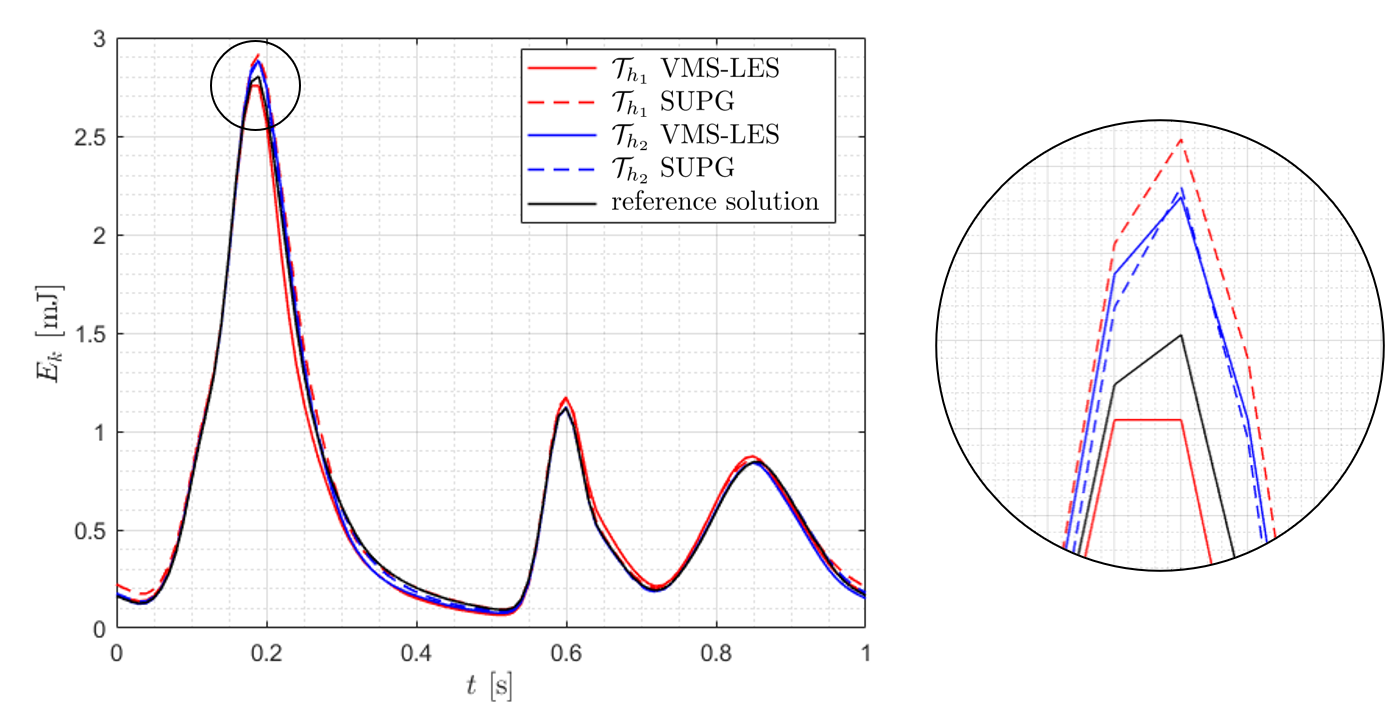}
	\caption{Total kinetic energy $E_{k}(\langle \bm u \rangle)$ using SUPG and VMS-LES methods on meshes $\mathcal{T}_{h_1}$ and $\mathcal{T}_{h_2}$ compared to the reference solution; zoom on the first peak.}
	\label{TKE}
\end{figure}

We define some turbulent indicators obtained integrating suitable variables over the whole domain that are then compared with reference data in order to validate the results of the numerical simulations. 
Specifically, we compute the total kinetic energy of the flow, by using the phase-averaged velocity (defined in Eq. \eqref{phase_averaged}), as
\begin{equation}
E_{k}\left ( \langle \bm u \rangle \right )
=
\frac{1}{2} \rho \int_{\Omega_t} |{\langle \bm u \rangle }|_2^2 d\Omega.
\end{equation}
Moreover, we define the enstrophy of the flow as \cite{UME_enst,LMDK_enst}
\begin{equation}
S\left ( \langle \bm u \rangle \right )=\frac{1}{2} \rho \int_{\Omega_t} |\nabla \times \langle \bm u \rangle |_2^2 d\Omega.
\end{equation} 
The latter is a
fluid dynamics indicator that can be used to identify a transitional flow \cite{UME_enst,LMDK_enst}. 
In Figure \ref{TKE}, we report $E_{k}$ computed on the reference solution and for the meshes $\mathcal{T}_{h_1}$ and $\mathcal{T}_{h_2}$ with SUPG and VMS-LES methods. The total kinetic energy presents three peaks in correspondence of E-wave, A-wave and systolic filling phase. Energy production is observed when high-speed blood flows arrive from the PVs. As the jets impact in the middle of the cardiac chamber, dissipation of the kinetic energy can be appreciated. All the methods and meshes share the same overall behaviour and coherent with the reference solution result. For the mesh $\mathcal{T}_{h_2}$, the results are almost always comparable, whereas small differences can be appreciated in correspondence of the first peak among VMS-LES and SUPG on the mesh $\mathcal{T}_{h_1}$: with a coarse level, we see how the VMS-LES method represents more accurately our reference solution, whereas SUPG overestimates it.  

\begin{figure}[!t]
	\centering
	\includegraphics[clip,width=0.7\textwidth]{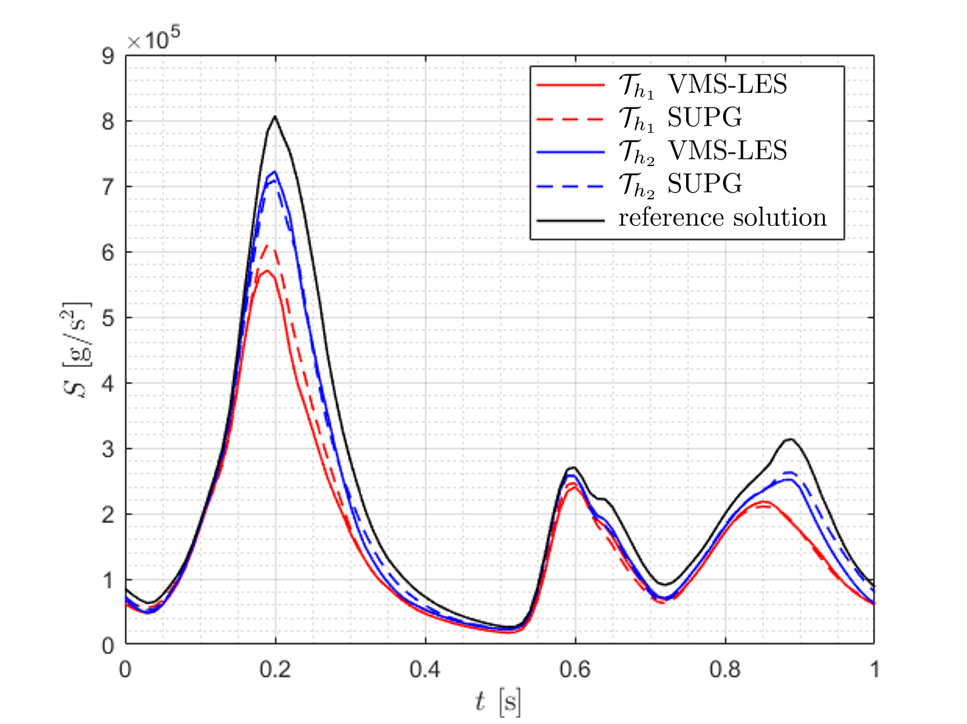}
	\caption{{Enstrophy $S(\langle \bm u \rangle)$ using SUPG and VMS-LES methods on meshes $\mathcal{T}_{h_1}$ and $\mathcal{T}_{h_2}$ compared to the reference solution.}}
	\label{Enstrophy}
\end{figure}

In Figure \ref{Enstrophy}, we show the enstrophy $S$ computed on the reference solution and for the meshes $\mathcal{T}_{h_1}$ and $\mathcal{T}_{h_2}$ with both SUPG and VMS-LES methods. As for the total kinetic energy $E_k$, we observe three main peaks during the hearbeat in correspondence of the production and consequent dissipation of vorticity. The solution largely depends on the underlying mesh and, as it is refined, the solution becomes more accurate and no remarkable differences among the methods can be appreciated. 

\begin{figure}[!t]
	\centering
	\includegraphics[trim={80 1 10 1}, clip,width=\textwidth]{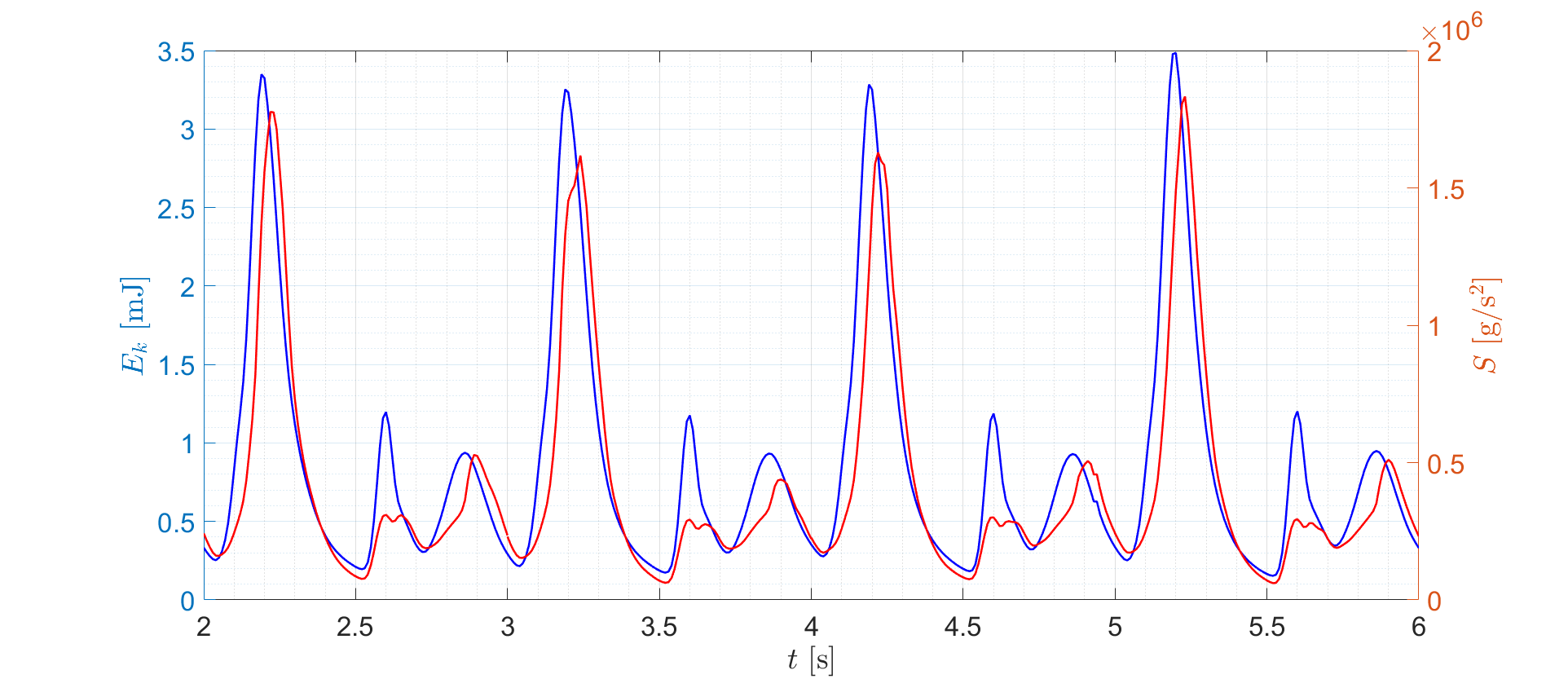}
	\caption{\color{black} Total kinetic energy $E_{k}(\bm u )$ and enstrophy $S( \bm u )$ of the reference solution during four heartbeats.}
	\label{TKE_Enstrophy}
\end{figure}

\begin{figure}[!t]
	\centering
	\includegraphics[clip,width=0.9\textwidth]{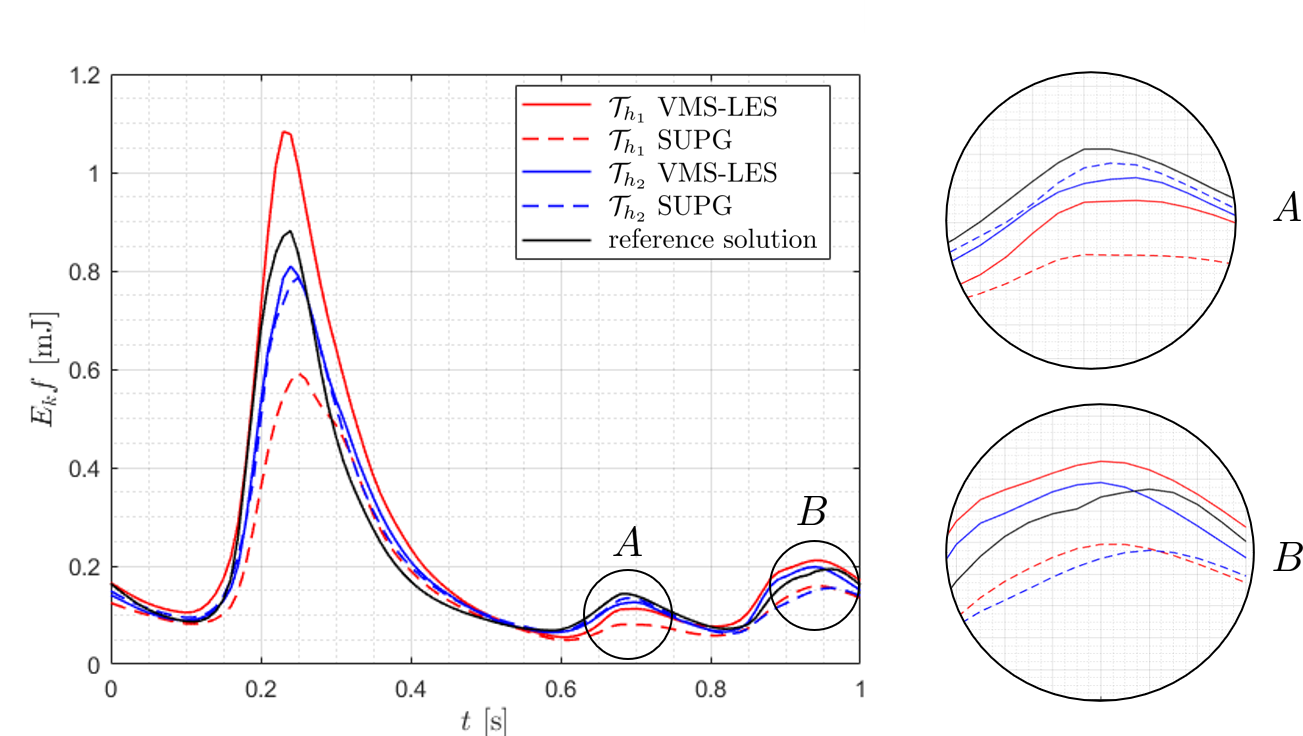}
	\caption{{Fluctuating kinetic energy $E_{kf}(\bm \sigma_{\bm u})$ using SUPG and VMS-LES methods on meshes $\mathcal{T}_{h_1}$ and $\mathcal{T}_{h_2}$ compared to the reference solution; zooms on the second and third peak.}}
	\label{FKE}
\end{figure}
\color{black}
In Figure \ref{TKE_Enstrophy}, we report $E_{k}(\bm u )$ and  $S( \bm u )$ computed on the reference solution during four heartbeats with the instantaneous velocity (not phase-averaged). The plot shows a high variability of the solution among heartbeats in terms of integral quantities: for instance, with respect to the last heartbeat, the E-wave peak shows a relative variation of about 6\% for $E_{k}$ and 12\% for  $S$. Thus, to quantify the large variation of the solution during different heart cycles, \color{black}we introduce the fluctuating kinetic energy of the flow as \cite{CMN_leftheart,TDQ_2d}
\begin{equation}
	E_{kf}\left ( \bm \sigma_{\bm u} \right )
	=
	\frac{1}{2} \rho \int_{\Omega_t} |\bm \sigma_{\bm u} |_2^2 d\Omega,
\end{equation}
being $\bm \sigma_{\bm u} = \left ( \sigma_{u_1}, \sigma_{u_2}, \sigma_{u_3}\right )^T$ a vector containing the standard deviation (i.e. the fluctuations) of each component $k$ of the velocity field with respect to the phase-averaged velocity. Its $k$--th component is defined as
\begin{equation}
	\sigma_{u_k}(\bm x, t) = \sqrt{\text{var}(u_k(\bm x, t))} = \sqrt{\langle {u_k} ^2(\bm x, t) \rangle  - \langle {u_k}(\bm x, t) \rangle ^2 }, \quad k = 1, 2, 3.
	\label{standard_deviation}
\end{equation}
The fluctuating kinetic energy is an important indicator of transition to turbulence but also provides informations on cycle-to-cycle variations. For this reason, it can be seen as one of the most characteristic indicator of transitional flow for hemodynamic applications \cite{CMN_leftheart,TDQ_2d}. 

\begin{figure}[!t]
	\centering
		\begin{subfigure}{.325\textwidth}
		\centering
		\includegraphics[trim={1 1 1 1 },clip,width=\textwidth]{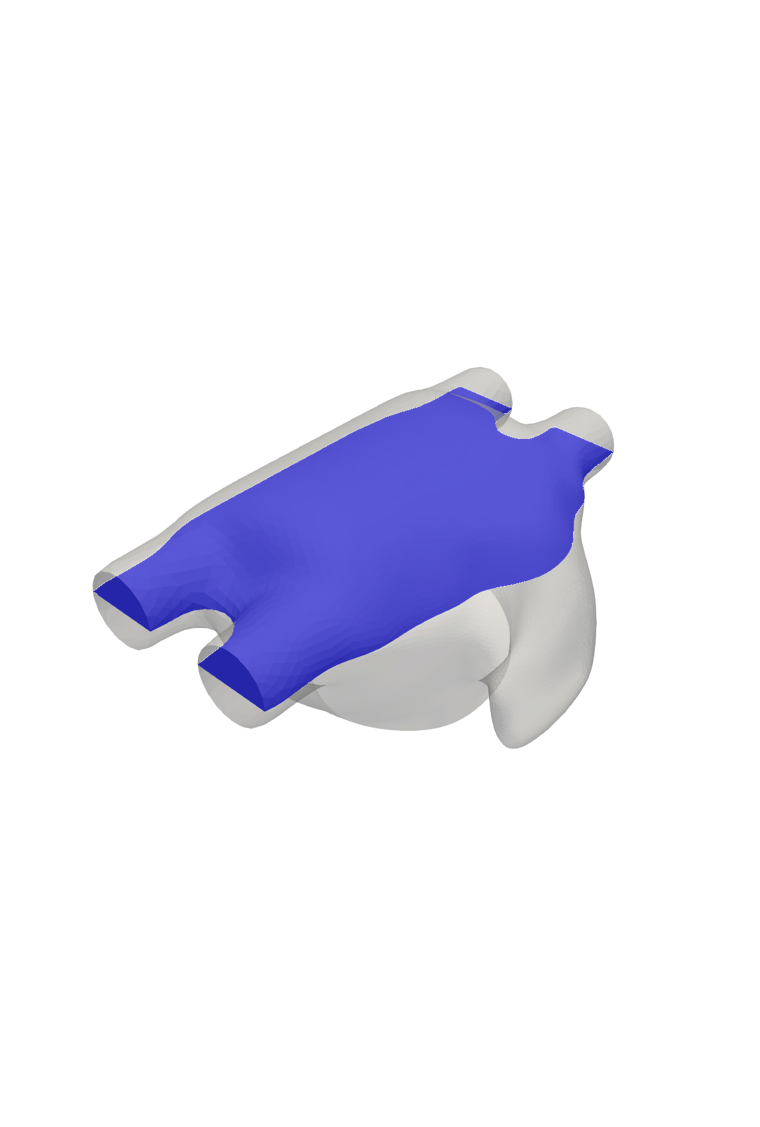}
		\caption{}
		\label{ekinfl_slice_view}
	\end{subfigure}
	\begin{subfigure}{.325\textwidth}
		\centering
		\includegraphics[trim={1 1 1 1 },clip,width=\textwidth]{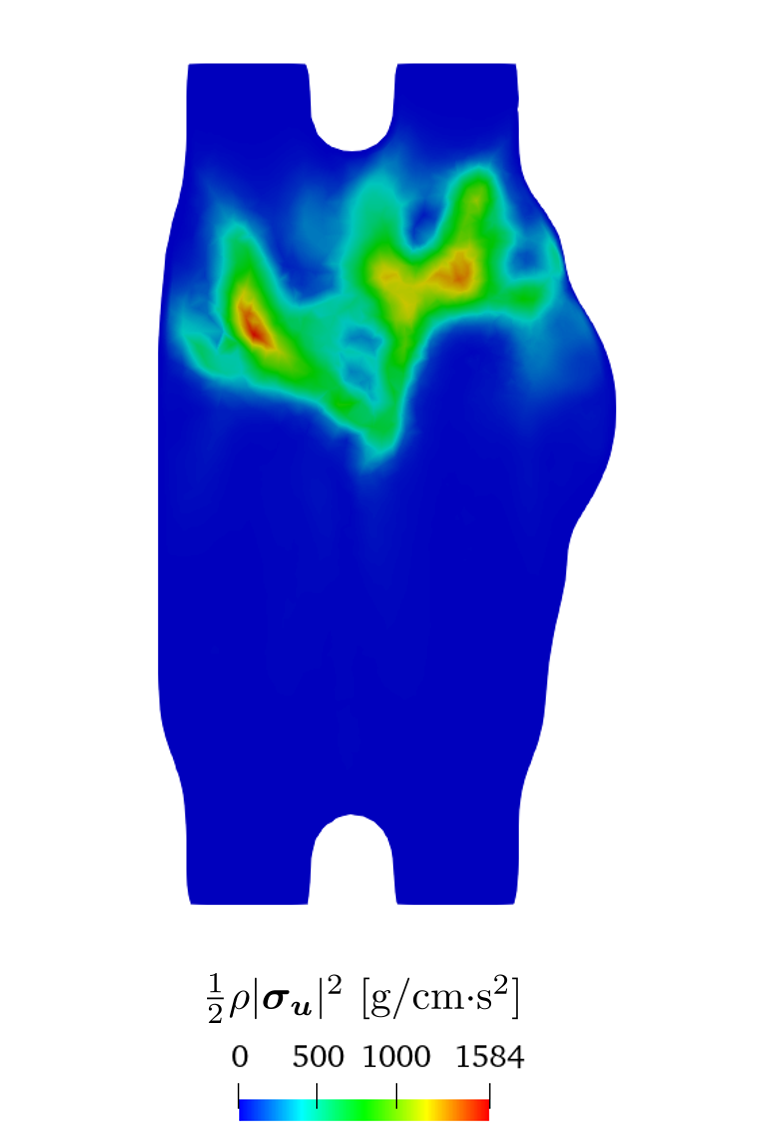}
		\caption{$\mathcal{T}_{h_1}$ SUPG}
		\label{ekinfl_coarseSUPG}
	\end{subfigure}
	\begin{subfigure}{.325\textwidth}
	\centering
	\includegraphics[trim={0 0 0 0 },clip,width=\textwidth]{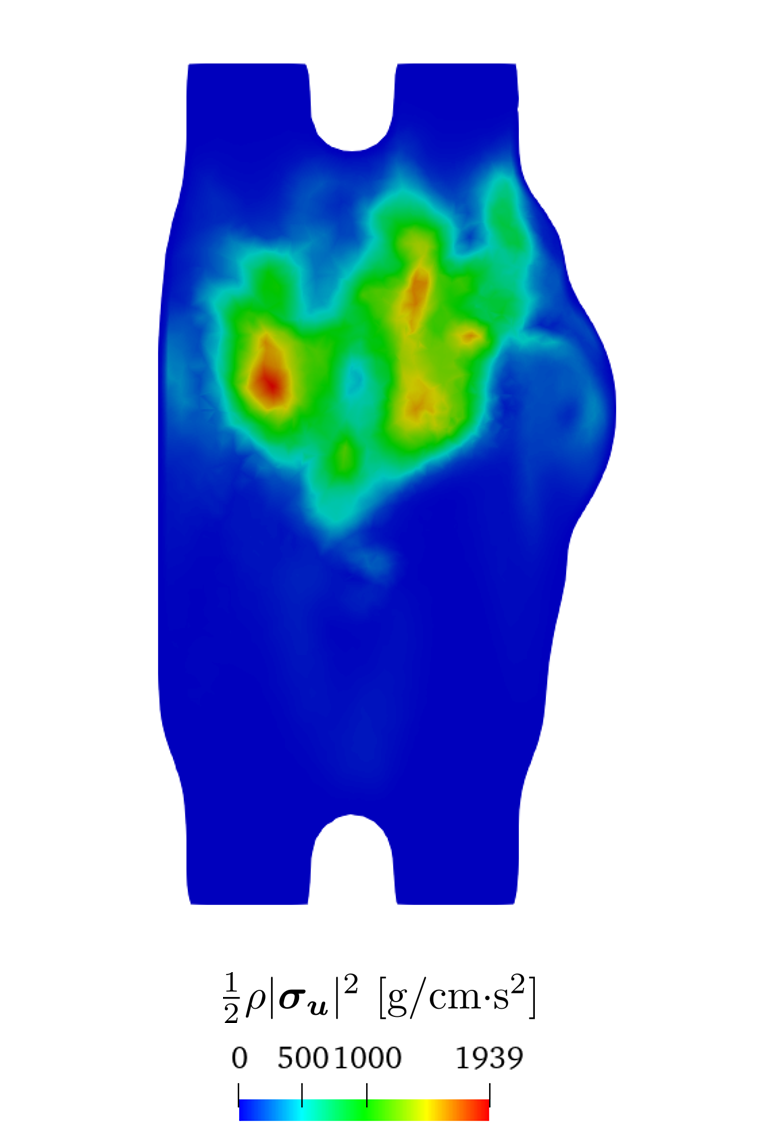}
	\caption{$\mathcal{T}_{h_1}$ VMS-LES}
	\label{ekinfl_coarseVMSLES}
\end{subfigure}

	\begin{subfigure}{.325\textwidth}
	\centering
	\includegraphics[trim={0 0 0 0 },clip,width=\textwidth]{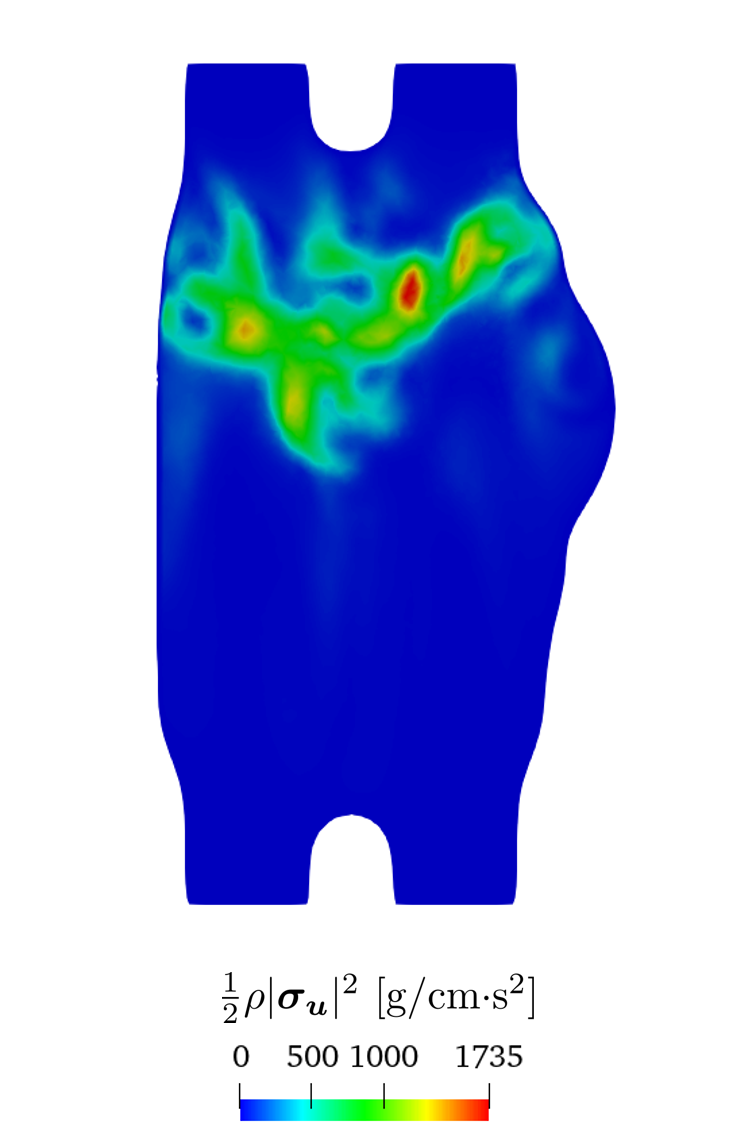}
	\caption{$\mathcal{T}_{h_2}$ SUPG}
	\label{ekinfl_mediumSUPG}
\end{subfigure}
	\begin{subfigure}{.325\textwidth}
	\centering
	\includegraphics[trim={0 0 0 0 },clip,width=\textwidth]{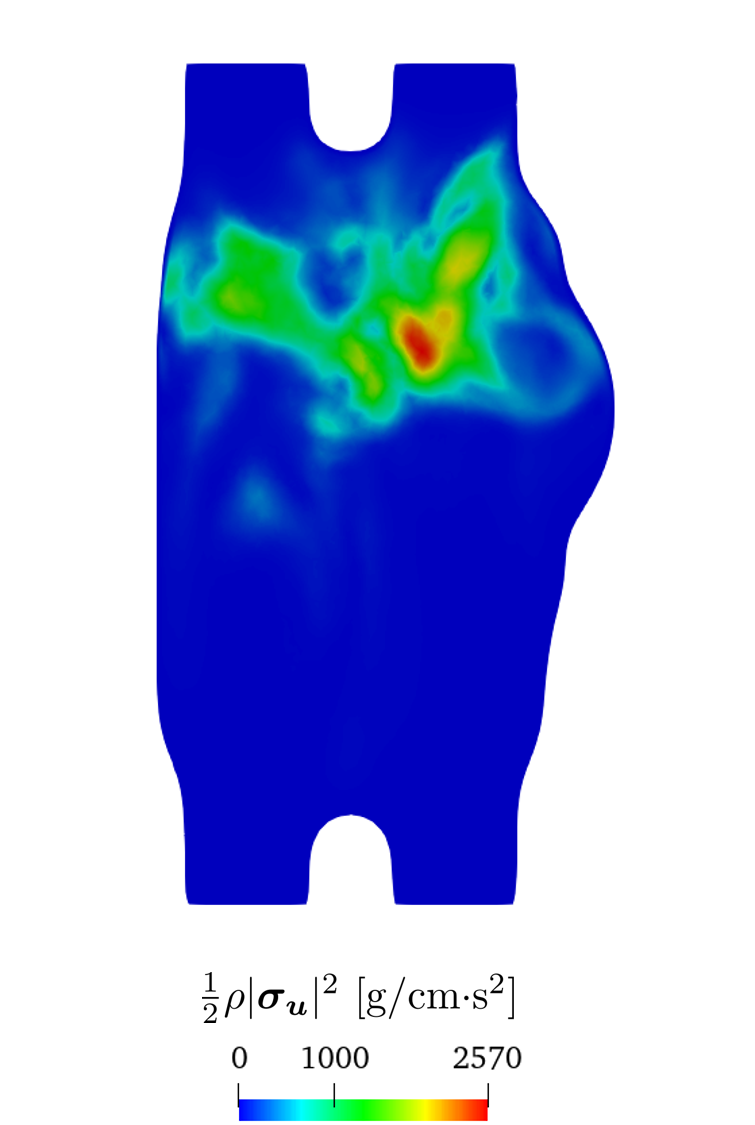}
	\caption{$\mathcal{T}_{h_2}$ VMS-LES}
	\label{ekinfl_mediumVMSLES}
\end{subfigure}
	\begin{subfigure}{.325\textwidth}
	\centering
	\includegraphics[trim={0 0 0 0 },clip,width=\textwidth]{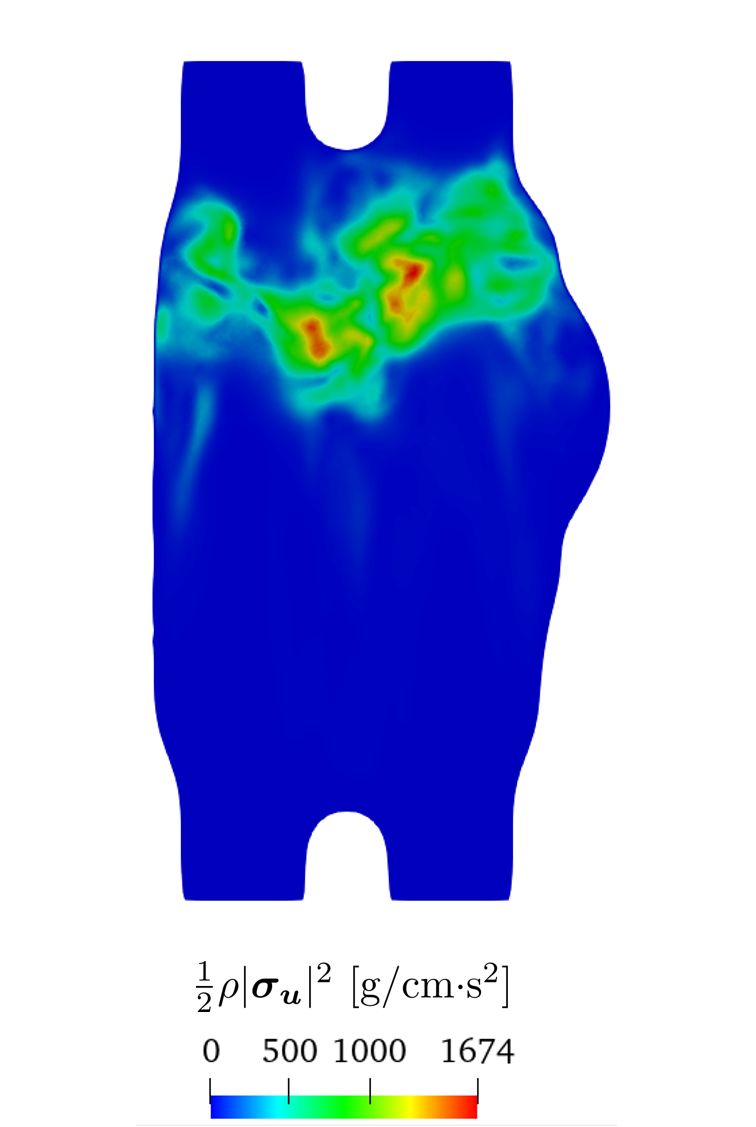}
	\caption{Reference solution}
	\label{ekinfl_REFERENCE}
\end{subfigure}
	\caption{Specific fluctuating kinetic energy $\frac{1}{2}\rho |\bm \sigma _{\bm u}|$ on a slice passing through the four PVs (see (a)) at time $t=0.25$ s using different meshes and methods. Large values of fluctuating velocities are observed in the region of impact among jets. }
	\label{ekinfl_slice}
\end{figure}

In Figure \ref{FKE}, we observe that the $E_{kf}$ shows a peak with a large amplitude immediately after the E-wave. This result suggests that velocity fluctuations $\bm \sigma_{\bm u}$ are higher during the first peak mainly due to small differences in the location of the shear layer and the vortical structures (where velocity gradients are high), as observed also in \cite{CMN_leftheart}. We show this by reporting in Figure \ref{ekinfl_slice} the specific fluctuating kinetic energy ($\frac{1}{2} \rho |\bm \sigma_{\bm u} |^2 $) on a slice passing through the four PVs at time $t=0.25$ s. It can be observed in fact that the largest values are obtained in the area where jets and vortical structures impact. \color{black} Large values of $E_{kf}$ also confirm the large variability among heartbeats previously observed in Figure \ref{TKE_Enstrophy}. 
In terms of meshes and methods, differently from the total kinetic energy $E_k$ and the enstrophy $S$, we found more noticeable differences in the fluctuating kinetic energy $E_{kf}$ as can be seen in Figure \ref{FKE}. 

On the one hand, analysing the results related to the mesh $\mathcal{T}_{h_2}$, we observe that the solutions obtained with SUPG and VMS-LES methods are very similar and both are close to the reference solution, except from the third peak (zoom B) when the VMS-LES method better predicts the reference solution. Moreover, to better quantify these differences, we report in Table \ref{table_e_kin_fl} the minimum, maximum and average discrepancy $e(t)$ achieved in terms of the fluctuating kinetic energy with respect to the reference solution $E_{kf}^\text{REF}(t)$, which shows that for the mesh $\mathcal{T}_{h_2}$, the VMS-LES method has a lower average and maximum discrepancy than SUPG. On the other hand, we found more remarkable differences for the mesh $\mathcal{T}_{h_1}$: as shown in Figure \ref{FKE}, the amplitude of the first peak is highly dependent on the stabilization method adopted; in particular, the VMS-LES solution produces a lower maximum error (that in this case coincides with the E-wave peak) than the one achieved with SUPG, as also confirmed in Table \ref{table_e_kin_fl} in terms of maximum discrepancy. Moreover, the VMS-LES solution on $\mathcal{T}_{h_1}$ better predicts the reference solution than the SUPG on the same mesh (as confirmed also by zooms A and B and then better quantified in Table \ref{table_e_kin_fl} in terms of mean discrepancy). 

\begin{figure}[!t]
	\centering
	\includegraphics[trim={4.5cm 0 4.5cm 0 },clip,width=\textwidth]{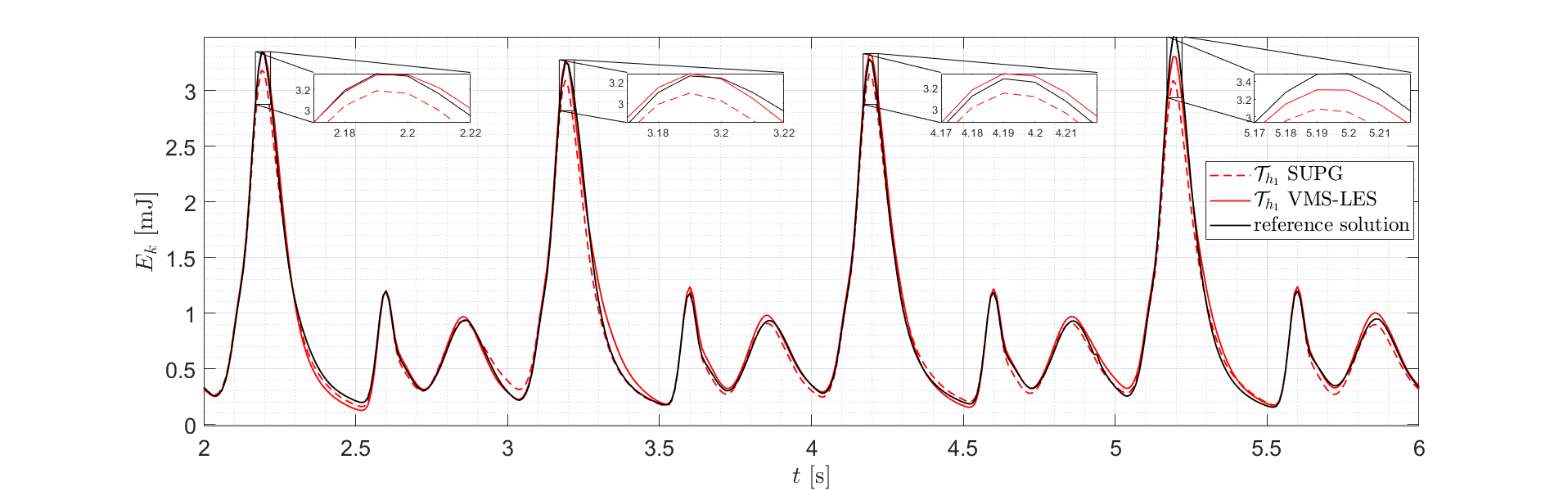}
	\caption{\color{black}Total kinetic energy $E_{k}(\bm u )$ during four heart-cycles with SUPG and VMS-LES on $\mathcal T_{h_1}$ compared with reference solution.}
	\label{EKIN_coarse_4HBs}
\end{figure}

We investigated how the result obtained in terms of difference among fluctuating kinetic energy with coarse meshes ($\mathcal{T}_{h_1}$) may affect the actual flow field: in Figure \ref{EKIN_coarse_4HBs} we report the total kinetic energy computed with the instantaneous velocity $\bm u$ on four heartbeats obtained with the reference solution and with the SUPG and VMS-LES methods on $\mathcal{T}_{h_1}$. We observe that both methods correctly represent $E_k(\bm u)$, but they both loose accuracy during energy dissipation stages, not revealing a clear trend among the two methods during these phases. On the contrary, the main difference between the two methods - and explained also by previous outcomes - is observed during the E-wave energy peak, which always shows that VMS-LES gives more accurate results, while SUPG method underestimates the peaks, revealing also how VMS-LES better predicts E-wave peaks variation from a cycle to another. This result shows that role of the VMS-LES method is more evident in the solution when larger Reynolds numbers are achieved, as during the E-wave, where at the MV section we measured a Reynolds number $Re_\text{MV} \approx 3800 $ and turbulence phenomena are more evident, as thoroughly detailed in Section \ref{RESU_reference}. We believe this justifies the use of additional stabilization terms in Eq. \eqref{discretevmsles} modelling also Reynolds stresses \cite{BCC_vmsles, FD_vmsles} in a LES fashion.

In the literature, we found few works that compare the SUPG and VMS-LES methods, and their conclusions go along different directions. In \cite{Behr_2019}, the stabilized formulations for the fully-implicit log-morphology equation is adopted and applied to the centrifugal ventricular assist device: it is shown that the VMS stabilized formulation has better convergence behaviour and superior stabilization properties compared to the SUPG one. On the other hand, in \cite{Ahmed_2019} the numerical tests carried out revealed that both SUPG and VMS-LES methods exhibit comparable accuracy and they conclude that for their case the SUPG stabilization method is accurate enough. However, in our experience,  we found that, as the mesh is refined, comparable results are achieved with SUPG and VMS-LES methods: the role of the turbulence model hence vanishes as the mesh becomes finer, which is coherent with the standard definition of a LES model. Thus, if sufficiently fine meshes are adopted, the SUPG method is accurate enough to predict transitional flows, and the use of the additional terms modelled by the VMS-LES do not yield additional benefits in terms of accuracy. On the contrary, the two methods show significant differences with coarser meshes in terms of fluctuating kinetic energy: VMS-LES produces a lower discrepancy with respect to the reference solution than with SUPG stabilization method; we also found that VMS-LES better predicts the E-wave kinetic energy peaks and their variations from a heartbeat to another. Thus, the VMS-LES method plays a significant role allowing to better catch transitional effects usually occurring in cardiac haemodynamics and cycle-to-cycle flow variations, fluid properties well described by the fluctuating kinetic energy. For this reason, when relatively coarse meshes are adopted, the use of a standard SUPG stabilization method might be not sufficient to correctly model cardiac haemodynamics.
\color{black}

\begin{table}[!t]
	\centering
	\begin{tabular}{|c|c|c|c|c|}
		\hline
		Mesh level & Method & $\min_t(e(t))$ [mJ] &  $\max_t(e(t))$ [mJ] & $\overline{e(t)}$ [mJ]    \\
		\hline
			$\mathcal{T}_{h_1}$ & SUPG &  0.0001  &  0.3408 &   0.0492  \\
		$\mathcal{T}_{h_1}$ & VMS-LES &  0.0009  &  0.2130  &  0.0438\\
		$\mathcal{T}_{h_2}$  & SUPG & 0.0003   & 0.1826 &   0.0266 \\
		$\mathcal{T}_{h_2}$ & VMS-LES & 0.0002 & 0.1490  &0.0256\\
		\hline
	\end{tabular}
	\caption{\color{black}Minimum, maximum and average discrepancy of fluctuating kinetic energy with respect to the reference solution $e(t)=|E_{kf}(t) - E_{kf}^\text{REF}(t)|$.\color{black}}
	\label{table_e_kin_fl}
\end{table}

\section{Conclusions}
\label{CONCLU}
In this paper, we simulated the hemodynamics of an idealized human LA with the goal of better characterizing and understanding the blood flow behavior in this little explored chamber.
We used the standard SUPG and the VMS-LES stabilization methods to yield stable, discrete formulations of the Navier-Stokes equations approximated by means of the Finite Element method and to take into account of turbulence modelling (in the case of VMS-LES). The ALE formulation with prescribed deformation of the computational domain has been considered in combination with the Navier-Stokes solver. 
We run simulations on a fine mesh for six heartbeats discarding the first two in order to forget the influence of the initial conditions. The result obtained plays the role of our reference solution and it shows some characteristic blood flow features in the LA. The formation of vortex rings from the PVs is the main process occurring in this chamber. The impact of flow jets from the PVs and vortices breakup induce blood mixing and large values of the WSS in the wall nearby the impact regions.  Large variability among the cardiac cycles is observed too. This, in combination with other fluid dynamics indicators, highlights that the blood flow in the LA (in these idealized physiological conditions) is definitely neither laminar nor fully turbulent, but rather transitional. Such transitional nature of the blood flow is also highlighted in the LV cavity as shown e.g. in \cite{CMN_leftheart,TDQ_3d}.
A further indication that we deduce from our study is that the blood velocity profile at the MV section considerably departs from that of a flat or a Poiseuille profile, an assumption that is often, but incorrectly made when simulating LV hemodynamics; this result is coherent with the findings of \cite{TDQ_2d,TDQ_3d}. As a matter of fact, we found that the formation of vortices above the MV section produces low velocities and recirculation regions.  We computed hemodynamic indicators and we deduced that a significant variation of WSS is observed in the bottom of the LAA and on the top of the LA. In particular, in the LAA low velocities and recirculation effects are observed, with consequent high values of RRT which suggests blood stasis. To quantify the latter, we
propose a method useful to compute the number of particles inside a chamber.
Finally, we present a mesh refinement study combined with an analysis on the numerical results obtained by means of the SUPG and VMS-LES methods. We compute total kinetic energy and enstrophy based on the velocity field phase-averaged on four heartbeats, and we compare these results with our reference solution. In terms of these turbulence indicators, we found that, as the mesh is refined, the solution is more accurate using both stabilization methods. In particular, discrepancies among methods become less evident as the mesh becomes finer. Furthermore, we compared our results in terms of fluctuating kinetic energy, based on the standard deviation of the velocity field. This is an important measure in hemodynamic applications since it represents an indicator of cycle-to-cycle variations and also of transitional flow regimes. We found that the position where jets and vortices impact is highly variable from cycle-to-cycle, producing hence high values of fluctuating kinetic energy. 
\color{black}
In terms of the latter, we found that when relatively coarse meshes are adopted, the SUPG shows a larger discrepancy with respect to the reference solution compared to VMS-LES. Moreover, in terms of turbulent kinetic energy computed with the instantaneous velocity, the VMS-LES better catches the E-wave peaks and their cycle-to-cycle variations. To conclude, we found that if sufficiently fine meshes are adopted, the two methods provide almost comparable numerical solutions, hence the SUPG method is accurate enough to correctly catch transitional effects in the LA; on the contrary, larger differences among the two methods are more evident if coarser meshes are adopted and when large Reynolds numbers are achieved during the heartbeat, as during the E-wave. For coarser meshes, the usage of VMS-LES becomes significant in these applications, even if a fully turbulent regime is never met during the heartbeat, allowing to correctly predict transitional effects that typically occur in cardiac flows, as in the haemodynamics of the LA in normal conditions.
\color{black}

\appendix	
\section{}
\label{appendix_particles}
\color{black}
\begin{figure}[!t]
	\centering
	\begin{subfigure}{.59 \textwidth}
		\includegraphics[trim={3 3 3 3 },clip,width=\textwidth]{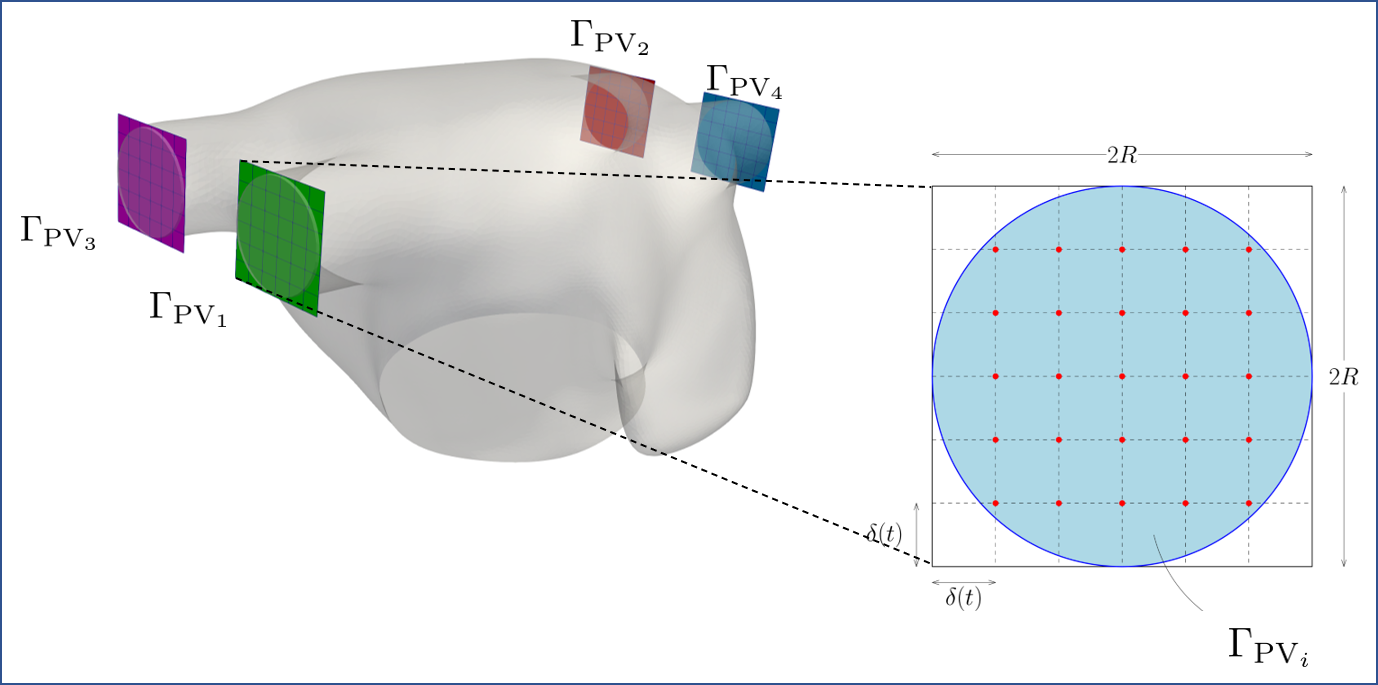}
		\caption{}
		\label{grids_veins}
	\end{subfigure}
	\begin{subfigure}{.39 \textwidth}
		\includegraphics[trim={0 0 25 0 },clip,width=\textwidth]{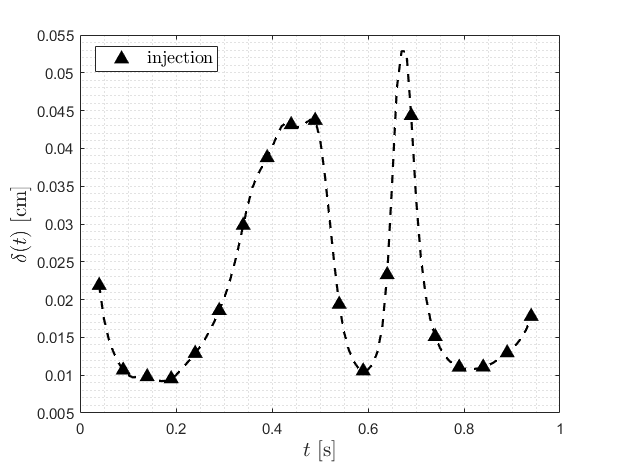}
		\caption{}
		\label{delta_injection}
	\end{subfigure}
	\caption{Details on the methodology adopted to estimate number of particles. Four grids built around the PVs of the LA with a focus on the grid (in red the $n(t)/4$ particles entering in each vein) (left). The behaviour of the grid element size $\delta(t)$ in time, with triangles, we denote the particles injection instants (every 0.05 s) (right).} 
	\label{grids_veins_delta}
\end{figure}

 We inject in the chamber a number of particles $n(t)$ that is proportional to the inlet flux $\Phi_{\text{in}}(t)= \sum\nolimits_{i=1}^{4} \Phi_{\text{PV}_i}(t)$ as:
\begin{equation}
	n(t) = N_p\dfrac{\Phi_{\text{in}}(t)}{\underset{t \in [0, T_\text{HB}]}{\max} \Phi_{\text{in}}(t)},
	\label{numberofparticles}
\end{equation}
being $N_p$ the maximum number particles injected at a specific time.
In particular, we populate the LA with $n(t)$ particles only during the first heartbeat, and, at the end of each heart-cycle, we count how many particles are left inside the LA. As shown in Figure \ref{grids_veins}, particles injection in the PVs is achieved by considering four squares of edge $2R$ (inlet sections diameter) discretized with a cartesian grid with element size $\delta(t)$. In a single PV, at time $t$, the number of particles entering in the LA is $\frac{n(t)}{4}$, which can be approximated as
\begin{equation}
	\frac{n(t)}{4} \approx \left ( \frac{2R}{\delta(t)} - 1 \right )^2.
\end{equation}
Using Eq. \eqref{numberofparticles}, the following expression of time-varying grid element size holds:
\begin{equation}
	\delta(t) = R \left (\sqrt{{N_p}\frac{\Phi_{\text{in}}(t)}{ \underset{t \in [0, T_\text{HB}]}{\max} \Phi_{\text{in}}(t)}} + 2 \right )^{-1},
\end{equation}
which suggests that high flow rates correspond to small grid elements and therefore more particles are introduced. In Figure \ref{delta_injection}, we report the behaviour of the grid element size in time.

\color{black}

\section*{Acknowledgments}
This work has been supported by the ERC Advanced Grant iHEART, ``An Integrated Heart Model for the simulation of the cardiac function'', 2017–2022,  P.I. A. Quarteroni (ERC–2016– ADG, project ID: 740132). 
We gratefully acknowledge the CINECA award under the ISCRA C initiative, for the availability of high performance computing resources and support under the project Computational Fluid Dynamics of Human Heart (CFDHH, P.I. A. Zingaro, 2020-2021). Finally, the authors acknowledge Dr. Davide Forti for fruitful discussions about this topic.

\end{document}